\documentclass[twocolumn]{aastex62}

\graphicspath{{./}{figures/}}

\submitjournal{AJ}

\shorttitle{Angular Diameters of O-type Stars}
\shortauthors{Gordon et al.}

\usepackage{graphicx}
\usepackage{epstopdf}
\usepackage{natbib}
\begin{document}

\title{Angular Sizes and Effective Temperatures of O-type Stars \\ 
 from Optical Interferometry with the CHARA Array}

\correspondingauthor{Kathryn D. Gordon}
\email{kgordon@astro.gsu.edu}

\author{Kathryn D. Gordon}
\affil{Center for High Angular Resolution Astronomy and  
 Department of Physics and Astronomy,\\ 
 Georgia State University, P. O. Box 5060, Atlanta, GA 30302-5060, USA}

\author{Douglas R. Gies}
\affil{Center for High Angular Resolution Astronomy and  
 Department of Physics and Astronomy,\\ 
 Georgia State University, P. O. Box 5060, Atlanta, GA 30302-5060, USA}
\author{Gail H. Schaefer}
\affil{The CHARA Array of Georgia State University,  
 Mount Wilson Observatory, Mount Wilson, CA 91023, USA} 
\author{Daniel Huber}
\affil{Institute for Astronomy, University of Hawai`i, 
 2680 Woodlawn Drive, Honolulu, HI 96822, USA}
\author{Michael Ireland}
\affil{Research School of Astronomy \& Astrophysics,  
 Australian National University, Canberra, ACT 2611, Australia}
\author{D. John Hillier}
\affil{Department of Physics and Astronomy and Pittsburgh Particle Physics,
Astrophysics, and Cosmology Center (PITT PACC), \\
University of Pittsburgh, Pittsburgh, PA 15260, USA}

\begin{abstract}

We present interferometric observations of six O-type stars that were made with the Precision Astronomical Visible Observations (PAVO) beam combiner at the Center for High Angular Resolution Astronomy (CHARA) Array.  
The observations include multiple brackets for three targets, $\lambda$~Ori~A, $\zeta$~Oph, and 10~Lac, but there are only preliminary, single observations of the other three stars, $\xi$~Per, $\alpha$~Cam, and $\zeta$~Ori~A. 
The stellar angular diameters range from 0.55 milliarcsec for $\zeta$~Ori~A down to 0.11 mas for 10~Lac, the smallest star yet resolved with the CHARA Array.  The rotational oblateness of the rapidly rotating star $\zeta$ Oph is directly measured for the first time.  
We assembled ultraviolet to infrared flux measurements for these stars, and then derived angular diameters and reddening estimates using model atmospheres and an effective temperature set by published results from analysis of the line spectrum.  The model-based angular diameters are in good agreement with observed angular diameters.  We also present estimates for the effective temperatures of these stars derived by setting the interferometric angular size and fitting the spectrophotometry. 
\end{abstract}


%%%%%%%%%%%%%%%%%%%%%%%%%%%%%%%%%%%%%%%%%%%%%%%%%%%%%%%%%%%%%%%%%%%%%%%%%%%%%%
\section{Introduction}
%%%%%%%%%%%%%%%%%%%%%%%%%%%%%%%%%%%%%%%%%%%%%%%%%%%%%%%%%%%%%%%%%%%%%%%%%%%%%%

The fundamental properties of \textbf{early O-type stars}, such as mass, radius, and temperature, are the essential parameters required to place stars in the Hertzsprung-Russell diagram for comparison with stellar evolutionary tracks.  The derivation of these quantities is usually reserved to members of eclipsing binary stars in which combined radial velocity and photometric light curve analysis can yield estimates of mass and radius \citep{Torres2010}.
However, some of these systems probably interacted at some point in their evolution, and comparisons of their properties with evolutionary tracks for single stars may not be entirely reliable.  Instead, most of the work on the fundamental parameters of individual \textbf{O-type stars} is based on analysis of the line spectrum, in which the temperature and gravity are derived from the He line ionization balance and the pressure broadening of the hydrogen lines. These studies rely upon detailed
atmospheric models that must take into account non-LTE populations, line-blanketing, and spherical winds. FASTWIND \citep{SantolayaRey1997, Puls2005}, TLUSTY \citep{Hubeny1995}, and CMFGEN \citep{Hillier1998} are all non-LTE models that can be applied to O-star atmospheres and use different methods to deal with the line-blanketing problem. FASTWIND and CMFGEN also take into account the effects of winds (important in more luminous stars) while TLUSTY does not.  These studies are often used to calibrate the MK-classifications of O-stars in terms of temperature and gravity \citep{Martins2005,Holgado2018}.  \textbf{However, although these calibrations serve as a general reference to connect the spectral classifications of a given star and their parameters, one has to use them carefully when studying the properties of a specific individual target of interest.} For example, \citet{SimonDiaz2014} applied spectral line fits using FASTWIND models to investigate the effective temperature relation with O-subtype. They found a significant scatter of effective temperatures among O-stars of similar spectral type that is partially due to the variation of $\log g$ within O-dwarf luminosity class.

\begin{deluxetable*}{cccccccccc}[hb]
\tabletypesize{\footnotesize}
\tablecolumns{10} 
\tablewidth{0pt}
 \tablecaption{Parameters of Target Stars
 \label{tab:physpara}}
 \tablehead{
 \colhead{} & 
 \colhead{Star} & 
 \colhead{HD} & 
 \colhead{Spectral} & 
 \colhead{$V$} &  
 \colhead{$B-V$} & 
 \colhead{$V-K$} & 
 \colhead{$T_{\rm eff}$} & 
 \colhead{$\log g$} &  
 \colhead{$V\sin i$} \\ 
 \colhead{Identifier} & 
 \colhead{Name} & 
 \colhead{Number} & 
 \colhead{Classification} & 
 \colhead{(mag)} & 
 \colhead{(mag)}& 
 \colhead{(mag)} & 
 \colhead{(kK)} & 
 \colhead{(c.g.s.)} & 
 \colhead{(km s$^{-1}$)} 
 } 
 \startdata 
 $a$ & $\xi$ Per & 24912 & O7.5 III(n)((f)) & 4.06 & \phs 0.02 & \phs 0.11 & 34.3$\pm$0.8 & 3.49$\pm$0.12  & 230\\ 
 $b$ & $\alpha$ Cam & 30614 & O9 Ia & 4.29 & \phs 0.05 & \phs 0.05 &  29.4$\pm$1.0 & 3.03$\pm$0.19 & 113\\ 
 $c$ & $\lambda$ Ori A & 36861 & O8 III((f)) & 3.47 & \phs 0.01 & $-0.56$ & 34.5$\pm$0.8 & 3.64$\pm$0.13 &  \phn 59\\
$d$ & $\zeta$ Ori A & 37742 & O9.2 Ib & 1.88 & $-0.11$ & $-0.44$ & 29.5$\pm$1.0 & 3.25$\pm$0.25 &  124\\
$e$ & $\zeta$ Oph & 149757 & O9.2 IVnn & 2.56 & \phs 0.02 & $-0.06$ & 32.1$\pm$1.3 & 3.66$\pm$0.13 &  311\\
$f$ & 10 Lac & 214680 & O9 V & 4.88 & $-0.21$ & $-0.62$ & 35.5$\pm$0.5 & 3.97$\pm$0.08  & \phn 16
 \enddata
 \tablecomments{Effective temperatures and gravities are average values taken from the sources listed in Table \ref{tab:littemps}. Spectral types from the Galactic O-Star Spectroscopic Survey \citep{Sota2011}.
Projected rotational velocities $V \sin i$ are from \citet{SimonDiaz2014b}. }
\end{deluxetable*}
\begin{deluxetable*}{cccccccc}[htb]
\tabletypesize{\footnotesize}
\tablecolumns{10}
\tablewidth{0pt}
\tablecaption{Literature Properties: $T_{\rm eff}$ (kK), $\log g$ (c.g.s.)
\label{tab:littemps}}
\tablehead{
\colhead{Source} & \colhead{$\xi$ Per} & \colhead{ $\alpha$ Cam} & \colhead{$\lambda$ Ori A} & \colhead{$\zeta$ Ori A} & \colhead{$\zeta$ Oph} & \colhead{10 Lac} & \colhead{Code}}
\startdata
 \citet{Bouret2008} & \nodata &  \nodata & \nodata & 29.5, 3.25 &  \nodata &  \nodata & CMFGEN\\
 \citet{Herrero2002} &  \nodata &  \nodata &  \nodata &  \nodata &  \nodata & 35.5, 3.95 & FASTWIND\\
 \citet{Holgado2018} & \nodata & 29.4, 2.90 & 35.2, 3.50 & \nodata & \nodata & 35.2, 3.90 & IACOB-GBAT\\
 \citet{Marcolino2009} &  \nodata &  \nodata &  \nodata &  \nodata & 32.0, 3.60 &  \nodata & TLUSTY\\
\citet{Markova2004} & 34.0, 3.35 & 31.0, 3.19 & 33.6, 3.56 &  \nodata &  \nodata &  \nodata & FASTWIND\\
\citet{Martins2012}    &   \nodata    &  \nodata     &   \nodata    & 29.5, 3.25 &   \nodata    &  \nodata & CMFGEN \\
 \citet{Martins2015} &  34.0, 3.60 &  29.5, 3.25 &  35.0, 3.75 &   \nodata &  31.0, 3.60 & 35.0, 4.05 & CMFGEN \\
 \citet{Martins2017}    & 33.5, 3.50 & \nodata & 35.0, 3.75 &  \nodata & \nodata & \nodata & CMFGEN\\
\citet{Mokiem2005} &  \nodata &  \nodata &  \nodata &  \nodata & 32.1, 3.62 & 36.0, 4.03 & FASTWIND\\
 \citet{Najarro2011} &  \nodata & 28.9, 3.01 & 34.5, 3.70 &  \nodata &  \nodata &  \nodata & CMFGEN\\
 \citet{Puls2006} &  35.0, 3.50 & 29.0, 3.00 & 33.6, 3.56 &  \nodata &  \nodata &  \nodata & FASTWIND\\
 \citet{Repolust2004} & 34.0, 3.50 & 29.0, 2.97 &  \nodata &  \nodata & 32.0, 3.65 &  \nodata & FASTWIND\\
 \citet{Repolust2005} &  \nodata & 29.0, 2.88 &  \nodata &  \nodata & 33.5, 3.85 &  \nodata & FASTWIND\\
\citet{SimonDiaz2006} &  \nodata &  \nodata &  \nodata &  \nodata &  \nodata & 36.0, 3.90 & FASTWIND
\enddata
\end{deluxetable*}

It is important to test the atmosphere models for O-stars because we now rely upon them so completely. The simplest approach in principle is to determine the stellar effective temperature $T_{\rm eff}$ from the relationship between the extinction corrected bolometric flux, angular diameter $\theta$, and effective temperature:
$f_{\rm bol}={1\over 4}\theta^2\sigma T_{\rm eff}^4$.
This method was pioneered by \citet{HanburyBrown1974} who used the Narrabri Stellar Intensity Interferometer (NSII) to measure the angular diameters of 32 stars in the spectral range from O5 to F8 (including two stars considered here). \citet{Code1976} used the early satellite observations of the UV flux of these stars to record their spectral energy distribution (SED) and determine effective temperature.  However, there were only three O-type stars in the sample observed by \citet{HanburyBrown1974}, and the great distances of the O-stars made other high angular resolution measurements impossible until now.  

Furthermore, there are a number of intrinsic problems in using the flux and angular diameter method that were initially discussed by \citet{Abbott1985} and \citet{Hummer1988}. The primary issue is that most O-stars radiate a large fraction of their flux in the extreme UV, at wavelengths smaller than the Lyman limit at 912 \AA . This part of the spectrum is totally obscured by the local interstellar medium in general, and consequently we must rely upon stellar atmosphere models to predict the flux at short wavelengths. The region of spectrum we can observe is in the Rayleigh-Jeans tail where the continuum shape is more or less independent of temperature, so that the intrinsic shape of the UV continuum \citep{Massa1985} and unreddened optical/IR colors \citep{Martins2006} are similar among all O-stars. The second problem is that the flux in the observed part of the spectrum has a gravity dependence.  The size of the Lyman jump in the SED depends on the ionization state of H, and the higher ratio of ionized to neutral H atoms among the supergiants results in a smaller Lyman jump and lower flux at longer wavelengths compared to that of the dwarf O-stars with the same $T_{\rm eff}$ (see Fig.~11 in \citealt{Lanz2003}). Consequently, the star's gravity must be known from other means in order to fit the observed continuum with an appropriate model.  Finally, most O-stars are very distant objects and their flux distributions are altered by interstellar extinction. Thus, fits of their observed flux must deal carefully with the details of dust extinction that can vary with our line of sight through the interstellar medium \citep{Fitzpatrick1999, MaizApellaniz2014}.

\begin{deluxetable*}{ccccccc}[htb]
\tabletypesize{\footnotesize}
\tablecolumns{10}
\tablewidth{0pt}
\tablecaption{Companions of Target Stars
\label{tab:companions}}
\tablehead{
 \colhead{Star} & \colhead{Separation} & \colhead{$\Delta m_V$} & \colhead{$T_{\rm eff}$} & \colhead{$\log g$} & \colhead{Spectrophotometric} & \colhead{Interferometric} \\
 \colhead{Name} & \colhead{(arcsec)} & \colhead{(mag)} & \colhead{(kK)} & \colhead{(c.g.s.)}
 & \colhead{Correction} & \colhead{Correction}}
\startdata
$\xi$ Per &  2.4${^a}$  &  9.8${^a}$  &    \nodata  &   \nodata & N & N  \\
$\lambda$ Ori A &  4.2${^a}$  &     2.3${^a}$    &    25.4${^b}$  &    4.21${^b}$ & Y & N\\
$\zeta$ Ori A & 0.037${^d}$ &  2.2${^d}$  &   26.7${^c}$  &   4.0${^c}$ & Y & Y\\
  & 2.4${^a}$ &  1.9${^a}$  & 31.0$^{e}$ & 3.8$^{e}$ & Y & N\\
10 Lac & 3.6${^a}$ &  9.9${^a}$ &  \nodata &  \nodata & N & N
\enddata
 \tablecomments{Last two columns indicate if it was necessary (Y = yes, N = no) to make corrections to the spectrophotometric or interferometric fits due to the companions.\\
$a$ - \citet{Turner2008}; $b$ - \citet{Lyubimkov2004}; $c$ - Typical values for a B1~IV star \citep{Hummel2013}; $d$ - \citet{Hummel2013}; $e$ - Typical values for a B0~III star \citep{Hummel2013}.}
\end{deluxetable*}

Four decades after the work of \citet{HanburyBrown1974}, we are now able to renew the flux and angular diameter method thanks to the advent of long-baseline optical interferometry. Using measurements made possible with the Center for High Angular Resolution Astronomy (CHARA) Array \citep{Theo2005}, we can determine the angular diameters of the brighter O-stars that have good SED measurements available and that have well-established gravities from detailed studies of their line spectra.  The angular diameter measurements need to be accurate in order to compare the continuum and line results. The relation between the emitted $F_\lambda$
and observed $f_\lambda$ monochromatic flux is
$$ F_\lambda(T_{\rm eff}, \log g, Z) = f_\lambda 10^{0.4 A_\lambda} / (\theta /2)^2$$
where $A_\lambda$ is the extinction (in magnitudes) and $\theta$ is the angular diameter (in radians). The monochromatic flux in the Rayleigh-Jeans tail varies with temperature as $F_\lambda \propto T_{\rm eff}$,
so for a given observed monochromatic flux and extinction, the derived temperature will vary as $T_{\rm eff} \propto \theta^{-2}$ and the fractional uncertainty will be $\triangle T_{\rm eff} / T_{\rm eff} \approx 2 \triangle\theta / \theta$. Thus, accurate, sub-milliarcsecond angular measurements are needed to provide an effective test of the spectral predictions of the current generation of atmospheric models.

Here we present the results of our angular diameter measurements of a sample of six O-type stars. Section 2 discusses the observations, calibration, and reduction methods to obtain the interferometric visibility as a function of baseline and position angle in the sky. In Section 3.1, we show fits of the visibility measurements to obtain a limb darkened angular diameter $\theta_{LD}$, and in Section 3.2 we collect the available spectrophotometry and make fits of the SED as a function of temperature, angular size, and reddening. In Section 4, we gather distance estimates in order to transform the angular diameters into linear radii, and we compare the angular size derived from interferometry with that from model fits of the continuum flux for published values of effective temperature. 

%%%%%%%%%%%%%%%%%%%%%%%%%%%%%%%%%%%%%%%%%%%%%%%%%%%%%%%%%%%%%%%%%%%%%%%
\section{Observations}
%%%%%%%%%%%%%%%%%%%%%%%%%%%%%%%%%%%%%%%%%%%%%%%%%%%%%%%%%%%%%%%%%%%%%%%

The stars in our sample consist of six O spectral type stars ranging from O7.5 to O9. There are two supergiants, two giants, one sub-giant and one dwarf. Table \ref{tab:physpara} lists the parameters of the objects. The effective 
temperature $T_{\rm eff}$ given in column 8 is the average of the recently published 
values that are gathered in Table \ref{tab:littemps}.  The gravity $\log g$ listed
in column 9 is likewise the average from those papers noted in Table \ref{tab:littemps}.

% Calibrators
\begin{deluxetable*}{cccccc}[htb]
\tabletypesize{\footnotesize}
\tablecolumns{10}
\tablewidth{0pt}
\tablecaption{Calibrator Stars
\label{tab:cals}}
\tablehead{
\colhead{Star} & 
\colhead{Spectral} & 
\colhead{Target} & 
\colhead{$\theta$$_{\text{$LD$}}$(JSDC)} & 
\colhead{$\theta$$_{\text{$LD$}}$(CADARS)} & 
\colhead{$\theta$$_{\text{$LD$}}$(Swihart)}\\
\colhead{Name} & 
\colhead{Class.} & 
\colhead{ID} & 
\colhead{(mas)} & 
\colhead{(mas)} & 
\colhead{(mas)}}
\startdata
HD 27777 & B8 V   & $a$ & 0.172$\pm$0.012 & 0.20  & 0.175$\pm$0.037 \\
HD 29646 & A2 V   & $d$ & 0.199$\pm$0.014 &\nodata& 0.249$\pm$0.019 \\
HD 34989 & B1 V   & $c$ & 0.132$\pm$0.009 & 0.10  & 0.118$\pm$0.026 \\
HD 37320 & B8 III & $c$ & 0.158$\pm$0.011 &\nodata& 0.163$\pm$0.035 \\
HD 38831 & A0 Vs  & $b$ & 0.149$\pm$0.010 & 0.17  & \nodata \\
HD 154445& B1 V   & $e$ & 0.280$\pm$0.019 & 0.16  & 0.180$\pm$0.041 \\
HD 204403& B3 V   & $f$ & 0.171$\pm$0.006 & 0.17  & 0.154$\pm$0.048 \\
HD 212978& B2 V   & $f$ & 0.106$\pm$0.004 & 0.10  & \nodata \\
HD 213272& A2 V   & $f$ & 0.175$\pm$0.005 & 0.16  & \nodata \\
\enddata
\tablecomments{Target ID is the star identifier given in Table \ref{tab:physpara}.}
\end{deluxetable*}

Four of the six O stars observed have close companions reported from speckle, adaptive optics, and other studies listed in the Washington Double Star Catalog\footnote{http://ad.usno.navy.mil/wds/}. 
Parameters of the companions are shown in Table \ref{tab:companions}. 
Most of these companions are faint and have separations greater than the interferometric field of view of $1\arcsec$, so their flux has no influence on our interferometric data. 
The exception is the close companion to $\zeta$~Ori~A that is discussed below. 
The companions will, however, contribute to the net flux measured in the spectral 
energy distributions (with the exception of the negligible contributions from 
the very faint companions of $\xi$~Per and 10~Lac), and we deal with this 
complication in Section 3.2 below.

Observations of our targets were made using the PAVO beam combiner \citep{Ireland2008} at the CHARA Array \citep{Theo2005}, located at Mount Wilson Observatory in California. The CHARA Array is an optical interferometer composed of six 1 meter telescopes arranged in a $Y$-shaped configuration. Combinations of the telescopes allow for 15 different usable baselines ranging in length from 34 meters to 331 meters. Combining the longest usable baseline currently available in the world and the operating wavelength range of the PAVO beam combiner (650 - 800 nm), we are able to achieve an extremely high angular resolution for our targets of about 0.2 milli-arcseconds (mas). 

The beams from the Array are fed into the PAVO combiner by a set of lenses and mirrors \citep{Ireland2008,Maestro2012}. The final set are three small movable prisms that guide the beams to the image plane. After being focused in the image plane, the beams pass through a three-hole mask that spatially filters the light (only two beams and two holes are used here). The beams then interfere and produce fringes that are spatially modulated in the pupil plane. Finally, a lenslet array divides the pupil into 16 segments and a prism disperses the fringes in each segment into 23 or more spectral channels. Throughout a night of observing the alignments of the image and pupil plane are checked for each new target, fringes are found, and adjustments made to optimize the longitudinal dispersion corrector position. 

Data for each target were taken using the standard `bracket' method where one bracket is three scans in order of: calibrator - target - calibrator. Observing in brackets allows the visibilities recorded for the target to be properly calibrated and eventually fit to obtain angular diameters. 
We obtained multiple brackets for $\lambda$~Ori~A, $\zeta$~Oph, and 10~Lac, but we were limited to a single bracket with only one calibrator observation for $\xi$~Per, $\alpha$~Cam, and $\zeta$~Ori~A, because of changing sky conditions over the timespan of the bracket. 
The results on these latter three targets are preliminary. 

Calibrators are chosen to be unresolved, single, slowly rotating stars that are close to the target in brightness and position in the sky. The angular diameters were adopted from the Jean-Marie Mariotti Center (JMMC) Stellar Diameter Catalog (JSDC\footnote{http://www.jmmc.fr/jsdc}) using values from Version 1 \citep{Lafrasse2010} for most and from Version 2 \citep{Bourges2014} for the three calibrator stars associated with the 10~Lac observations.
The JSDC diameters are listed in Table \ref{tab:cals} and are estimated by making a polynomial fit of the differential surface brightness of a star as a function of spectral type \citep{Chelli2016}. The JSDC diameters are generally in good agreement with independent estimates from CADARS \citep{Pasinetti2001} and \citet{Swihart2017} (see Table \ref{tab:cals}).  The exception is the calibrator for $\zeta$~Oph, HD~154445, which has a larger size in the JSDC Catalog. 
The angular diameter of HD~154445 is given as 0.16 mas by CADARS, 0.18 mas by \citet{Swihart2017}, and a spectroscopic study by \citet{Lyubimkov2002} implies a diameter of 0.21 mas. These diameter estimates are all significantly smaller than the JSDC value of 0.28 mas, so we adopted the angular diameter found by \citet{Swihart2017} of 0.18 mas for this calibrator. 

Observations of our targets with CHARA were accomplished from 2013 November to 2017 June. For all observations only one baseline, or two telescopes, was used at a time. Observing data, including dates, baselines, and calibrated visibilities, are given in Table \ref{calibvis} (given in full in the electronic version). \textbf{Column 4 of Table \ref{calibvis} lists the spatial frequency of the observation, or baseline divided by wavelength. Columns 5 and 6 give the positions in the $u,v$ plane of spatial frequency for each observation. Column 8 gives the visibility squared and column 9 gives the uncertainty associated with each $V^2$ measurement. } 

The data were reduced using the standard data reduction and fitting pipeline written for use with the PAVO instrument \citep{Ireland2008,Maestro2012}. This pipeline first takes the raw data through a routine that allows for the rejection of bad data frames. This allows the user to make cuts based on S/N values and loss of lock on fringes. The data undergo background subtraction and photon-bias subtraction, with foreground frames, ratio frames, and dark frames taken during a shutter sequence that runs after fringe data are saved. During a foreground frame all shutters are open but fringe tracking is turned off, and during a ratio frame only one shutter is left open at a time. 
%The user can also use the wavelength color coding of the plotted data to make certain that no improper dispersion effects are present. 
The processed data are then sent through a second routine that calculates the projected baseline vectors for each observation, calibrates the brackets, and outputs the calibrated visibilities. This routine allows the user to define estimates and uncertainties on the calibrator diameters, as well as provide a limb-darkening coefficient for a linear limb-darkening law.

We generally assume spherical symmetry for the targets so that the angular diameter is independent of the position angle of the baseline projected on the sky.  However, $\zeta$ Oph is a special case as it is very rapidly rotating with a projected rotational velocity $V\sin i = 348$ km~s$^{-1}$, and it will have a rotationally distorted shape. We observed $\zeta$ Oph on two nearly orthogonal baselines to obtain a range in $(u,v)$ coverage and to ascertain the angular size at different position angles across the star. 

The target $\zeta$~Ori~A has a close companion with an orbit measured by \citet{Hummel2013}, and the predicted angular separation was 33 mas at the time of our observation with a $V$-band magnitude difference of 2.2 mag. This companion is close enough and bright enough to contribute extra light and a periodic modulation to the observed visibility measurements, which, if not accounted for, will make the fitted angular size incorrect.  The companion flux was taken into account in our interferometric and spectrophotometric fitting of the $\zeta$~Ori~A measurements. Using our knowledge of the position of the companion based on the measured orbit, we computed a binary model to fit the visibilities (see Section 4.1).

%\clearpage
\startlongtable
\begin{deluxetable*}{ccccccccc}
%\centering
\tablecaption{Calibrated Visibilities \label{calibvis}}
%\tabletypesize{\scriptsize}
\tablewidth{0pt}
%\rotate
%\tablenum{3}
%\toprule
\tablehead{
\colhead{HD} & 
\colhead{} & 
\colhead{Telescope} & 
\colhead{$10^{-6} B/\lambda$} &
\colhead{$u$} & 
\colhead{$v$} & 
\colhead{Baseline} & 
\colhead{} & 
\colhead{}  \\
\colhead{Number} & 
\colhead{MJD} &
\colhead{Pair} & 
\colhead{(rad$^{-1}$)} &
\colhead{(arcsec$^{-1}$)} & 
\colhead{(arcsec$^{-1}$)} & 
\colhead{(m)} & 
\colhead{$V^2$} & 
\colhead{$\Delta V^2$} 
}
\colnumbers
\startdata
%%%%%%%%%%%%%%
24912	&	57340	&	W1E1	&	391.03	&	-1857.16	&	-380.64	&	310.28	&	0.668	&	0.025	\\
24912	&	57340	&	W1E1	&	394.91	&	-1875.60	&	-384.42	&	310.28	&	0.674	&	0.052	\\
24912	&	57340	&	W1E1	&	398.87	&	-1894.40	&	-388.28	&	310.28	&	0.675	&	0.035	\\
24912	&	57340	&	W1E1	&	402.65	&	-1912.35	&	-391.96	&	310.28	&	0.666	&	0.031	\\
24912	&	57340	&	W1E1	&	406.50	&	-1930.64	&	-395.70	&	310.28	&	0.585	&	0.024	\\
24912	&	57340	&	W1E1	&	410.21	&	-1948.25	&	-399.31	&	310.28	&	0.600	&	0.039	\\
24912	&	57340	&	W1E1	&	413.93	&	-1965.92	&	-402.94	&	310.28	&	0.662	&	0.061	\\
24912	&	57340	&	W1E1	&	417.72	&	-1983.92	&	-406.62	&	310.28	&	0.720	&	0.057	\\
24912	&	57340	&	W1E1	&	421.58	&	-2002.25	&	-410.38	&	310.28	&	0.752	&	0.057	\\
24912	&	57340	&	W1E1	&	425.51	&	-2020.92	&	-414.21	&	310.28	&	0.672	&	0.051	\\
24912	&	57340	&	W1E1	&	429.58	&	-2040.23	&	-418.17	&	310.28	&	0.628	&	0.048	\\
24912	&	57340	&	W1E1	&	433.42	&	-2058.47	&	-421.90	&	310.28	&	0.618	&	0.059	\\
24912	&	57340	&	W1E1	&	437.33	&	-2077.03	&	-425.71	&	310.28	&	0.661	&	0.062	\\
24912	&	57340	&	W1E1	&	441.05	&	-2094.75	&	-429.34	&	310.28	&	0.659	&	0.050	\\
24912	&	57340	&	W1E1	&	444.72	&	-2112.16	&	-432.91	&	310.28	&	0.630	&	0.056	\\
24912	&	57340	&	W1E1	&	448.51	&	-2130.18	&	-436.60	&	310.28	&	0.639	&	0.064	\\
24912	&	57340	&	W1E1	&	452.31	&	-2148.19	&	-440.29	&	310.28	&	0.618	&	0.053	\\
24912	&	57340	&	W1E1	&	456.03	&	-2165.87	&	-443.92	&	310.28	&	0.613	&	0.044	\\
24912	&	57340	&	W1E1	&	459.68	&	-2183.19	&	-447.47	&	310.28	&	0.564	&	0.050	\\
24912	&	57340	&	W1E1	&	463.32	&	-2200.47	&	-451.01	&	310.28	&	0.541	&	0.040	\\
24912	&	57340	&	W1E1	&	467.08	&	-2218.36	&	-454.68	&	310.28	&	0.550	&	0.039	\\
24912	&	57340	&	W1E1	&	470.84	&	-2236.20	&	-458.33	&	310.28	&	0.574	&	0.061	\\
24912	&	57340	&	W1E1	&	474.73	&	-2254.67	&	-462.12	&	310.28	&	0.583	&	0.072	\\
30614	&	56605	&	S1E1	&	375.13	&	313.24	&	1791.48	&	297.66	&	0.685	&	0.091	\\
30614	&	56605	&	S1E1	&	378.85	&	316.35	&	1809.26	&	297.66	&	0.656	&	0.144	\\
30614	&	56605	&	S1E1	&	382.65	&	319.52	&	1827.41	&	297.66	&	0.710	&	0.176	\\
30614	&	56605	&	S1E1	&	386.27	&	322.54	&	1844.72	&	297.66	&	0.750	&	0.194	\\
30614	&	56605	&	S1E1	&	389.97	&	325.63	&	1862.36	&	297.66	&	0.786	&	0.209	\\
30614	&	56605	&	S1E1	&	393.52	&	328.60	&	1879.35	&	297.66	&	0.715	&	0.174	\\
30614	&	56605	&	S1E1	&	397.09	&	331.58	&	1896.40	&	297.66	&	0.590	&	0.070	\\
30614	&	56605	&	S1E1	&	400.73	&	334.62	&	1913.76	&	297.66	&	0.517	&	0.028	\\
30614	&	56605	&	S1E1	&	404.43	&	337.71	&	1931.44	&	297.66	&	0.652	&	0.056	\\
30614	&	56605	&	S1E1	&	408.20	&	340.86	&	1949.45	&	297.66	&	0.610	&	0.099	\\
30614	&	56605	&	S1E1	&	412.10	&	344.11	&	1968.07	&	297.66	&	0.643	&	0.172	\\
30614	&	56605	&	S1E1	&	415.79	&	347.19	&	1985.67	&	297.66	&	0.627	&	0.149	\\
30614	&	56605	&	S1E1	&	419.54	&	350.32	&	2003.58	&	297.66	&	0.709	&	0.178	\\
30614	&	56605	&	S1E1	&	423.12	&	353.31	&	2020.67	&	297.66	&	0.721	&	0.153	\\
30614	&	56605	&	S1E1	&	426.63	&	356.25	&	2037.46	&	297.66	&	0.647	&	0.104	\\
30614	&	56605	&	S1E1	&	430.27	&	359.28	&	2054.84	&	297.66	&	0.631	&	0.088	\\
30614	&	56605	&	S1E1	&	433.91	&	362.32	&	2072.21	&	297.66	&	0.670	&	0.127	\\
30614	&	56605	&	S1E1	&	437.48	&	365.30	&	2089.27	&	297.66	&	0.689	&	0.165	\\
30614	&	56605	&	S1E1	&	440.98	&	368.23	&	2105.98	&	297.66	&	0.763	&	0.231	\\
30614	&	56605	&	S1E1	&	444.47	&	371.14	&	2122.65	&	297.66	&	0.767	&	0.254	\\
30614	&	56605	&	S1E1	&	448.08	&	374.16	&	2139.91	&	297.66	&	0.680	&	0.178	\\
30614	&	56605	&	S1E1	&	451.69	&	377.17	&	2157.12	&	297.66	&	0.613	&	0.094	\\
30614	&	56605	&	S1E1	&	455.42	&	380.28	&	2174.94	&	297.66	&	0.584	&	0.052	\\
36861	&	56757	&	W1E2	&	211.86	&	896.33	&	501.55	&	168.11	&	0.895	&	0.120	\\
36861	&	56757	&	W1E2	&	213.96	&	905.23	&	506.53	&	168.11	&	0.809	&	0.094	\\
36861	&	56757	&	W1E2	&	216.11	&	914.30	&	511.61	&	168.11	&	0.857	&	0.100	\\
36861	&	56757	&	W1E2	&	218.15	&	922.96	&	516.46	&	168.11	&	0.766	&	0.093	\\
36861	&	56757	&	W1E2	&	220.24	&	931.79	&	521.40	&	168.11	&	0.729	&	0.083	\\
36861	&	56757	&	W1E2	&	222.25	&	940.29	&	526.16	&	168.11	&	0.818	&	0.090	\\
36861	&	56757	&	W1E2	&	224.26	&	948.82	&	530.93	&	168.11	&	0.862	&	0.095	\\
36861	&	56757	&	W1E2	&	226.32	&	957.51	&	535.79	&	168.11	&	0.784	&	0.080	\\
36861	&	56757	&	W1E2	&	228.41	&	966.35	&	540.74	&	168.11	&	0.858	&	0.077	\\
36861	&	56757	&	W1E2	&	230.54	&	975.37	&	545.78	&	168.11	&	0.878	&	0.075	\\
36861	&	56757	&	W1E2	&	232.74	&	984.68	&	551.00	&	168.11	&	0.893	&	0.080	\\
36861	&	56757	&	W1E2	&	234.82	&	993.49	&	555.92	&	168.11	&	0.910	&	0.096	\\
36861	&	56757	&	W1E2	&	240.95	&	1019.40	&	570.42	&	168.11	&	0.950	&	0.109	\\
36861	&	56757	&	W1E2	&	243.00	&	1028.10	&	575.29	&	168.11	&	0.943	&	0.097	\\
36861	&	56757	&	W1E2	&	247.07	&	1045.32	&	584.93	&	168.11	&	0.995	&	0.094	\\
36861	&	56757	&	W1E2	&	249.05	&	1053.68	&	589.61	&	168.11	&	0.959	&	0.089	\\
36861	&	56757	&	W1E2	&	251.02	&	1062.02	&	594.27	&	168.11	&	0.844	&	0.082	\\
36861	&	56757	&	W1E2	&	253.06	&	1070.66	&	599.10	&	168.11	&	0.851	&	0.090	\\
36861	&	56757	&	W1E2	&	255.10	&	1079.27	&	603.92	&	168.11	&	0.724	&	0.088	\\
36861	&	56757	&	W1E2	&	257.20	&	1088.18	&	608.91	&	168.11	&	0.585	&	0.067	\\
36861	&	56757	&	W1E2	&	200.02	&	824.68	&	510.16	&	158.72	&	0.872	&	0.116	\\
36861	&	56757	&	W1E2	&	202.00	&	832.86	&	515.23	&	158.72	&	0.877	&	0.099	\\
36861	&	56757	&	W1E2	&	204.03	&	841.21	&	520.39	&	158.72	&	0.927	&	0.105	\\
36861	&	56757	&	W1E2	&	205.96	&	849.18	&	525.32	&	158.72	&	0.798	&	0.096	\\
36861	&	56757	&	W1E2	&	207.93	&	857.30	&	530.35	&	158.72	&	0.765	&	0.085	\\
36861	&	56757	&	W1E2	&	209.83	&	865.12	&	535.18	&	158.72	&	0.877	&	0.094	\\
36861	&	56757	&	W1E2	&	211.73	&	872.97	&	540.04	&	158.72	&	0.888	&	0.096	\\
36861	&	56757	&	W1E2	&	213.67	&	880.96	&	544.98	&	158.72	&	0.832	&	0.084	\\
36861	&	56757	&	W1E2	&	215.65	&	889.10	&	550.02	&	158.72	&	0.895	&	0.076	\\
36861	&	56757	&	W1E2	&	217.66	&	897.39	&	555.15	&	158.72	&	0.901	&	0.072	\\
36861	&	56757	&	W1E2	&	219.74	&	905.97	&	560.45	&	158.72	&	0.907	&	0.077	\\
36861	&	56757	&	W1E2	&	221.70	&	914.07	&	565.46	&	158.72	&	0.961	&	0.097	\\
36861	&	56757	&	W1E2	&	236.99	&	977.12	&	604.47	&	158.72	&	0.922	&	0.084	\\
36861	&	56757	&	W1E2	&	238.92	&	985.07	&	609.38	&	158.72	&	0.936	&	0.095	\\
36861	&	56757	&	W1E2	&	240.84	&	992.99	&	614.28	&	158.72	&	0.815	&	0.095	\\
36861	&	56757	&	W1E2	&	242.83	&	1001.19	&	619.36	&	158.72	&	0.709	&	0.075	\\
36861	&	56757	&	W1E2	&	380.87	&	740.68	&	1691.45	&	302.22	&	0.598	&	0.202	\\
36861	&	56757	&	W1E2	&	384.65	&	748.04	&	1708.24	&	302.22	&	0.652	&	0.215	\\
36861	&	56757	&	W1E2	&	388.51	&	755.54	&	1725.37	&	302.22	&	0.795	&	0.235	\\
36861	&	56757	&	W1E2	&	392.19	&	762.69	&	1741.71	&	302.22	&	0.626	&	0.158	\\
36861	&	56757	&	W1E2	&	395.94	&	769.99	&	1758.37	&	302.22	&	0.545	&	0.105	\\
36861	&	56757	&	W1E2	&	399.55	&	777.01	&	1774.41	&	302.22	&	0.602	&	0.166	\\
36861	&	56757	&	W1E2	&	403.18	&	784.06	&	1790.51	&	302.22	&	0.622	&	0.271	\\
36861	&	56757	&	W1E2	&	406.87	&	791.24	&	1806.90	&	302.22	&	0.617	&	0.313	\\
36861	&	56757	&	W1E2	&	410.63	&	798.55	&	1823.59	&	302.22	&	0.606	&	0.166	\\
36861	&	56757	&	W1E2	&	414.46	&	806.00	&	1840.60	&	302.22	&	0.584	&	0.106	\\
36861	&	56757	&	W1E2	&	418.41	&	813.70	&	1858.18	&	302.22	&	0.599	&	0.107	\\
36861	&	56757	&	W1E2	&	422.16	&	820.97	&	1874.79	&	302.22	&	0.616	&	0.118	\\
36861	&	56757	&	W1E2	&	425.96	&	828.37	&	1891.70	&	302.22	&	0.554	&	0.071	\\
36861	&	56757	&	W1E2	&	429.60	&	835.44	&	1907.84	&	302.22	&	0.534	&	0.060	\\
36861	&	56757	&	W1E2	&	433.17	&	842.39	&	1923.70	&	302.22	&	0.546	&	0.059	\\
36861	&	56757	&	W1E2	&	436.86	&	849.57	&	1940.10	&	302.22	&	0.513	&	0.049	\\
36861	&	56757	&	W1E2	&	440.56	&	856.75	&	1956.51	&	302.22	&	0.496	&	0.048	\\
36861	&	56757	&	W1E2	&	444.18	&	863.80	&	1972.61	&	302.22	&	0.509	&	0.052	\\
36861	&	56757	&	W1E2	&	447.73	&	870.71	&	1988.39	&	302.22	&	0.477	&	0.051	\\
36861	&	56757	&	W1E2	&	451.28	&	877.60	&	2004.13	&	302.22	&	0.508	&	0.059	\\
36861	&	56757	&	W1E2	&	454.95	&	884.74	&	2020.42	&	302.22	&	0.539	&	0.071	\\
36861	&	56757	&	W1E2	&	458.61	&	891.85	&	2036.67	&	302.22	&	0.511	&	0.063	\\
36861	&	56757	&	W1E2	&	462.39	&	899.22	&	2053.49	&	302.22	&	0.486	&	0.060	\\
36861	&	56602	&	E1S1	&	374.97	&	1779.71	&	370.70	&	297.54	&	0.392	&	0.114	\\
36861	&	56602	&	E1S1	&	378.69	&	1797.38	&	374.38	&	297.54	&	0.404	&	0.107	\\
36861	&	56602	&	E1S1	&	382.49	&	1815.40	&	378.13	&	297.54	&	0.464	&	0.141	\\
36861	&	56602	&	E1S1	&	386.11	&	1832.60	&	381.71	&	297.54	&	0.505	&	0.151	\\
36861	&	56602	&	E1S1	&	389.81	&	1850.13	&	385.36	&	297.54	&	0.519	&	0.148	\\
36861	&	56602	&	E1S1	&	393.36	&	1867.01	&	388.88	&	297.54	&	0.453	&	0.093	\\
36861	&	56602	&	E1S1	&	396.93	&	1883.94	&	392.41	&	297.54	&	0.558	&	0.139	\\
36861	&	56602	&	E1S1	&	400.56	&	1901.19	&	396.00	&	297.54	&	0.566	&	0.131	\\
36861	&	56602	&	E1S1	&	404.27	&	1918.75	&	399.66	&	297.54	&	0.549	&	0.121	\\
36861	&	56602	&	E1S1	&	408.04	&	1936.65	&	403.38	&	297.54	&	0.603	&	0.125	\\
36861	&	56602	&	E1S1	&	411.93	&	1955.15	&	407.24	&	297.54	&	0.567	&	0.085	\\
36861	&	56602	&	E1S1	&	415.62	&	1972.63	&	410.88	&	297.54	&	0.636	&	0.092	\\
36861	&	56602	&	E1S1	&	419.36	&	1990.42	&	414.58	&	297.54	&	0.688	&	0.140	\\
36861	&	56602	&	E1S1	&	422.94	&	2007.40	&	418.12	&	297.54	&	0.845	&	0.217	\\
36861	&	56602	&	E1S1	&	426.46	&	2024.08	&	421.60	&	297.54	&	0.853	&	0.213	\\
36861	&	56602	&	E1S1	&	430.09	&	2041.35	&	425.19	&	297.54	&	0.814	&	0.173	\\
36861	&	56602	&	E1S1	&	433.73	&	2058.60	&	428.79	&	297.54	&	0.836	&	0.161	\\
36861	&	56602	&	E1S1	&	437.30	&	2075.55	&	432.32	&	297.54	&	0.715	&	0.110	\\
36861	&	56602	&	E1S1	&	440.80	&	2092.15	&	435.77	&	297.54	&	0.705	&	0.096	\\
36861	&	56602	&	E1S1	&	444.29	&	2108.71	&	439.22	&	297.54	&	0.645	&	0.091	\\
36861	&	56602	&	E1S1	&	447.90	&	2125.85	&	442.79	&	297.54	&	0.631	&	0.093	\\
36861	&	56602	&	E1S1	&	451.50	&	2142.95	&	446.35	&	297.54	&	0.685	&	0.116	\\
36861	&	56602	&	E1S1	&	455.23	&	2160.65	&	450.04	&	297.54	&	0.661	&	0.112	\\
36861	&	56603	&	W1E1	&	391.00	&	1846.64	&	428.21	&	310.26	&	0.492	&	0.144	\\
36861	&	56603	&	W1E1	&	394.89	&	1864.98	&	432.46	&	310.26	&	0.532	&	0.141	\\
36861	&	56603	&	W1E1	&	398.85	&	1883.68	&	436.79	&	310.26	&	0.593	&	0.182	\\
36861	&	56603	&	W1E1	&	402.62	&	1901.52	&	440.93	&	310.26	&	0.673	&	0.202	\\
36861	&	56603	&	W1E1	&	406.47	&	1919.71	&	445.15	&	310.26	&	0.732	&	0.209	\\
36861	&	56603	&	W1E1	&	410.18	&	1937.22	&	449.21	&	310.26	&	0.628	&	0.128	\\
36861	&	56603	&	W1E1	&	413.90	&	1954.79	&	453.28	&	310.26	&	0.714	&	0.179	\\
36861	&	56603	&	W1E1	&	417.69	&	1972.69	&	457.43	&	310.26	&	0.626	&	0.148	\\
36861	&	56603	&	W1E1	&	421.55	&	1990.91	&	461.66	&	310.26	&	0.572	&	0.130	\\
36861	&	56603	&	W1E1	&	425.48	&	2009.48	&	465.97	&	310.26	&	0.685	&	0.146	\\
36861	&	56603	&	W1E1	&	429.55	&	2028.67	&	470.42	&	310.26	&	0.656	&	0.103	\\
36861	&	56603	&	W1E1	&	433.39	&	2046.81	&	474.62	&	310.26	&	0.737	&	0.111	\\
36861	&	56603	&	W1E1	&	437.30	&	2065.27	&	478.90	&	310.26	&	0.704	&	0.147	\\
36861	&	56603	&	W1E1	&	441.03	&	2082.89	&	482.99	&	310.26	&	0.726	&	0.191	\\
36861	&	56603	&	W1E1	&	444.69	&	2100.20	&	487.00	&	310.26	&	0.681	&	0.176	\\
36861	&	56603	&	W1E1	&	448.48	&	2118.11	&	491.16	&	310.26	&	0.699	&	0.153	\\
36861	&	56603	&	W1E1	&	452.28	&	2136.02	&	495.31	&	310.26	&	0.777	&	0.154	\\
36861	&	56603	&	W1E1	&	456.00	&	2153.60	&	499.39	&	310.26	&	0.706	&	0.113	\\
36861	&	56603	&	W1E1	&	459.65	&	2170.83	&	503.38	&	310.26	&	0.729	&	0.103	\\
36861	&	56603	&	W1E1	&	463.28	&	2188.01	&	507.36	&	310.26	&	0.727	&	0.105	\\
36861	&	56603	&	W1E1	&	467.05	&	2205.80	&	511.49	&	310.26	&	0.712	&	0.107	\\
36861	&	56603	&	W1E1	&	470.81	&	2223.54	&	515.60	&	310.26	&	0.679	&	0.118	\\
36861	&	56603	&	W1E1	&	474.70	&	2241.91	&	519.86	&	310.26	&	0.530	&	0.093	\\
36861	&	56603	&	W1E1	&	394.07	&	1856.24	&	452.17	&	312.70	&	0.469	&	0.136	\\
36861	&	56603	&	W1E1	&	397.98	&	1874.67	&	456.65	&	312.70	&	0.486	&	0.128	\\
36861	&	56603	&	W1E1	&	401.97	&	1893.46	&	461.23	&	312.70	&	0.555	&	0.171	\\
36861	&	56603	&	W1E1	&	405.78	&	1911.40	&	465.60	&	312.70	&	0.642	&	0.194	\\
36861	&	56603	&	W1E1	&	409.66	&	1929.68	&	470.06	&	312.70	&	0.687	&	0.197	\\
36861	&	56603	&	W1E1	&	413.40	&	1947.28	&	474.34	&	312.70	&	0.587	&	0.121	\\
36861	&	56603	&	W1E1	&	417.15	&	1964.95	&	478.65	&	312.70	&	0.658	&	0.166	\\
36861	&	56603	&	W1E1	&	420.97	&	1982.94	&	483.03	&	312.70	&	0.542	&	0.131	\\
36861	&	56603	&	W1E1	&	424.86	&	2001.26	&	487.49	&	312.70	&	0.497	&	0.117	\\
36861	&	56603	&	W1E1	&	428.82	&	2019.92	&	492.04	&	312.70	&	0.621	&	0.135	\\
36861	&	56603	&	W1E1	&	432.92	&	2039.21	&	496.74	&	312.70	&	0.577	&	0.093	\\
36861	&	56603	&	W1E1	&	436.79	&	2057.44	&	501.18	&	312.70	&	0.643	&	0.099	\\
36861	&	56603	&	W1E1	&	440.73	&	2076.00	&	505.70	&	312.70	&	0.605	&	0.127	\\
36861	&	56603	&	W1E1	&	444.49	&	2093.71	&	510.01	&	312.70	&	0.575	&	0.154	\\
36861	&	56603	&	W1E1	&	448.18	&	2111.11	&	514.25	&	312.70	&	0.566	&	0.149	\\
36861	&	56603	&	W1E1	&	452.00	&	2129.12	&	518.64	&	312.70	&	0.624	&	0.140	\\
36861	&	56603	&	W1E1	&	455.83	&	2147.12	&	523.02	&	312.70	&	0.674	&	0.137	\\
36861	&	56603	&	W1E1	&	459.58	&	2164.79	&	527.33	&	312.70	&	0.617	&	0.102	\\
36861	&	56603	&	W1E1	&	463.25	&	2182.11	&	531.55	&	312.70	&	0.639	&	0.093	\\
36861	&	56603	&	W1E1	&	466.92	&	2199.38	&	535.75	&	312.70	&	0.646	&	0.095	\\
36861	&	56603	&	W1E1	&	470.72	&	2217.26	&	540.11	&	312.70	&	0.620	&	0.095	\\
36861	&	56603	&	W1E1	&	474.50	&	2235.09	&	544.45	&	312.70	&	0.581	&	0.102	\\
36861	&	56603	&	W1E1	&	478.42	&	2253.56	&	548.95	&	312.70	&	0.424	&	0.077	\\
36861	&	56756	&	W1E2	&	216.72	&	925.29	&	497.76	&	171.97	&	0.985	&	0.273	\\
36861	&	56756	&	W1E2	&	218.87	&	934.47	&	502.70	&	171.97	&	0.841	&	0.143	\\
36861	&	56756	&	W1E2	&	221.06	&	943.84	&	507.74	&	171.97	&	0.846	&	0.146	\\
36861	&	56756	&	W1E2	&	223.16	&	952.78	&	512.55	&	171.97	&	0.841	&	0.181	\\
36861	&	56756	&	W1E2	&	225.29	&	961.90	&	517.46	&	171.97	&	0.797	&	0.167	\\
36861	&	56756	&	W1E2	&	227.35	&	970.67	&	522.18	&	171.97	&	0.755	&	0.124	\\
36861	&	56756	&	W1E2	&	229.41	&	979.48	&	526.91	&	171.97	&	0.779	&	0.104	\\
36861	&	56756	&	W1E2	&	231.51	&	988.44	&	531.74	&	171.97	&	0.911	&	0.140	\\
36861	&	56756	&	W1E2	&	233.65	&	997.57	&	536.65	&	171.97	&	0.933	&	0.135	\\
36861	&	56756	&	W1E2	&	235.83	&	1006.88	&	541.65	&	171.97	&	0.883	&	0.095	\\
36861	&	56756	&	W1E2	&	238.08	&	1016.50	&	546.83	&	171.97	&	0.848	&	0.080	\\
36861	&	56756	&	W1E2	&	240.21	&	1025.58	&	551.72	&	171.97	&	0.902	&	0.086	\\
36861	&	56756	&	W1E2	&	242.38	&	1034.83	&	556.69	&	171.97	&	0.934	&	0.108	\\
36861	&	56756	&	W1E2	&	250.68	&	1070.28	&	575.76	&	171.97	&	0.981	&	0.103	\\
36861	&	56756	&	W1E2	&	252.74	&	1079.09	&	580.50	&	171.97	&	0.885	&	0.080	\\
36861	&	56756	&	W1E2	&	254.76	&	1087.73	&	585.15	&	171.97	&	0.790	&	0.066	\\
36861	&	56756	&	W1E2	&	256.78	&	1096.33	&	589.78	&	171.97	&	0.779	&	0.061	\\
36861	&	56756	&	W1E2	&	258.87	&	1105.25	&	594.57	&	171.97	&	0.754	&	0.062	\\
36861	&	56756	&	W1E2	&	260.95	&	1114.13	&	599.35	&	171.97	&	0.869	&	0.097	\\
36861	&	56756	&	W1E2	&	263.11	&	1123.34	&	604.31	&	171.97	&	0.880	&	0.118	\\
36861	&	56756	&	W1E2	&	203.55	&	846.24	&	507.68	&	161.52	&	1.000	&	0.277	\\
36861	&	56756	&	W1E2	&	205.57	&	854.64	&	512.72	&	161.52	&	0.835	&	0.142	\\
36861	&	56756	&	W1E2	&	207.63	&	863.21	&	517.86	&	161.52	&	0.782	&	0.135	\\
36861	&	56756	&	W1E2	&	209.60	&	871.39	&	522.77	&	161.52	&	0.816	&	0.174	\\
36861	&	56756	&	W1E2	&	211.61	&	879.72	&	527.77	&	161.52	&	0.824	&	0.172	\\
36861	&	56756	&	W1E2	&	213.54	&	887.75	&	532.58	&	161.52	&	0.805	&	0.131	\\
36861	&	56756	&	W1E2	&	215.47	&	895.80	&	537.41	&	161.52	&	0.759	&	0.102	\\
36861	&	56756	&	W1E2	&	217.45	&	904.00	&	542.33	&	161.52	&	0.840	&	0.131	\\
36861	&	56756	&	W1E2	&	219.45	&	912.35	&	547.35	&	161.52	&	0.882	&	0.129	\\
36861	&	56756	&	W1E2	&	221.50	&	920.86	&	552.45	&	161.52	&	0.886	&	0.097	\\
36861	&	56756	&	W1E2	&	223.62	&	929.66	&	557.73	&	161.52	&	0.895	&	0.085	\\
36861	&	56756	&	W1E2	&	225.62	&	937.97	&	562.71	&	161.52	&	0.904	&	0.088	\\
36861	&	56756	&	W1E2	&	227.65	&	946.43	&	567.79	&	161.52	&	0.919	&	0.108	\\
36861	&	56756	&	W1E2	&	235.45	&	978.85	&	587.24	&	161.52	&	0.969	&	0.103	\\
36861	&	56756	&	W1E2	&	237.39	&	986.91	&	592.07	&	161.52	&	0.890	&	0.082	\\
36861	&	56756	&	W1E2	&	239.29	&	994.80	&	596.81	&	161.52	&	0.883	&	0.074	\\
36861	&	56756	&	W1E2	&	241.18	&	1002.68	&	601.53	&	161.52	&	0.865	&	0.069	\\
36861	&	56756	&	W1E2	&	243.14	&	1010.83	&	606.42	&	161.52	&	0.808	&	0.067	\\
36861	&	56756	&	W1E2	&	245.10	&	1018.96	&	611.30	&	161.52	&	0.894	&	0.100	\\
36861	&	56756	&	W1E2	&	247.12	&	1027.38	&	616.35	&	161.52	&	0.842	&	0.114	\\
37742	&	57340	&	W1E1	&	348.55	&	-1631.83	&	-438.96	&	276.58	&	0.056	&	0.022	\\
37742	&	57340	&	W1E1	&	352.01	&	-1648.03	&	-443.32	&	276.58	&	0.054	&	0.020	\\
37742	&	57340	&	W1E1	&	355.54	&	-1664.55	&	-447.77	&	276.58	&	0.057	&	0.021	\\
37742	&	57340	&	W1E1	&	358.91	&	-1680.32	&	-452.01	&	276.58	&	0.049	&	0.016	\\
37742	&	57340	&	W1E1	&	362.34	&	-1696.39	&	-456.33	&	276.58	&	0.045	&	0.015	\\
37742	&	57340	&	W1E1	&	365.65	&	-1711.87	&	-460.49	&	276.58	&	0.045	&	0.015	\\
37742	&	57340	&	W1E1	&	368.97	&	-1727.40	&	-464.67	&	276.58	&	0.052	&	0.019	\\
37742	&	57340	&	W1E1	&	372.34	&	-1743.21	&	-468.92	&	276.58	&	0.062	&	0.024	\\
37742	&	57340	&	W1E1	&	375.78	&	-1759.31	&	-473.26	&	276.58	&	0.056	&	0.021	\\
37742	&	57340	&	W1E1	&	379.29	&	-1775.72	&	-477.67	&	276.58	&	0.044	&	0.013	\\
37742	&	57340	&	W1E1	&	382.91	&	-1792.68	&	-482.23	&	276.58	&	0.032	&	0.008	\\
37742	&	57340	&	W1E1	&	386.34	&	-1808.71	&	-486.54	&	276.58	&	0.030	&	0.008	\\
37742	&	57340	&	W1E1	&	389.82	&	-1825.03	&	-490.93	&	276.58	&	0.036	&	0.010	\\
37742	&	57340	&	W1E1	&	393.15	&	-1840.59	&	-495.12	&	276.58	&	0.037	&	0.009	\\
37742	&	57340	&	W1E1	&	396.41	&	-1855.89	&	-499.24	&	276.58	&	0.035	&	0.007	\\
37742	&	57340	&	W1E1	&	399.79	&	-1871.72	&	-503.49	&	276.58	&	0.033	&	0.007	\\
37742	&	57340	&	W1E1	&	403.17	&	-1887.54	&	-507.75	&	276.58	&	0.029	&	0.006	\\
37742	&	57340	&	W1E1	&	406.49	&	-1903.08	&	-511.93	&	276.58	&	0.026	&	0.005	\\
37742	&	57340	&	W1E1	&	409.74	&	-1918.30	&	-516.03	&	276.58	&	0.024	&	0.005	\\
37742	&	57340	&	W1E1	&	412.99	&	-1933.49	&	-520.11	&	276.58	&	0.026	&	0.006	\\
37742	&	57340	&	W1E1	&	416.34	&	-1949.20	&	-524.34	&	276.58	&	0.023	&	0.006	\\
37742	&	57340	&	W1E1	&	419.69	&	-1964.88	&	-528.55	&	276.58	&	0.016	&	0.004	\\
37742	&	57340	&	W1E1	&	423.16	&	-1981.11	&	-532.92	&	276.58	&	0.018	&	0.005	\\
149757	&	57933	&	S2E2	&	295.58	&	1187.43	&	-802.22	&	234.54	&	0.165	&	0.091	\\
149757	&	57933	&	S2E2	&	298.52	&	1199.21	&	-810.18	&	234.54	&	0.129	&	0.071	\\
149757	&	57933	&	S2E2	&	301.51	&	1211.24	&	-818.31	&	234.54	&	0.114	&	0.082	\\
149757	&	57933	&	S2E2	&	304.36	&	1222.71	&	-826.06	&	234.54	&	0.107	&	0.057	\\
149757	&	57933	&	S2E2	&	307.28	&	1234.41	&	-833.96	&	234.54	&	0.115	&	0.037	\\
149757	&	57933	&	S2E2	&	310.08	&	1245.67	&	-841.57	&	234.54	&	0.148	&	0.027	\\
149757	&	57933	&	S2E2	&	312.89	&	1256.97	&	-849.20	&	234.54	&	0.162	&	0.048	\\
149757	&	57933	&	S2E2	&	315.76	&	1268.47	&	-856.98	&	234.54	&	0.153	&	0.055	\\
149757	&	57933	&	S2E2	&	318.67	&	1280.19	&	-864.89	&	234.54	&	0.110	&	0.064	\\
149757	&	57933	&	S2E2	&	321.65	&	1292.13	&	-872.96	&	234.54	&	0.108	&	0.063	\\
149757	&	57933	&	S2E2	&	324.72	&	1304.48	&	-881.30	&	234.54	&	0.105	&	0.046	\\
149757	&	57933	&	S2E2	&	327.62	&	1316.14	&	-889.18	&	234.54	&	0.123	&	0.051	\\
149757	&	57933	&	S2E2	&	330.58	&	1328.01	&	-897.20	&	234.54	&	0.129	&	0.068	\\
149757	&	57933	&	S2E2	&	333.40	&	1339.34	&	-904.85	&	234.54	&	0.122	&	0.036	\\
149757	&	57933	&	S2E2	&	336.17	&	1350.47	&	-912.37	&	234.54	&	0.130	&	0.042	\\
149757	&	57933	&	S2E2	&	339.03	&	1361.99	&	-920.15	&	234.54	&	0.132	&	0.060	\\
149757	&	57933	&	S2E2	&	341.90	&	1373.50	&	-927.93	&	234.54	&	0.130	&	0.054	\\
149757	&	57933	&	S2E2	&	344.71	&	1384.81	&	-935.57	&	234.54	&	0.141	&	0.043	\\
149757	&	57933	&	S2E2	&	347.47	&	1395.89	&	-943.05	&	234.54	&	0.148	&	0.050	\\
149757	&	57933	&	S2E2	&	350.22	&	1406.93	&	-950.52	&	234.54	&	0.150	&	0.035	\\
149757	&	57933	&	S2E2	&	353.07	&	1418.37	&	-958.24	&	234.54	&	0.152	&	0.027	\\
149757	&	57933	&	S2E2	&	355.91	&	1429.78	&	-965.95	&	234.54	&	0.135	&	0.042	\\
149757	&	57933	&	S2E2	&	358.85	&	1441.59	&	-973.93	&	234.54	&	0.115	&	0.056	\\
149757	&	57933	&	S2E2	&	306.60	&	1228.85	&	-836.35	&	243.29	&	0.218	&	0.047	\\
149757	&	57933	&	S2E2	&	309.65	&	1241.05	&	-844.65	&	243.29	&	0.198	&	0.031	\\
149757	&	57933	&	S2E2	&	312.75	&	1253.50	&	-853.12	&	243.29	&	0.226	&	0.036	\\
149757	&	57933	&	S2E2	&	315.72	&	1265.37	&	-861.20	&	243.29	&	0.239	&	0.039	\\
149757	&	57933	&	S2E2	&	318.73	&	1277.47	&	-869.44	&	243.29	&	0.195	&	0.029	\\
149757	&	57933	&	S2E2	&	321.64	&	1289.13	&	-877.37	&	243.29	&	0.171	&	0.023	\\
149757	&	57933	&	S2E2	&	324.56	&	1300.82	&	-885.33	&	243.29	&	0.172	&	0.029	\\
149757	&	57933	&	S2E2	&	327.53	&	1312.73	&	-893.43	&	243.29	&	0.180	&	0.018	\\
149757	&	57933	&	S2E2	&	330.56	&	1324.86	&	-901.69	&	243.29	&	0.186	&	0.025	\\
149757	&	57933	&	S2E2	&	333.64	&	1337.21	&	-910.10	&	243.29	&	0.202	&	0.030	\\
149757	&	57933	&	S2E2	&	336.83	&	1349.99	&	-918.79	&	243.29	&	0.188	&	0.022	\\
149757	&	57933	&	S2E2	&	339.84	&	1362.06	&	-927.00	&	243.29	&	0.182	&	0.018	\\
149757	&	57933	&	S2E2	&	342.90	&	1374.34	&	-935.37	&	243.29	&	0.179	&	0.027	\\
149757	&	57933	&	S2E2	&	345.83	&	1386.06	&	-943.34	&	243.29	&	0.184	&	0.024	\\
149757	&	57933	&	S2E2	&	348.70	&	1397.59	&	-951.19	&	243.29	&	0.177	&	0.029	\\
149757	&	57933	&	S2E2	&	351.68	&	1409.51	&	-959.30	&	243.29	&	0.137	&	0.032	\\
149757	&	57933	&	S2E2	&	354.65	&	1421.42	&	-967.41	&	243.29	&	0.128	&	0.042	\\
149757	&	57933	&	S2E2	&	357.57	&	1433.12	&	-975.37	&	243.29	&	0.125	&	0.046	\\
149757	&	57933	&	S2E2	&	360.43	&	1444.59	&	-983.17	&	243.29	&	0.124	&	0.046	\\
149757	&	57933	&	S2E2	&	363.28	&	1456.02	&	-990.95	&	243.29	&	0.121	&	0.052	\\
149757	&	57933	&	S2E2	&	366.24	&	1467.85	&	-999.01	&	243.29	&	0.114	&	0.027	\\
149757	&	57933	&	S2E2	&	369.18	&	1479.66	&	-1007.04	&	243.29	&	0.124	&	0.041	\\
149757	&	57933	&	S2E2	&	372.23	&	1491.88	&	-1015.36	&	243.29	&	0.120	&	0.030	\\
149757	&	57933	&	S2E2	&	310.66	&	1239.40	&	-855.75	&	246.51	&	0.218	&	0.088	\\
149757	&	57933	&	S2E2	&	313.75	&	1251.70	&	-864.25	&	246.51	&	0.209	&	0.080	\\
149757	&	57933	&	S2E2	&	316.89	&	1264.26	&	-872.91	&	246.51	&	0.211	&	0.036	\\
149757	&	57933	&	S2E2	&	319.89	&	1276.23	&	-881.18	&	246.51	&	0.226	&	0.058	\\
149757	&	57933	&	S2E2	&	322.95	&	1288.44	&	-889.61	&	246.51	&	0.188	&	0.046	\\
149757	&	57933	&	S2E2	&	325.90	&	1300.19	&	-897.73	&	246.51	&	0.155	&	0.028	\\
149757	&	57933	&	S2E2	&	328.86	&	1311.99	&	-905.87	&	246.51	&	0.136	&	0.023	\\
149757	&	57933	&	S2E2	&	331.87	&	1324.00	&	-914.16	&	246.51	&	0.143	&	0.017	\\
149757	&	57933	&	S2E2	&	334.93	&	1336.23	&	-922.61	&	246.51	&	0.165	&	0.024	\\
149757	&	57933	&	S2E2	&	338.06	&	1348.69	&	-931.21	&	246.51	&	0.174	&	0.053	\\
149757	&	57933	&	S2E2	&	341.28	&	1361.57	&	-940.11	&	246.51	&	0.161	&	0.055	\\
149757	&	57933	&	S2E2	&	344.34	&	1373.75	&	-948.51	&	246.51	&	0.149	&	0.023	\\
149757	&	57933	&	S2E2	&	347.44	&	1386.14	&	-957.07	&	246.51	&	0.144	&	0.015	\\
149757	&	57933	&	S2E2	&	350.40	&	1397.96	&	-965.23	&	246.51	&	0.143	&	0.020	\\
149757	&	57933	&	S2E2	&	353.32	&	1409.58	&	-973.25	&	246.51	&	0.128	&	0.022	\\
149757	&	57933	&	S2E2	&	356.33	&	1421.60	&	-981.55	&	246.51	&	0.116	&	0.018	\\
149757	&	57933	&	S2E2	&	359.34	&	1433.62	&	-989.85	&	246.51	&	0.105	&	0.012	\\
149757	&	57933	&	S2E2	&	362.30	&	1445.42	&	-998.00	&	246.51	&	0.091	&	0.026	\\
149757	&	57933	&	S2E2	&	365.20	&	1456.98	&	-1005.98	&	246.51	&	0.081	&	0.029	\\
149757	&	57933	&	S2E2	&	368.09	&	1468.51	&	-1013.95	&	246.51	&	0.079	&	0.032	\\
149757	&	57933	&	S2E2	&	371.08	&	1480.45	&	-1022.19	&	246.51	&	0.066	&	0.030	\\
149757	&	57933	&	S2E2	&	374.07	&	1492.36	&	-1030.41	&	246.51	&	0.082	&	0.026	\\
149757	&	57933	&	S2E2	&	377.16	&	1504.69	&	-1038.92	&	246.51	&	0.084	&	0.015	\\
149757	&	57933	&	S2E2	&	246.69	&	-612.39	&	-1027.28	&	195.75	&	0.411	&	0.090	\\
149757	&	57933	&	S2E2	&	249.13	&	-618.47	&	-1037.48	&	195.75	&	0.377	&	0.185	\\
149757	&	57933	&	S2E2	&	251.63	&	-624.67	&	-1047.89	&	195.75	&	0.368	&	0.170	\\
149757	&	57933	&	S2E2	&	254.02	&	-630.59	&	-1057.81	&	195.75	&	0.320	&	0.115	\\
149757	&	57933	&	S2E2	&	256.45	&	-636.62	&	-1067.93	&	195.75	&	0.273	&	0.078	\\
149757	&	57933	&	S2E2	&	258.79	&	-642.43	&	-1077.67	&	195.75	&	0.271	&	0.082	\\
149757	&	57933	&	S2E2	&	261.13	&	-648.25	&	-1087.45	&	195.75	&	0.297	&	0.084	\\
149757	&	57933	&	S2E2	&	263.52	&	-654.19	&	-1097.40	&	195.75	&	0.311	&	0.078	\\
149757	&	57933	&	S2E2	&	265.96	&	-660.23	&	-1107.54	&	195.75	&	0.314	&	0.054	\\
149757	&	57933	&	S2E2	&	268.44	&	-666.39	&	-1117.87	&	195.75	&	0.292	&	0.048	\\
149757	&	57933	&	S2E2	&	271.00	&	-672.76	&	-1128.55	&	195.75	&	0.241	&	0.031	\\
149757	&	57933	&	S2E2	&	273.43	&	-678.77	&	-1138.64	&	195.75	&	0.233	&	0.032	\\
149757	&	57933	&	S2E2	&	275.89	&	-684.89	&	-1148.91	&	195.75	&	0.255	&	0.044	\\
149757	&	57933	&	S2E2	&	278.24	&	-690.73	&	-1158.71	&	195.75	&	0.268	&	0.049	\\
149757	&	57933	&	S2E2	&	280.56	&	-696.48	&	-1168.34	&	195.75	&	0.235	&	0.038	\\
149757	&	57933	&	S2E2	&	282.95	&	-702.42	&	-1178.30	&	195.75	&	0.205	&	0.030	\\
149757	&	57933	&	S2E2	&	285.34	&	-708.35	&	-1188.27	&	195.75	&	0.184	&	0.028	\\
149757	&	57933	&	S2E2	&	287.69	&	-714.18	&	-1198.05	&	195.75	&	0.160	&	0.030	\\
149757	&	57933	&	S2E2	&	289.99	&	-719.90	&	-1207.63	&	195.75	&	0.143	&	0.040	\\
149757	&	57933	&	S2E2	&	292.29	&	-725.60	&	-1217.19	&	195.75	&	0.109	&	0.040	\\
149757	&	57933	&	S2E2	&	294.66	&	-731.49	&	-1227.08	&	195.75	&	0.153	&	0.052	\\
149757	&	57933	&	S2E2	&	297.03	&	-737.38	&	-1236.95	&	195.75	&	0.170	&	0.034	\\
149757	&	57933	&	S2E2	&	299.49	&	-743.47	&	-1247.17	&	195.75	&	0.168	&	0.026	\\
149757	&	57573	&	S2W1	&	235.14	&	-519.63	&	-1014.65	&	186.58	&	0.310	&	0.082	\\
149757	&	57573	&	S2W1	&	237.47	&	-524.79	&	-1024.72	&	186.58	&	0.326	&	0.094	\\
149757	&	57573	&	S2W1	&	239.85	&	-530.06	&	-1035.00	&	186.58	&	0.317	&	0.098	\\
149757	&	57573	&	S2W1	&	242.12	&	-535.08	&	-1044.80	&	186.58	&	0.318	&	0.062	\\
149757	&	57573	&	S2W1	&	244.44	&	-540.19	&	-1054.79	&	186.58	&	0.276	&	0.068	\\
149757	&	57573	&	S2W1	&	246.67	&	-545.12	&	-1064.41	&	186.58	&	0.245	&	0.063	\\
149757	&	57573	&	S2W1	&	248.91	&	-550.07	&	-1074.07	&	186.58	&	0.260	&	0.054	\\
149757	&	57573	&	S2W1	&	251.18	&	-555.10	&	-1083.90	&	186.58	&	0.264	&	0.051	\\
149757	&	57573	&	S2W1	&	253.51	&	-560.23	&	-1093.92	&	186.58	&	0.276	&	0.037	\\
149757	&	57573	&	S2W1	&	255.87	&	-565.46	&	-1104.12	&	186.58	&	0.254	&	0.039	\\
149757	&	57573	&	S2W1	&	258.31	&	-570.86	&	-1114.67	&	186.58	&	0.230	&	0.027	\\
149757	&	57573	&	S2W1	&	260.62	&	-575.96	&	-1124.63	&	186.58	&	0.220	&	0.028	\\
149757	&	57573	&	S2W1	&	262.97	&	-581.16	&	-1134.78	&	186.58	&	0.223	&	0.048	\\
149757	&	57573	&	S2W1	&	265.22	&	-586.11	&	-1144.45	&	186.58	&	0.233	&	0.043	\\
149757	&	57573	&	S2W1	&	267.42	&	-590.98	&	-1153.97	&	186.58	&	0.211	&	0.037	\\
149757	&	57573	&	S2W1	&	269.70	&	-596.02	&	-1163.81	&	186.58	&	0.181	&	0.027	\\
149757	&	57573	&	S2W1	&	271.98	&	-601.06	&	-1173.65	&	186.58	&	0.147	&	0.025	\\
149757	&	57573	&	S2W1	&	274.22	&	-606.01	&	-1183.31	&	186.58	&	0.134	&	0.027	\\
149757	&	57573	&	S2W1	&	276.41	&	-610.86	&	-1192.77	&	186.58	&	0.108	&	0.035	\\
149757	&	57573	&	S2W1	&	278.60	&	-615.69	&	-1202.21	&	186.58	&	0.107	&	0.062	\\
149757	&	57573	&	S2W1	&	280.87	&	-620.70	&	-1211.99	&	186.58	&	0.129	&	0.040	\\
149757	&	57573	&	S2W1	&	283.13	&	-625.69	&	-1221.73	&	186.58	&	0.155	&	0.029	\\
149757	&	57573	&	S2W1	&	285.47	&	-630.86	&	-1231.83	&	186.58	&	0.145	&	0.023	\\
149757	&	57573	&	S2W1	&	226.83	&	-443.23	&	-1006.41	&	179.99	&	0.349	&	0.059	\\
149757	&	57573	&	S2W1	&	229.08	&	-447.63	&	-1016.40	&	179.99	&	0.399	&	0.067	\\
149757	&	57573	&	S2W1	&	231.38	&	-452.12	&	-1026.59	&	179.99	&	0.406	&	0.081	\\
149757	&	57573	&	S2W1	&	233.57	&	-456.40	&	-1036.32	&	179.99	&	0.431	&	0.093	\\
149757	&	57573	&	S2W1	&	235.80	&	-460.77	&	-1046.23	&	179.99	&	0.351	&	0.088	\\
149757	&	57573	&	S2W1	&	237.95	&	-464.97	&	-1055.77	&	179.99	&	0.273	&	0.071	\\
149757	&	57573	&	S2W1	&	240.11	&	-469.19	&	-1065.35	&	179.99	&	0.290	&	0.048	\\
149757	&	57573	&	S2W1	&	242.31	&	-473.48	&	-1075.10	&	179.99	&	0.340	&	0.085	\\
149757	&	57573	&	S2W1	&	244.55	&	-477.86	&	-1085.03	&	179.99	&	0.359	&	0.052	\\
149757	&	57573	&	S2W1	&	246.83	&	-482.31	&	-1095.15	&	179.99	&	0.315	&	0.039	\\
149757	&	57573	&	S2W1	&	249.19	&	-486.92	&	-1105.61	&	179.99	&	0.282	&	0.034	\\
149757	&	57573	&	S2W1	&	251.41	&	-491.28	&	-1115.50	&	179.99	&	0.250	&	0.042	\\
149757	&	57573	&	S2W1	&	253.68	&	-495.71	&	-1125.56	&	179.99	&	0.271	&	0.069	\\
149757	&	57573	&	S2W1	&	255.85	&	-499.93	&	-1135.16	&	179.99	&	0.285	&	0.045	\\
149757	&	57573	&	S2W1	&	257.97	&	-504.09	&	-1144.60	&	179.99	&	0.258	&	0.038	\\
149757	&	57573	&	S2W1	&	260.17	&	-508.39	&	-1154.36	&	179.99	&	0.226	&	0.033	\\
149757	&	57573	&	S2W1	&	262.37	&	-512.69	&	-1164.12	&	179.99	&	0.186	&	0.041	\\
149757	&	57573	&	S2W1	&	264.53	&	-516.91	&	-1173.70	&	179.99	&	0.161	&	0.030	\\
149757	&	57573	&	S2W1	&	266.65	&	-521.04	&	-1183.09	&	179.99	&	0.151	&	0.042	\\
149757	&	57573	&	S2W1	&	268.76	&	-525.17	&	-1192.45	&	179.99	&	0.144	&	0.087	\\
149757	&	57573	&	S2W1	&	270.94	&	-529.44	&	-1202.15	&	179.99	&	0.156	&	0.058	\\
149757	&	57573	&	S2W1	&	273.12	&	-533.69	&	-1211.81	&	179.99	&	0.173	&	0.073	\\
149757	&	57573	&	S2W1	&	275.38	&	-538.10	&	-1221.83	&	179.99	&	0.183	&	0.032	\\
149757	&	57573	&	S2W1	&	216.74	&	-331.19	&	-997.22	&	171.98	&	0.376	&	0.064	\\
149757	&	57573	&	S2W1	&	218.89	&	-334.48	&	-1007.12	&	171.98	&	0.381	&	0.075	\\
149757	&	57573	&	S2W1	&	221.09	&	-337.83	&	-1017.22	&	171.98	&	0.413	&	0.091	\\
149757	&	57573	&	S2W1	&	223.18	&	-341.03	&	-1026.86	&	171.98	&	0.421	&	0.137	\\
149757	&	57573	&	S2W1	&	225.31	&	-344.29	&	-1036.68	&	171.98	&	0.326	&	0.085	\\
149757	&	57573	&	S2W1	&	227.37	&	-347.43	&	-1046.14	&	171.98	&	0.288	&	0.091	\\
149757	&	57573	&	S2W1	&	229.43	&	-350.58	&	-1055.63	&	171.98	&	0.333	&	0.089	\\
149757	&	57573	&	S2W1	&	231.53	&	-353.79	&	-1065.29	&	171.98	&	0.371	&	0.063	\\
149757	&	57573	&	S2W1	&	233.67	&	-357.06	&	-1075.13	&	171.98	&	0.355	&	0.080	\\
149757	&	57573	&	S2W1	&	235.85	&	-360.39	&	-1085.16	&	171.98	&	0.325	&	0.042	\\
149757	&	57573	&	S2W1	&	238.10	&	-363.83	&	-1095.53	&	171.98	&	0.327	&	0.067	\\
149757	&	57573	&	S2W1	&	240.23	&	-367.09	&	-1105.32	&	171.98	&	0.296	&	0.046	\\
149757	&	57573	&	S2W1	&	242.40	&	-370.40	&	-1115.29	&	171.98	&	0.323	&	0.072	\\
149757	&	57573	&	S2W1	&	244.47	&	-373.56	&	-1124.80	&	171.98	&	0.303	&	0.071	\\
149757	&	57573	&	S2W1	&	246.50	&	-376.66	&	-1134.15	&	171.98	&	0.278	&	0.045	\\
149757	&	57573	&	S2W1	&	248.60	&	-379.87	&	-1143.82	&	171.98	&	0.239	&	0.050	\\
149757	&	57573	&	S2W1	&	250.70	&	-383.09	&	-1153.50	&	171.98	&	0.196	&	0.038	\\
149757	&	57573	&	S2W1	&	252.77	&	-386.24	&	-1162.99	&	171.98	&	0.172	&	0.054	\\
149757	&	57573	&	S2W1	&	254.79	&	-389.33	&	-1172.29	&	171.98	&	0.163	&	0.060	\\
149757	&	57573	&	S2W1	&	256.81	&	-392.41	&	-1181.57	&	171.98	&	0.108	&	0.061	\\
149757	&	57573	&	S2W1	&	258.89	&	-395.60	&	-1191.18	&	171.98	&	0.182	&	0.122	\\
149757	&	57573	&	S2W1	&	260.98	&	-398.78	&	-1200.76	&	171.98	&	0.184	&	0.091	\\
149757	&	57573	&	S2W1	&	263.13	&	-402.08	&	-1210.68	&	171.98	&	0.209	&	0.050	\\
149757	&	57573	&	S2W1	&	210.46	&	-239.23	&	-991.91	&	167.00	&	0.323	&	0.055	\\
149757	&	57573	&	S2W1	&	212.55	&	-241.60	&	-1001.76	&	167.00	&	0.298	&	0.056	\\
149757	&	57573	&	S2W1	&	214.68	&	-244.02	&	-1011.81	&	167.00	&	0.346	&	0.056	\\
149757	&	57573	&	S2W1	&	216.72	&	-246.34	&	-1021.39	&	167.00	&	0.334	&	0.052	\\
149757	&	57573	&	S2W1	&	218.79	&	-248.69	&	-1031.16	&	167.00	&	0.285	&	0.058	\\
149757	&	57573	&	S2W1	&	220.79	&	-250.96	&	-1040.57	&	167.00	&	0.287	&	0.063	\\
149757	&	57573	&	S2W1	&	222.79	&	-253.24	&	-1050.01	&	167.00	&	0.313	&	0.052	\\
149757	&	57573	&	S2W1	&	224.83	&	-255.56	&	-1059.62	&	167.00	&	0.321	&	0.046	\\
149757	&	57573	&	S2W1	&	226.91	&	-257.92	&	-1069.41	&	167.00	&	0.296	&	0.059	\\
149757	&	57573	&	S2W1	&	229.02	&	-260.32	&	-1079.38	&	167.00	&	0.278	&	0.040	\\
149757	&	57573	&	S2W1	&	231.21	&	-262.81	&	-1089.69	&	167.00	&	0.266	&	0.054	\\
149757	&	57573	&	S2W1	&	233.28	&	-265.16	&	-1099.43	&	167.00	&	0.237	&	0.036	\\
149757	&	57573	&	S2W1	&	235.38	&	-267.55	&	-1109.35	&	167.00	&	0.222	&	0.037	\\
149757	&	57573	&	S2W1	&	237.39	&	-269.83	&	-1118.81	&	167.00	&	0.205	&	0.034	\\
149757	&	57573	&	S2W1	&	239.36	&	-272.08	&	-1128.11	&	167.00	&	0.237	&	0.037	\\
149757	&	57573	&	S2W1	&	241.40	&	-274.40	&	-1137.73	&	167.00	&	0.213	&	0.039	\\
149757	&	57573	&	S2W1	&	243.44	&	-276.72	&	-1147.35	&	167.00	&	0.190	&	0.042	\\
149757	&	57573	&	S2W1	&	245.45	&	-278.99	&	-1156.80	&	167.00	&	0.189	&	0.040	\\
149757	&	57573	&	S2W1	&	247.41	&	-281.23	&	-1166.05	&	167.00	&	0.201	&	0.066	\\
149757	&	57573	&	S2W1	&	249.37	&	-283.45	&	-1175.28	&	167.00	&	0.179	&	0.057	\\
149757	&	57573	&	S2W1	&	251.40	&	-285.75	&	-1184.83	&	167.00	&	0.250	&	0.069	\\
149757	&	57573	&	S2W1	&	253.42	&	-288.05	&	-1194.36	&	167.00	&	0.200	&	0.037	\\
149757	&	57573	&	S2W1	&	255.51	&	-290.43	&	-1204.23	&	167.00	&	0.189	&	0.031	\\
149757	&	57573	&	S2W1	&	205.89	&	-140.66	&	-988.22	&	163.37	&	0.419	&	0.081	\\
149757	&	57573	&	S2W1	&	207.93	&	-142.06	&	-998.03	&	163.37	&	0.362	&	0.062	\\
149757	&	57573	&	S2W1	&	210.02	&	-143.48	&	-1008.04	&	163.37	&	0.364	&	0.062	\\
149757	&	57573	&	S2W1	&	212.01	&	-144.84	&	-1017.59	&	163.37	&	0.370	&	0.058	\\
149757	&	57573	&	S2W1	&	214.04	&	-146.23	&	-1027.32	&	163.37	&	0.385	&	0.075	\\
149757	&	57573	&	S2W1	&	215.99	&	-147.56	&	-1036.69	&	163.37	&	0.441	&	0.091	\\
149757	&	57573	&	S2W1	&	217.95	&	-148.90	&	-1046.10	&	163.37	&	0.439	&	0.080	\\
149757	&	57573	&	S2W1	&	219.94	&	-150.26	&	-1055.68	&	163.37	&	0.384	&	0.055	\\
149757	&	57573	&	S2W1	&	221.98	&	-151.65	&	-1065.43	&	163.37	&	0.358	&	0.054	\\
149757	&	57573	&	S2W1	&	224.05	&	-153.07	&	-1075.36	&	163.37	&	0.373	&	0.047	\\
149757	&	57573	&	S2W1	&	226.19	&	-154.53	&	-1085.64	&	163.37	&	0.359	&	0.044	\\
149757	&	57573	&	S2W1	&	228.21	&	-155.91	&	-1095.34	&	163.37	&	0.369	&	0.069	\\
149757	&	57573	&	S2W1	&	230.27	&	-157.32	&	-1105.22	&	163.37	&	0.356	&	0.058	\\
149757	&	57573	&	S2W1	&	232.23	&	-158.66	&	-1114.65	&	163.37	&	0.351	&	0.078	\\
149757	&	57573	&	S2W1	&	234.16	&	-159.98	&	-1123.92	&	163.37	&	0.369	&	0.051	\\
149757	&	57573	&	S2W1	&	236.16	&	-161.34	&	-1133.50	&	163.37	&	0.343	&	0.047	\\
149757	&	57573	&	S2W1	&	238.15	&	-162.70	&	-1143.08	&	163.37	&	0.343	&	0.106	\\
149757	&	57573	&	S2W1	&	240.11	&	-164.04	&	-1152.49	&	163.37	&	0.341	&	0.090	\\
149757	&	57573	&	S2W1	&	242.04	&	-165.36	&	-1161.71	&	163.37	&	0.307	&	0.164	\\
149757	&	57573	&	S2W1	&	243.95	&	-166.66	&	-1170.91	&	163.37	&	0.287	&	0.080	\\
149757	&	57573	&	S2W1	&	245.93	&	-168.02	&	-1180.42	&	163.37	&	0.344	&	0.085	\\
149757	&	57573	&	S2W1	&	247.91	&	-169.37	&	-1189.92	&	163.37	&	0.302	&	0.044	\\
149757	&	57573	&	S2W1	&	249.96	&	-170.77	&	-1199.75	&	163.37	&	0.307	&	0.042	\\
149757	&	57573	&	S2W1	&	203.46	&	-11.78	&	-986.32	&	161.44	&	0.493	&	0.108	\\
149757	&	57573	&	S2W1	&	205.48	&	-11.89	&	-996.11	&	161.44	&	0.468	&	0.100	\\
149757	&	57573	&	S2W1	&	207.54	&	-12.01	&	-1006.10	&	161.44	&	0.490	&	0.100	\\
149757	&	57573	&	S2W1	&	209.50	&	-12.13	&	-1015.63	&	161.44	&	0.453	&	0.090	\\
149757	&	57573	&	S2W1	&	211.51	&	-12.24	&	-1025.34	&	161.44	&	0.405	&	0.101	\\
149757	&	57573	&	S2W1	&	213.44	&	-12.36	&	-1034.70	&	161.44	&	0.455	&	0.152	\\
149757	&	57573	&	S2W1	&	215.37	&	-12.47	&	-1044.08	&	161.44	&	0.462	&	0.099	\\
149757	&	57573	&	S2W1	&	217.34	&	-12.58	&	-1053.64	&	161.44	&	0.446	&	0.078	\\
149757	&	57573	&	S2W1	&	219.35	&	-12.70	&	-1063.38	&	161.44	&	0.419	&	0.072	\\
149757	&	57573	&	S2W1	&	221.40	&	-12.82	&	-1073.29	&	161.44	&	0.412	&	0.061	\\
149757	&	57573	&	S2W1	&	223.51	&	-12.94	&	-1083.54	&	161.44	&	0.378	&	0.053	\\
149757	&	57573	&	S2W1	&	225.51	&	-13.05	&	-1093.23	&	161.44	&	0.374	&	0.056	\\
149757	&	57573	&	S2W1	&	227.55	&	-13.17	&	-1103.09	&	161.44	&	0.379	&	0.063	\\
149757	&	57573	&	S2W1	&	229.49	&	-13.28	&	-1112.50	&	161.44	&	0.430	&	0.084	\\
149757	&	57573	&	S2W1	&	231.39	&	-13.39	&	-1121.75	&	161.44	&	0.377	&	0.064	\\
149757	&	57573	&	S2W1	&	233.37	&	-13.51	&	-1131.32	&	161.44	&	0.319	&	0.050	\\
149757	&	57573	&	S2W1	&	235.34	&	-13.62	&	-1140.88	&	161.44	&	0.271	&	0.048	\\
149757	&	57573	&	S2W1	&	237.28	&	-13.74	&	-1150.27	&	161.44	&	0.245	&	0.053	\\
149757	&	57573	&	S2W1	&	239.18	&	-13.85	&	-1159.47	&	161.44	&	0.177	&	0.051	\\
149757	&	57573	&	S2W1	&	241.07	&	-13.95	&	-1168.65	&	161.44	&	0.210	&	0.082	\\
149757	&	57573	&	S2W1	&	243.03	&	-14.07	&	-1178.15	&	161.44	&	0.294	&	0.088	\\
149757	&	57573	&	S2W1	&	244.98	&	-14.18	&	-1187.62	&	161.44	&	0.308	&	0.056	\\
149757	&	57573	&	S2W1	&	247.01	&	-14.30	&	-1197.44	&	161.44	&	0.311	&	0.056	\\
214680	&	56637	&	S1E1	&	395.84	&	-1255.82	&	-1451.16	&	314.10	&	0.934	&	0.071	\\
214680	&	56637	&	S1E1	&	399.77	&	-1268.29	&	-1465.56	&	314.10	&	0.966	&	0.100	\\
214680	&	56637	&	S1E1	&	403.78	&	-1281.00	&	-1480.26	&	314.10	&	0.923	&	0.078	\\
214680	&	56637	&	S1E1	&	407.61	&	-1293.14	&	-1494.28	&	314.10	&	0.900	&	0.084	\\
214680	&	56637	&	S1E1	&	411.50	&	-1305.51	&	-1508.57	&	314.10	&	0.923	&	0.106	\\
214680	&	56637	&	S1E1	&	415.26	&	-1317.42	&	-1522.33	&	314.10	&	0.881	&	0.072	\\
214680	&	56637	&	S1E1	&	419.02	&	-1329.37	&	-1536.14	&	314.10	&	0.945	&	0.072	\\
214680	&	56637	&	S1E1	&	422.86	&	-1341.54	&	-1550.21	&	314.10	&	0.975	&	0.066	\\
214680	&	56637	&	S1E1	&	426.77	&	-1353.93	&	-1564.53	&	314.10	&	0.969	&	0.063	\\
214680	&	56637	&	S1E1	&	430.75	&	-1366.56	&	-1579.12	&	314.10	&	0.939	&	0.055	\\
214680	&	56637	&	S1E1	&	434.86	&	-1379.61	&	-1594.20	&	314.10	&	0.930	&	0.057	\\
214680	&	56637	&	S1E1	&	438.75	&	-1391.94	&	-1608.46	&	314.10	&	0.865	&	0.059	\\
214680	&	56637	&	S1E1	&	442.71	&	-1404.50	&	-1622.96	&	314.10	&	0.871	&	0.080	\\
214680	&	56637	&	S1E1	&	446.48	&	-1416.48	&	-1636.81	&	314.10	&	0.910	&	0.088	\\
214680	&	56637	&	S1E1	&	450.19	&	-1428.25	&	-1650.41	&	314.10	&	0.951	&	0.057	\\
214680	&	56637	&	S1E1	&	454.03	&	-1440.44	&	-1664.49	&	314.10	&	0.945	&	0.062	\\
214680	&	56637	&	S1E1	&	457.87	&	-1452.61	&	-1678.56	&	314.10	&	1.009	&	0.122	\\
214680	&	56637	&	S1E1	&	461.64	&	-1464.57	&	-1692.38	&	314.10	&	0.982	&	0.115	\\
214680	&	56637	&	S1E1	&	465.33	&	-1476.29	&	-1705.92	&	314.10	&	0.944	&	0.197	\\
214680	&	56637	&	S1E1	&	469.02	&	-1487.97	&	-1719.42	&	314.10	&	0.827	&	0.119	\\
214680	&	56637	&	S1E1	&	472.83	&	-1500.06	&	-1733.39	&	314.10	&	0.898	&	0.070	\\
214680	&	56637	&	S1E1	&	476.63	&	-1512.13	&	-1747.33	&	314.10	&	0.937	&	0.083	\\
214680	&	56637	&	S1E1	&	480.57	&	-1524.62	&	-1761.77	&	314.10	&	0.913	&	0.058	\\
214680	&	57934	&	S1E1	&	401.03	&	-1214.93	&	-1517.89	&	318.21	&	0.877	&	0.087	\\
214680	&	57934	&	S1E1	&	405.01	&	-1226.99	&	-1532.95	&	318.21	&	0.847	&	0.088	\\
214680	&	57934	&	S1E1	&	409.07	&	-1239.29	&	-1548.33	&	318.21	&	0.866	&	0.077	\\
214680	&	57934	&	S1E1	&	412.94	&	-1251.03	&	-1562.99	&	318.21	&	0.914	&	0.112	\\
214680	&	57934	&	S1E1	&	416.89	&	-1263.00	&	-1577.94	&	318.21	&	0.861	&	0.114	\\
214680	&	57934	&	S1E1	&	420.70	&	-1274.52	&	-1592.33	&	318.21	&	0.789	&	0.084	\\
214680	&	57934	&	S1E1	&	424.51	&	-1286.08	&	-1606.78	&	318.21	&	0.748	&	0.062	\\
214680	&	57934	&	S1E1	&	428.40	&	-1297.85	&	-1621.49	&	318.21	&	0.796	&	0.059	\\
214680	&	57934	&	S1E1	&	432.36	&	-1309.85	&	-1636.47	&	318.21	&	0.833	&	0.054	\\
214680	&	57934	&	S1E1	&	436.39	&	-1322.06	&	-1651.73	&	318.21	&	0.868	&	0.050	\\
214680	&	57934	&	S1E1	&	440.56	&	-1334.69	&	-1667.51	&	318.21	&	0.889	&	0.055	\\
214680	&	57934	&	S1E1	&	444.50	&	-1346.62	&	-1682.42	&	318.21	&	0.814	&	0.056	\\
214680	&	57934	&	S1E1	&	448.50	&	-1358.77	&	-1697.59	&	318.21	&	0.755	&	0.056	\\
214680	&	57934	&	S1E1	&	452.33	&	-1370.36	&	-1712.07	&	318.21	&	0.835	&	0.090	\\
214680	&	57934	&	S1E1	&	456.09	&	-1381.75	&	-1726.30	&	318.21	&	0.900	&	0.077	\\
214680	&	57934	&	S1E1	&	459.98	&	-1393.53	&	-1741.03	&	318.21	&	0.973	&	0.100	\\
214680	&	57934	&	S1E1	&	463.87	&	-1405.32	&	-1755.75	&	318.21	&	1.021	&	0.214	\\
214680	&	57934	&	S1E1	&	467.69	&	-1416.88	&	-1770.20	&	318.21	&	1.181	&	0.323	\\
214680	&	57934	&	S1E1	&	471.43	&	-1428.22	&	-1784.36	&	318.21	&	1.196	&	0.307	\\
214680	&	57934	&	S1E1	&	475.16	&	-1439.52	&	-1798.48	&	318.21	&	1.108	&	0.403	\\
214680	&	57934	&	S1E1	&	479.02	&	-1451.22	&	-1813.10	&	318.21	&	0.866	&	0.075	\\
214680	&	57934	&	S1E1	&	482.87	&	-1462.89	&	-1827.68	&	318.21	&	0.834	&	0.081	\\
214680	&	57934	&	S1E1	&	486.86	&	-1474.98	&	-1842.78	&	318.21	&	0.808	&	0.055	\\
214680	&	57934	&	S1E1	&	405.17	&	-1166.71	&	-1580.31	&	321.50	&	0.943	&	0.108	\\
214680	&	57934	&	S1E1	&	409.20	&	-1178.30	&	-1596.00	&	321.50	&	0.911	&	0.155	\\
214680	&	57934	&	S1E1	&	413.30	&	-1190.11	&	-1612.01	&	321.50	&	0.981	&	0.187	\\
214680	&	57934	&	S1E1	&	417.21	&	-1201.38	&	-1627.28	&	321.50	&	0.902	&	0.103	\\
214680	&	57934	&	S1E1	&	421.20	&	-1212.87	&	-1642.84	&	321.50	&	0.729	&	0.101	\\
214680	&	57934	&	S1E1	&	425.05	&	-1223.94	&	-1657.83	&	321.50	&	0.832	&	0.099	\\
214680	&	57934	&	S1E1	&	428.90	&	-1235.04	&	-1672.86	&	321.50	&	0.850	&	0.063	\\
214680	&	57934	&	S1E1	&	432.83	&	-1246.35	&	-1688.18	&	321.50	&	0.907	&	0.087	\\
214680	&	57934	&	S1E1	&	436.83	&	-1257.86	&	-1703.78	&	321.50	&	0.876	&	0.080	\\
214680	&	57934	&	S1E1	&	440.90	&	-1269.59	&	-1719.66	&	321.50	&	0.864	&	0.088	\\
214680	&	57934	&	S1E1	&	445.11	&	-1281.72	&	-1736.09	&	321.50	&	0.886	&	0.066	\\
214680	&	57934	&	S1E1	&	449.09	&	-1293.18	&	-1751.61	&	321.50	&	0.905	&	0.094	\\
214680	&	57934	&	S1E1	&	453.14	&	-1304.84	&	-1767.41	&	321.50	&	0.820	&	0.060	\\
214680	&	57934	&	S1E1	&	457.01	&	-1315.97	&	-1782.49	&	321.50	&	0.863	&	0.078	\\
214680	&	57934	&	S1E1	&	460.81	&	-1326.91	&	-1797.30	&	321.50	&	0.901	&	0.060	\\
214680	&	57934	&	S1E1	&	464.74	&	-1338.23	&	-1812.63	&	321.50	&	0.875	&	0.072	\\
214680	&	57934	&	S1E1	&	468.67	&	-1349.54	&	-1827.96	&	321.50	&	0.919	&	0.116	\\
214680	&	57934	&	S1E1	&	472.52	&	-1360.65	&	-1843.00	&	321.50	&	1.072	&	0.159	\\
214680	&	57934	&	S1E1	&	476.30	&	-1371.54	&	-1857.75	&	321.50	&	0.939	&	0.258	\\
214680	&	57934	&	S1E1	&	480.07	&	-1382.39	&	-1872.45	&	321.50	&	0.861	&	0.259	\\
214680	&	57934	&	S1E1	&	483.98	&	-1393.63	&	-1887.67	&	321.50	&	0.796	&	0.075	\\
214680	&	57934	&	S1E1	&	487.87	&	-1404.84	&	-1902.85	&	321.50	&	0.855	&	0.106	\\
214680	&	57934	&	S1E1	&	491.90	&	-1416.44	&	-1918.57	&	321.50	&	0.842	&	0.064	\\
214680	&	57934	&	S1E1	&	408.36	&	-1113.61	&	-1636.91	&	324.04	&	1.016	&	0.134	\\
214680	&	57934	&	S1E1	&	412.42	&	-1124.67	&	-1653.16	&	324.04	&	1.049	&	0.104	\\
214680	&	57934	&	S1E1	&	416.55	&	-1135.95	&	-1669.74	&	324.04	&	1.066	&	0.124	\\
214680	&	57934	&	S1E1	&	420.50	&	-1146.71	&	-1685.56	&	324.04	&	0.914	&	0.127	\\
214680	&	57934	&	S1E1	&	424.52	&	-1157.67	&	-1701.68	&	324.04	&	0.814	&	0.242	\\
214680	&	57934	&	S1E1	&	428.39	&	-1168.23	&	-1717.20	&	324.04	&	0.792	&	0.237	\\
214680	&	57934	&	S1E1	&	432.28	&	-1178.83	&	-1732.78	&	324.04	&	0.889	&	0.180	\\
214680	&	57934	&	S1E1	&	436.24	&	-1189.62	&	-1748.64	&	324.04	&	0.967	&	0.143	\\
214680	&	57934	&	S1E1	&	440.27	&	-1200.61	&	-1764.80	&	324.04	&	0.951	&	0.192	\\
214680	&	57934	&	S1E1	&	444.37	&	-1211.81	&	-1781.25	&	324.04	&	0.977	&	0.261	\\
214680	&	57934	&	S1E1	&	448.62	&	-1223.39	&	-1798.27	&	324.04	&	0.990	&	0.258	\\
214680	&	57934	&	S1E1	&	452.63	&	-1234.32	&	-1814.35	&	324.04	&	0.974	&	0.181	\\
214680	&	57934	&	S1E1	&	456.71	&	-1245.46	&	-1830.71	&	324.04	&	0.944	&	0.248	\\
214680	&	57934	&	S1E1	&	460.61	&	-1256.08	&	-1846.33	&	324.04	&	0.980	&	0.294	\\
214680	&	57934	&	S1E1	&	464.44	&	-1266.52	&	-1861.67	&	324.04	&	1.007	&	0.252	\\
214680	&	57934	&	S1E1	&	468.40	&	-1277.32	&	-1877.55	&	324.04	&	0.975	&	0.309	\\
214680	&	57934	&	S1E1	&	472.36	&	-1288.12	&	-1893.43	&	324.04	&	1.001	&	0.377	\\
214680	&	57934	&	S1E1	&	476.24	&	-1298.72	&	-1909.01	&	324.04	&	1.123	&	0.464	\\
214680	&	57934	&	S1E1	&	480.05	&	-1309.11	&	-1924.28	&	324.04	&	1.025	&	0.312	\\
214680	&	57934	&	S1E1	&	483.85	&	-1319.47	&	-1939.51	&	324.04	&	0.920	&	0.242	\\
214680	&	57934	&	S1E1	&	487.79	&	-1330.20	&	-1955.28	&	324.04	&	0.966	&	0.288	\\
214680	&	57934	&	S1E1	&	491.71	&	-1340.90	&	-1971.00	&	324.04	&	0.988	&	0.269	\\
214680	&	57934	&	S1E1	&	495.77	&	-1351.98	&	-1987.29	&	324.04	&	0.951	&	0.221	\\
214680	&	57934	&	S1E1	&	412.93	&	-984.84	&	-1742.92	&	327.66	&	1.003	&	0.121	\\
214680	&	57934	&	S1E1	&	417.03	&	-994.62	&	-1760.22	&	327.66	&	1.005	&	0.129	\\
214680	&	57934	&	S1E1	&	421.21	&	-1004.59	&	-1777.87	&	327.66	&	0.984	&	0.138	\\
214680	&	57934	&	S1E1	&	425.20	&	-1014.11	&	-1794.72	&	327.66	&	0.862	&	0.129	\\
214680	&	57934	&	S1E1	&	429.26	&	-1023.81	&	-1811.88	&	327.66	&	0.712	&	0.120	\\
214680	&	57934	&	S1E1	&	433.18	&	-1033.15	&	-1828.41	&	327.66	&	0.858	&	0.124	\\
214680	&	57934	&	S1E1	&	437.11	&	-1042.52	&	-1844.99	&	327.66	&	0.902	&	0.104	\\
214680	&	57934	&	S1E1	&	441.11	&	-1052.06	&	-1861.89	&	327.66	&	0.931	&	0.093	\\
214680	&	57934	&	S1E1	&	445.19	&	-1061.78	&	-1879.09	&	327.66	&	0.935	&	0.088	\\
214680	&	57934	&	S1E1	&	449.34	&	-1071.68	&	-1896.61	&	327.66	&	1.007	&	0.096	\\
214680	&	57934	&	S1E1	&	453.63	&	-1081.92	&	-1914.73	&	327.66	&	0.982	&	0.097	\\
214680	&	57934	&	S1E1	&	457.68	&	-1091.59	&	-1931.85	&	327.66	&	1.114	&	0.120	\\
214680	&	57934	&	S1E1	&	461.81	&	-1101.44	&	-1949.27	&	327.66	&	1.246	&	0.153	\\
214680	&	57934	&	S1E1	&	465.75	&	-1110.83	&	-1965.90	&	327.66	&	1.145	&	0.128	\\
214680	&	57934	&	S1E1	&	469.62	&	-1120.07	&	-1982.24	&	327.66	&	1.049	&	0.111	\\
214680	&	57934	&	S1E1	&	473.63	&	-1129.62	&	-1999.14	&	327.66	&	0.953	&	0.098	\\
214680	&	57934	&	S1E1	&	477.63	&	-1139.17	&	-2016.05	&	327.66	&	0.805	&	0.092	\\
214680	&	57934	&	S1E1	&	481.56	&	-1148.55	&	-2032.64	&	327.66	&	0.752	&	0.118	\\
214680	&	57934	&	S1E1	&	485.42	&	-1157.74	&	-2048.90	&	327.66	&	0.493	&	0.133	\\
214680	&	57934	&	S1E1	&	489.26	&	-1166.90	&	-2065.12	&	327.66	&	1.005	&	0.348	\\
214680	&	57934	&	S1E1	&	493.24	&	-1176.38	&	-2081.90	&	327.66	&	1.086	&	0.158	\\
214680	&	57934	&	S1E1	&	497.20	&	-1185.85	&	-2098.65	&	327.66	&	1.084	&	0.113	\\
214680	&	57934	&	S1E1	&	501.31	&	-1195.64	&	-2115.99	&	327.66	&	1.077	&	0.111	\\
214680	&	57934	&	S1E1	&	416.48	&	58.15	&	-2018.29	&	330.47	&	0.806	&	0.111	\\
214680	&	57934	&	S1E1	&	420.61	&	58.73	&	-2038.33	&	330.47	&	0.865	&	0.154	\\
214680	&	57934	&	S1E1	&	424.83	&	59.32	&	-2058.77	&	330.47	&	0.787	&	0.175	\\
214680	&	57934	&	S1E1	&	428.85	&	59.88	&	-2078.27	&	330.47	&	0.683	&	0.096	\\
214680	&	57934	&	S1E1	&	432.95	&	60.45	&	-2098.15	&	330.47	&	0.767	&	0.147	\\
214680	&	57934	&	S1E1	&	436.90	&	61.00	&	-2117.29	&	330.47	&	0.721	&	0.122	\\
214680	&	57934	&	S1E1	&	440.87	&	61.56	&	-2136.49	&	330.47	&	0.734	&	0.064	\\
214680	&	57934	&	S1E1	&	444.90	&	62.12	&	-2156.05	&	330.47	&	0.748	&	0.073	\\
214680	&	57934	&	S1E1	&	449.01	&	62.69	&	-2175.97	&	330.47	&	0.778	&	0.071	\\
214680	&	57934	&	S1E1	&	453.20	&	63.28	&	-2196.26	&	330.47	&	0.775	&	0.080	\\
214680	&	57934	&	S1E1	&	457.53	&	63.88	&	-2217.24	&	330.47	&	0.760	&	0.034	\\
214680	&	57934	&	S1E1	&	461.62	&	64.46	&	-2237.06	&	330.47	&	0.783	&	0.053	\\
214680	&	57934	&	S1E1	&	465.78	&	65.04	&	-2257.24	&	330.47	&	0.791	&	0.045	\\
214680	&	57934	&	S1E1	&	469.76	&	65.59	&	-2276.50	&	330.47	&	0.788	&	0.062	\\
214680	&	57934	&	S1E1	&	473.66	&	66.14	&	-2295.42	&	330.47	&	0.706	&	0.053	\\
214680	&	57934	&	S1E1	&	477.70	&	66.70	&	-2315.00	&	330.47	&	0.776	&	0.061	\\
214680	&	57934	&	S1E1	&	481.74	&	67.26	&	-2334.57	&	330.47	&	0.808	&	0.030	\\
214680	&	57934	&	S1E1	&	485.70	&	67.82	&	-2353.78	&	330.47	&	0.807	&	0.030	\\
214680	&	57934	&	S1E1	&	489.59	&	68.36	&	-2372.61	&	330.47	&	0.871	&	0.038	\\
214680	&	57934	&	S1E1	&	493.46	&	68.90	&	-2391.39	&	330.47	&	0.846	&	0.045	\\
214680	&	57934	&	S1E1	&	497.48	&	69.46	&	-2410.83	&	330.47	&	0.823	&	0.035	\\
214680	&	57934	&	S1E1	&	501.48	&	70.02	&	-2430.22	&	330.47	&	0.727	&	0.091	\\
214680	&	57934	&	S1E1	&	505.62	&	70.60	&	-2450.30	&	330.47	&	0.779	&	0.090	\\
\enddata
\tablecomments{The complete table is provided in the on-line version.}
\end{deluxetable*}

%%%%%%%%%%%%%%%%%%%%%%%%%%%%%%%%%%%%%%%%%%%%%%%%%%%%%%%%%%%%%%%%%%%%%%%%
\section{Stellar Parameters}
%%%%%%%%%%%%%%%%%%%%%%%%%%%%%%%%%%%%%%%%%%%%%%%%%%%%%%%%%%%%%%%%%%%%%%%%

\subsection{Interferometry}

\begin{figure*}[htb]
\centering
%  \begin{tabular}{@{}ccc@{}}
%    \includegraphics[width=.25\textwidth, angle=90]{HD24912_cal2.eps} &
%    \includegraphics[width=.25\textwidth, angle=90]{HD30614_cal1.eps} &
%    \includegraphics[width=.25\textwidth, angle=90]{HD36861.eps} \\
%    \includegraphics[width=.25\textwidth, angle=90]{HD37742_cal1.eps}   &
%    \includegraphics[width=.25\textwidth, angle=90]{HD149757v2.eps} &
%    \includegraphics[width=.25\textwidth, angle=90]{HD149757v2_2.eps} \\
%    \multicolumn{3}{c}{\includegraphics[width=.25\textwidth, angle=90]{HD214680_all.eps}}
  \begin{tabular}{@{}cc@{}}
    \includegraphics[trim=2cm 2cm 2cm 2cm,width=.45\textwidth]{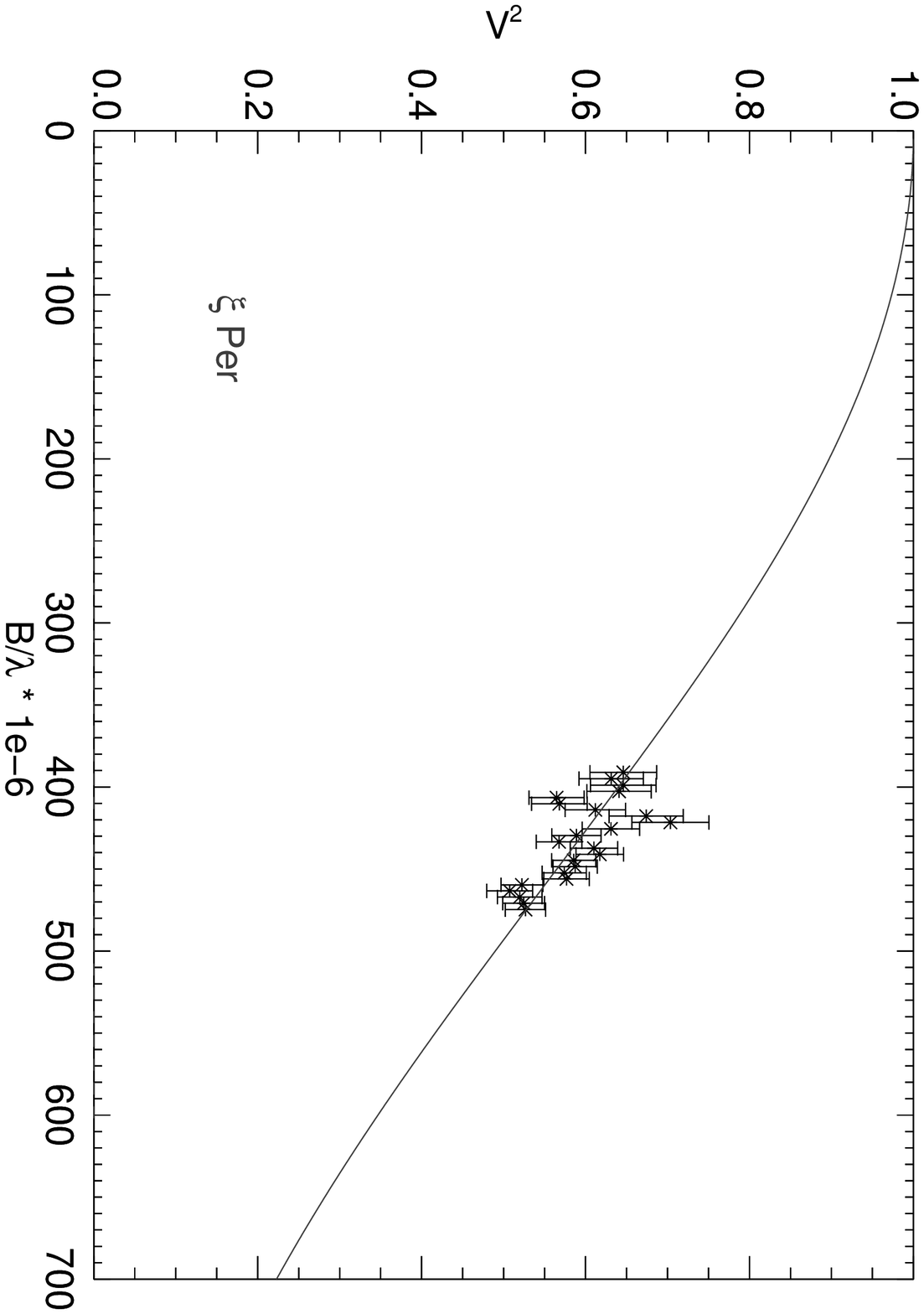} &
    \includegraphics[trim=2cm 2cm 2cm 2cm,width=.45\textwidth]{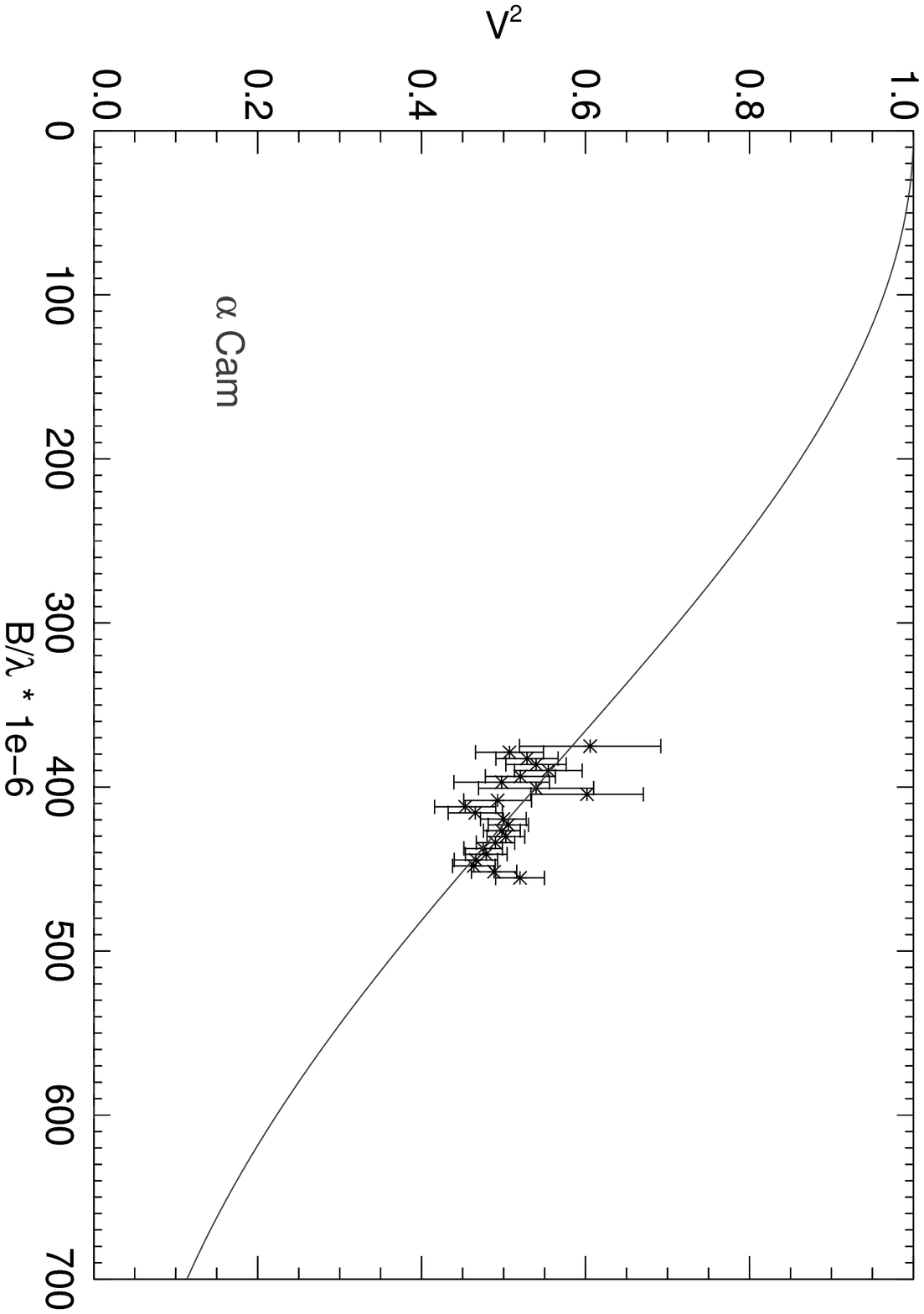} \\
    \includegraphics[trim=2cm 2cm 2cm 2cm,width=.45\textwidth]{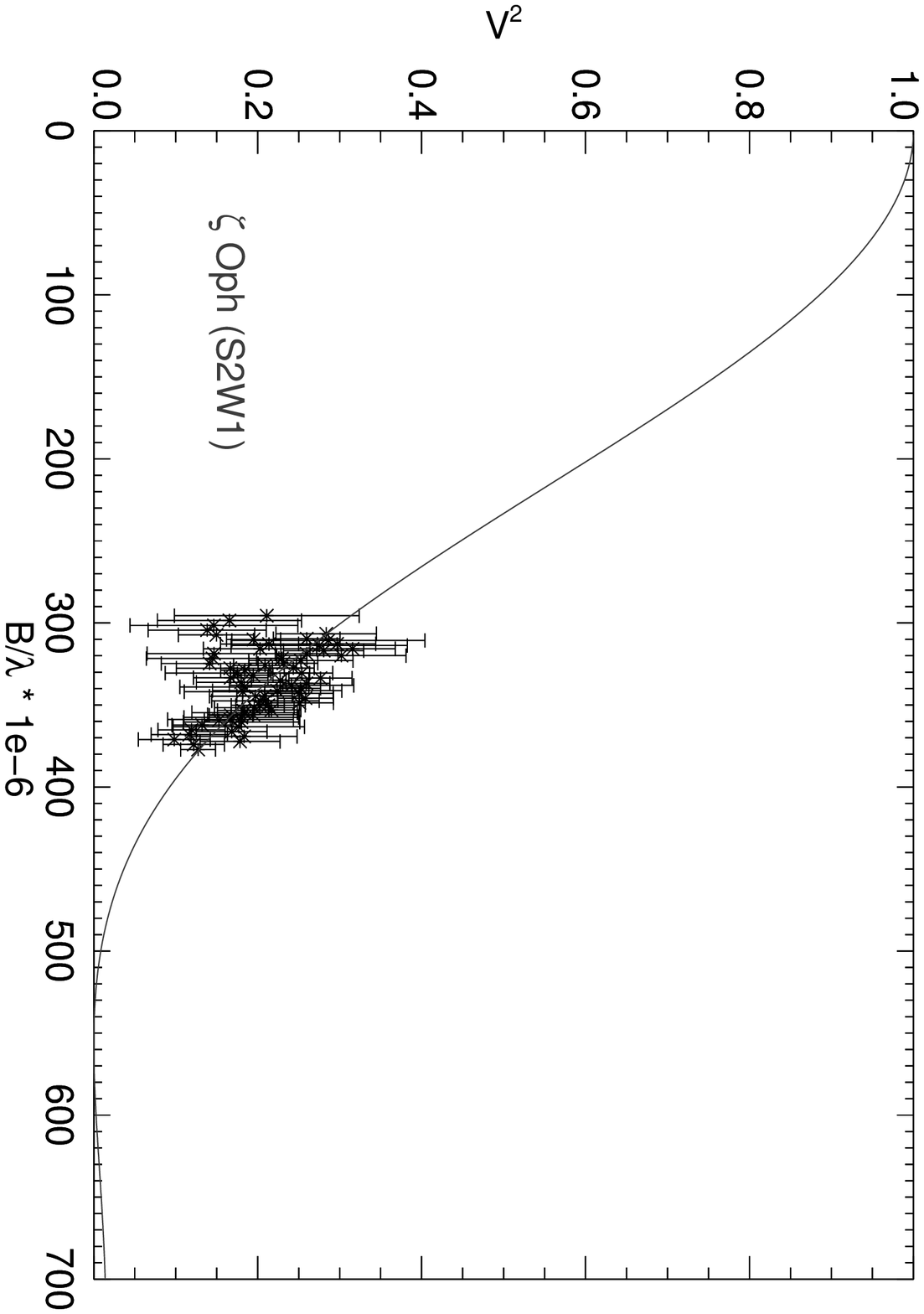} &
    \includegraphics[trim=2cm 2cm 2cm 2cm,width=.45\textwidth]{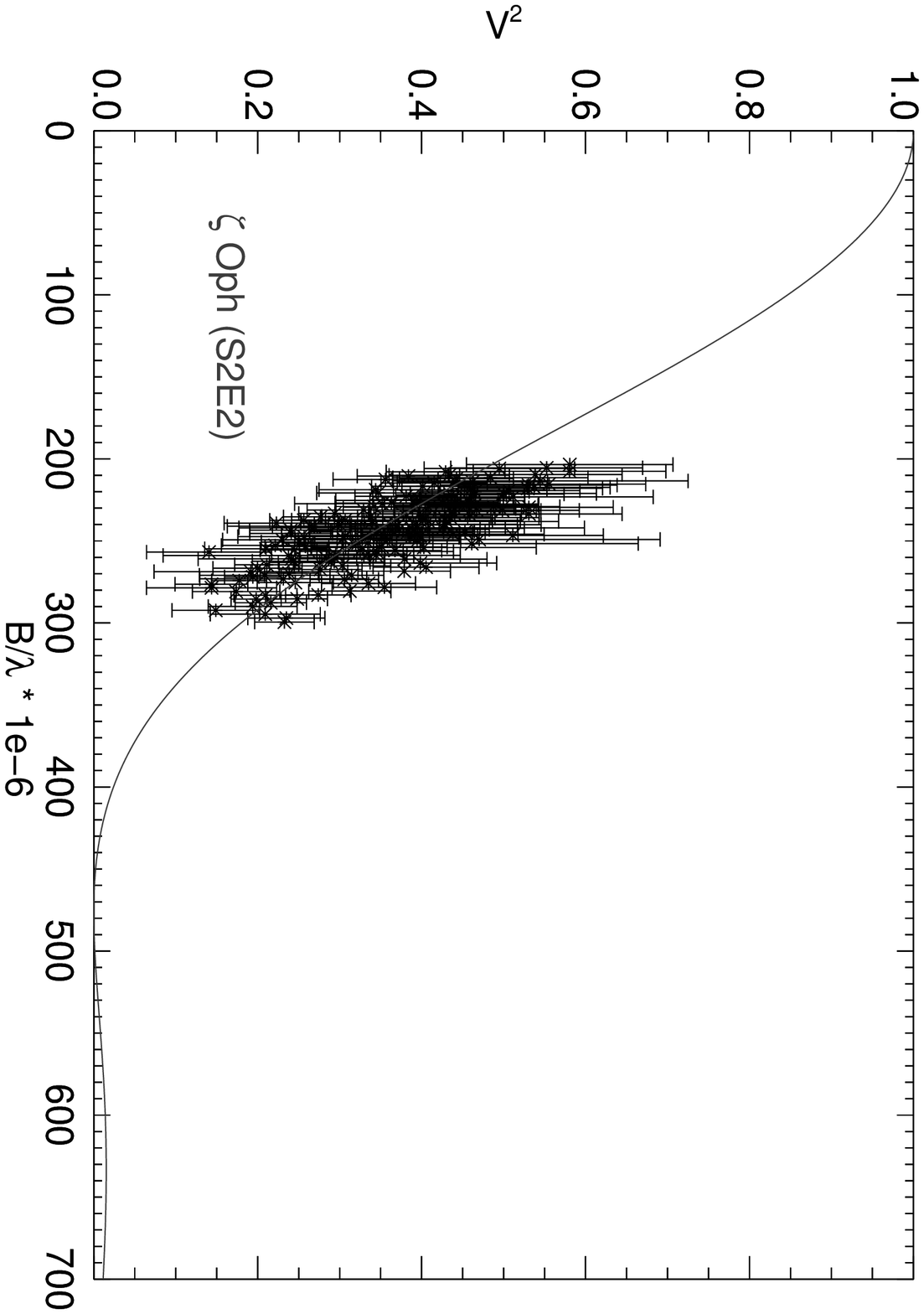} \\
    \includegraphics[trim=2cm 2cm 2cm 2cm,width=.45\textwidth]{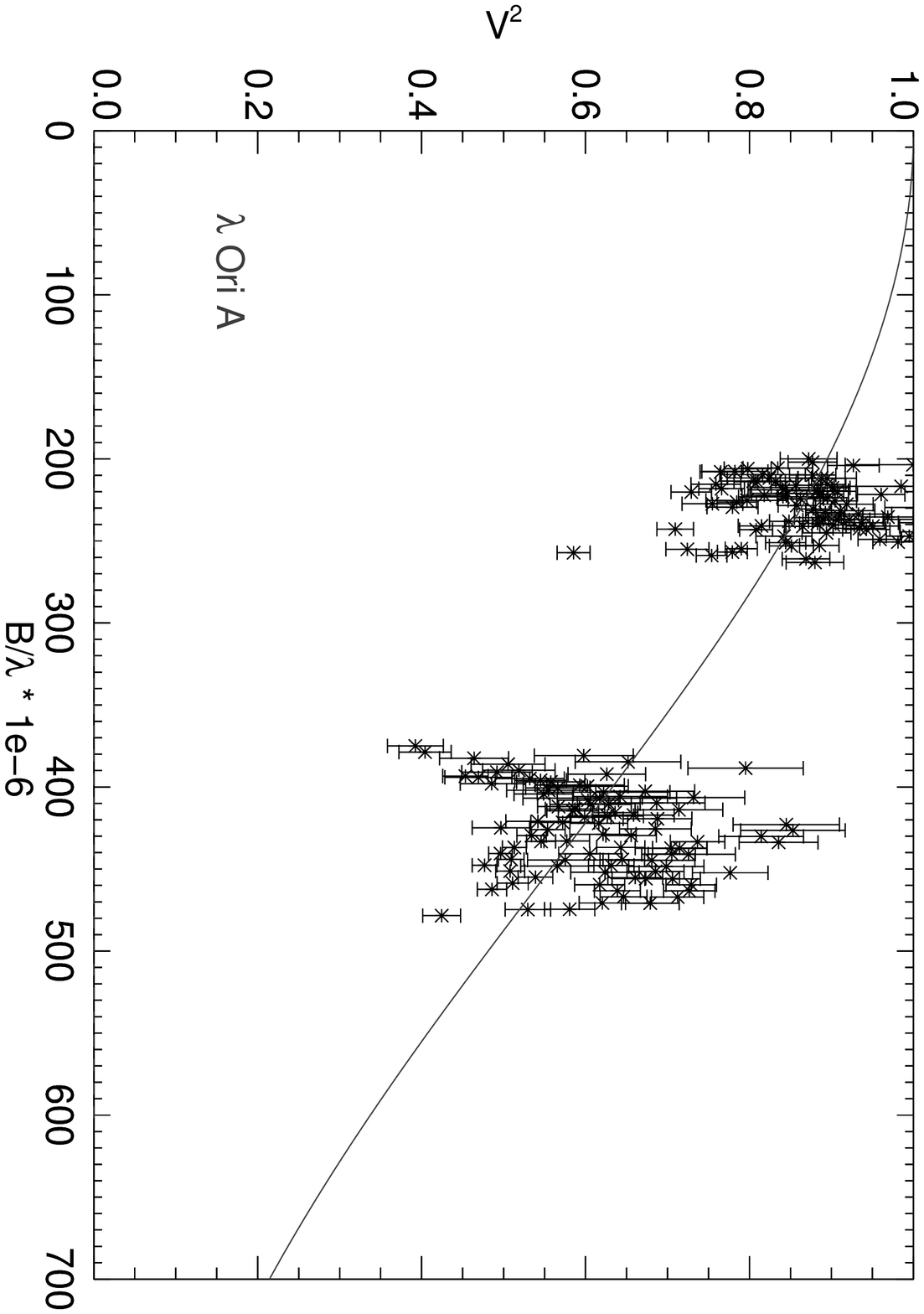} &
    \includegraphics[trim=2cm 2cm 2cm 2cm,width=.45\textwidth]{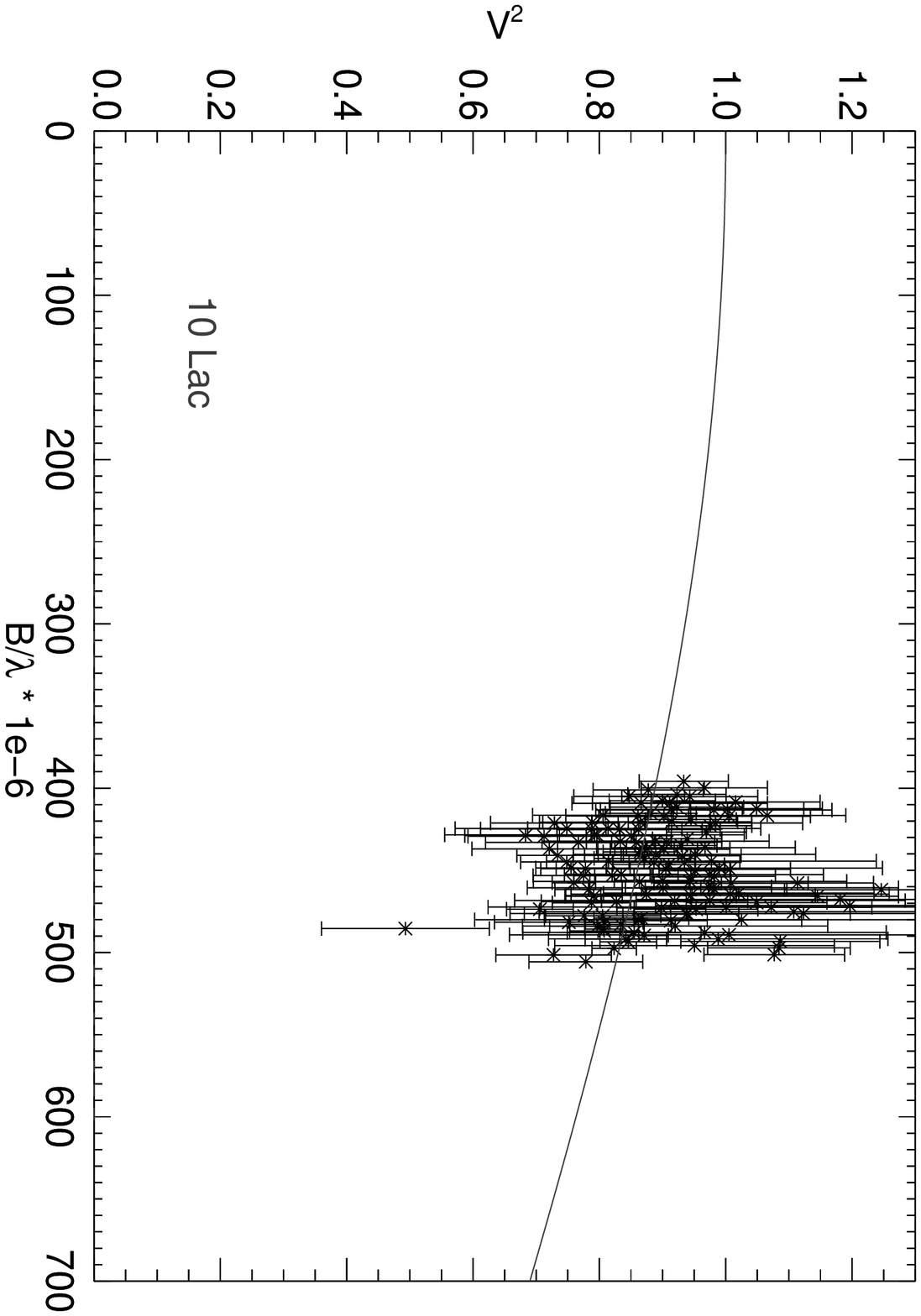}
  \end{tabular}
  \caption{Squared visibility versus spatial frequency for five target stars. The solid line indicates the best fit for a single star limb-darkened disk model.}
  \label{fig:visibility}
\end{figure*}

\begin{figure*}[htb]
\centering
 \includegraphics[width=1\textwidth]{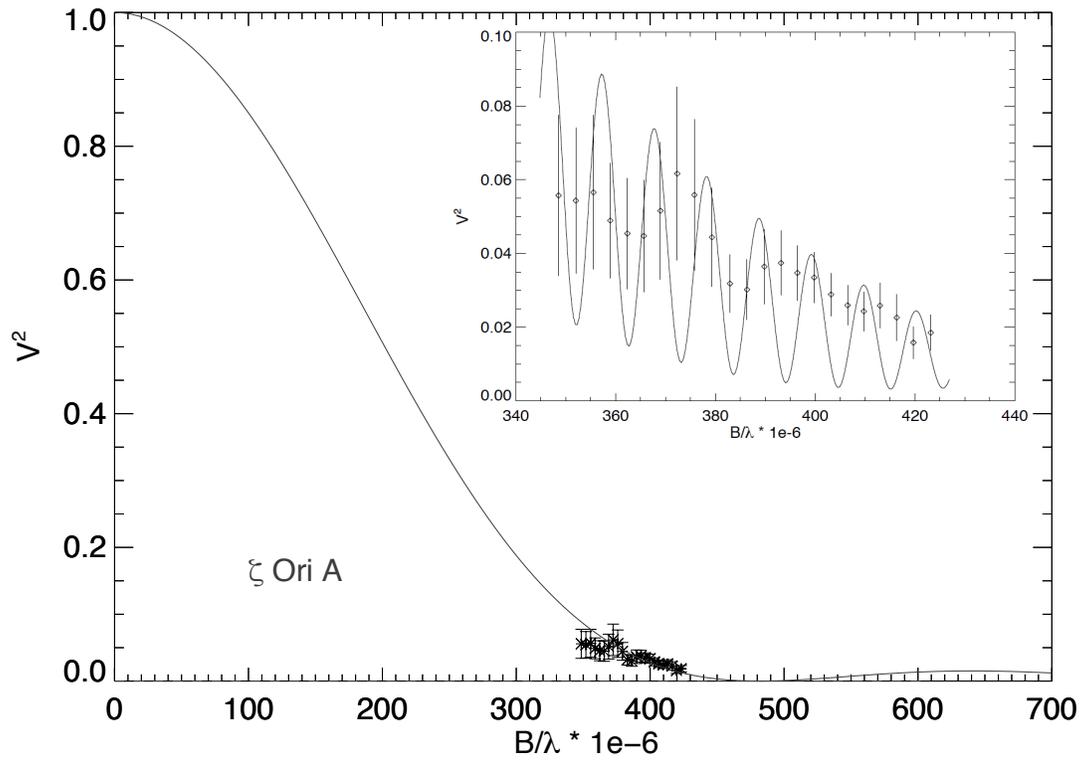}
 \caption{Visibility measurements for $\zeta$ Ori A with the solid line showing an error weighted single star fit. An error weighted binary fit to the visibility data is shown inset.}
 \label{fig:zetaori}
\end{figure*}

The visibility measurements we obtained for each star from our interferometric observations were fitted with a limb-darkened, single star, disk model. Linear limb-darkening coefficients in the $R$-band were interpolated from the tables available in \citet{Claret2011} using the stellar parameters given in Table \ref{tab:physpara}.  These limb darkening coefficients were calculated for model atmospheres that adopt a solar metallicity and a microturbulent velocity of 2 km~s$^{-1}$, and they were derived to maintain flux conservation. \textbf{The effect of limb-darkening is minimal in O-type stars and any uncertainty in the fitted size due to uncertainty in the adopted limb-darkening coefficient will be much smaller than our other sources of uncertainty discussed below.} Figures \ref{fig:visibility} and \ref{fig:zetaori} show the visibility measurements for each star plus an error-weighted fit of all the data with a limb-darkened disk model.  Table \ref{tab:diameters} lists the derived uniform disk (UD) and limb-darkened (LD) disk angular diameters $\theta$, the latter calculated for a linear limb-darkening coefficient $\mu$. 

\begin{deluxetable*}{cccccccccc}[htb]
\tabletypesize{\footnotesize}
%\tablecolumns{10}
\tablewidth{0pt}
\tablecaption{Angular Diameters
\label{tab:diameters}}
\tablehead{
\colhead{Star} & 
\colhead{Telescope} & 
\colhead{} & 
\colhead{$\theta$$_{\text{$UD$}}$} & 
\colhead{} & 
\colhead{$\theta _{LD}$} &
\colhead{$\theta _{LD}$(U79)} &
\colhead{$\theta _{LD}$(HB74)} &
\colhead{$\theta _{LD}$(CADARS)} &
\colhead{$\theta _{LD}$($T_{\rm eff}$)} \\
\colhead{Name} & 
\colhead{Pair} & 
\colhead{$N_{\text{$V^2$}}$} & 
\colhead{(mas)} & 
\colhead{$\mu$} & 
\colhead{(mas)} & 
\colhead{(mas)} & 
\colhead{(mas)} & 
\colhead{(mas)} &
\colhead{(mas)}}
\startdata
$\xi$ Per       & W1E1      &  23 & 0.216$\pm$0.016  & 0.174 & 0.218$\pm$0.016$^1$  & \nodata         & \nodata       & 0.26 & 0.245$\pm$0.010 \\
$\alpha$ Cam    & S1E1      &  23 & 0.250$\pm$0.014 & 0.250 & 0.256$\pm$0.014$^1$ & 0.292$\pm$0.003 & \nodata       & 0.29 & 0.245$\pm$0.010 \\
$\lambda$ Ori A & S1E1,W1E1 & 168 & 0.219$\pm$0.015 & 0.253 & 0.226$\pm$0.015 & 0.235$\pm$0.003 & \nodata       & 0.24 & 0.228$\pm$0.009 \\
$\zeta$ Ori A   & W1E1      &  23 & 0.546$\pm$0.029 & 0.203 & 0.556$\pm$0.029$^1$ & 0.527$\pm$0.010 & 0.48$\pm$0.04 & 0.47 & 0.485$\pm$0.019 \\
$\zeta$ Oph     & S2W1      &  69 & 0.454$\pm$0.010 & 0.204 & 0.462$\pm$0.010 & 0.494$\pm$0.003 & 0.51$\pm$0.05 & 0.54 & 0.539$\pm$0.021 \\
                & S2E2      & 161 & 0.532$\pm$0.010 & 0.204 & 0.540$\pm$0.010 &                 &               &      &                  \\
10 Lac          & S1E1      & 119 & 0.11$\pm$0.02  & 0.183 & 0.11$\pm$0.02  & 0.123$\pm$0.002 & \nodata       & 0.13 & 0.121$\pm$0.005
\enddata
\tablecomments{$^{1}$ $\theta$$_{\text{$LD$}}$ represents preliminary results that are based upon only a single data bracket. $N_{V^2}$ = number of visibility measurements. U79 = \citet{Underhill1979}, HB74 = \citet{HanburyBrown1974}, CADARS = \citet{Pasinetti2001}, $T_{\rm eff}$ = diameter derived from temperature and SED.}
\end{deluxetable*}

The fitting scheme assigned an uncertainty to the angular diameter based upon the size of the residuals to the fit.  However, multiple-night observations of stellar diameters with PAVO show an external night-to-night scatter that is larger than indicated by the uncertainty from measurements within a night by about 5\% \citep{Maestro2013}.  We checked this inter-night variation with larger sample of 25 B stars observed with the CHARA Array (to be presented in a forthcoming paper) and found the average error to be consistent with the 5\% quoted by \citet{Maestro2013} for PAVO data.  Thus, we have applied a 5\% night-to-night scatter to our error budget. The other source of uncertainty is related to the diameters of the calibrator stars. Because some of our targets are very small (only 0.11 mas for 10 Lac), the calibrator stars may have sizes comparable to the targets. These calibrators can still be used in the analysis but the error in their sizes will play a much bigger role in the error budget than is usually the case where the calibrators are much smaller than the corresponding targets. To account for this effect we fit the data for each star after adjusting the calibrator sizes by plus and minus one sigma, and then derived the range in the solutions.  The final uncertainties given in Table \ref{tab:diameters} are the quadratic sum of the uncertainties from the internal fit uncertainties, the night-to-night external error and the half-range from varying the calibrator size.  

The observations from different nights were not averaged for $\zeta$ Oph as there is a true physical difference in size measured along different baselines due to the star's rotational distortion. The results in Table \ref{tab:diameters} show that the angular size varies by $15\%$ between the S2W1 and S2E2 baselines.  Below in Section 4.1 we present an ellipsoidal fit of the angular diameter as a function of position angle.

\subsection{Spectrophotometry}

The goal of our spectrophotometric analysis was to compare our results from directly measured angular sizes and observed spectra to predictions of parameters from a stellar atmospheric model. Given the spectral flux across a wavelength range and an estimate of the stellar effective temperature $T_{\rm eff}$ from spectral line studies,  the models can predict what the angular size should be.  We can then compare this against our interferometrically determined angular sizes to test the consistency of model line and flux predictions.

We used fluxes from multiple sources for our targets to create spectral energy distributions (SEDs) that span the wavelength range from ultraviolet to infrared. Sources used for each part of the spectrum are given in Table \ref{tab:spectra}. Ultraviolet spectra obtained from the {\it International Ultraviolet Explorer} satellite were recalibrated with a routine by \citet{Massa2000} to correct the flux values. Longer wavelength infrared fluxes were omitted for the supergiant and giant O stars because their winds create an excess flux in the far infrared that is absent from models without winds.  We rebinned the UV and optical spectra to a low resolving power of $R=\lambda / \triangle\lambda = 60$ on a $\log \lambda$ grid in order to better balance the sampling across the whole spectrum.  
All flux values in the spectra were assigned a uniform 3\% error to ensure our fitting program fit all points equally and did not give more weight to any one part of the spectrum. The exception to this was the case of 10~Lac, a target which has very good data available in the ultraviolet and optical from HST/STIS \citep{Bohlin2017}. The errors on these flux values were lower than 3\% and the original error values were used for the fitting.

\begin{deluxetable*}{ccccc}[htb]
\tabletypesize{\footnotesize}
\tablecolumns{10}
\tablewidth{0pt}
\tablecaption{Spectrophotometry Sources
\label{tab:spectra}}
\tablehead{
\colhead{Star} & \colhead{Far UV} & \colhead{Near UV} & \colhead{Optical} &  \colhead{IR}}
\startdata
$\xi$ Per &  SWP45474  &  LWP23809  &    SP1  &   2MASS  \\
$\alpha$ Cam &  HUT  &   LWP17592   &   SP2 &    2MASS   \\
$\lambda$ Ori A &  OAO   &     OAO    &    SP1  &    2MASS  \\
$\zeta$ Ori A & SWP33049, SWP33050 &  LWP11671, LWP12826   &   SP3  &   2MASS  \\
$\zeta$ Oph &  SWP06776, SWP18252 &  LWP12637, LWR14381  &    SP1  &   2MASS,  WISE, AKARI, Spitzer, IRAS \\
10 Lac & SWP*+STIS &  HST/STIS &  HST/STIS &  2MASS, WISE, AKARI, IRAS
\enddata
 \tablecomments{The UV spectra are primarily from the archive of the {\it International Ultraviolet Explorer} (low dispersion, large aperture) where the file number is related to the camera: SWP = Short Wavelength Prime, LWP = Long Wavelength Prime, and LWR = Long Wavelength Redundant. SWP* refers to the average of 52 SWP spectra covering the 1160 -- 1646 \AA ~range.  All the fluxes were corrected using the algorithm from \citet{Massa2000}.  Other UV fluxes are from  HUT = Hopkins Ultraviolet Telescope \citep{Buss1995}, OAO = Orbiting Astronomical Observatory 2 \citep{Code1979}, and HST/STIS from the CALSPEC database \citep{Bohlin2017}. Optical spectrophotometry sources are coded as SP1 = \citet{Burnashev1985}, SP2 = \citet{Kharitonov1988}, and SP3 = \citet{Krisciunas2017}. IR fluxes are from 2MASS \citep{Cutri2003}, WISE \citep{Cutri2012}, AKARI \citep{Ishihara2010}, Spitzer \citep{Ardilla2010}, and IRAS \citep{IRAS1988}.}
\end{deluxetable*}

The spectra were compared to the TLUSTY OSTAR2002 stellar atmosphere models that adopt solar metallicity and a microturbulent velocity of 10 km~s$^{-1}$ \citep{Hubeny1995,Lanz2003}. Our fitting routine used a grid search method to fit the SED using three parameters: limb-darkened angular size ($\theta_{LD}$), effective temperature ($T_{\rm eff}$), and reddening ($E(B-V)$). For any given set of parameters, a spectrum was extracted from the O star grid through interpolation in $T_{\rm eff}$ and $\log g$ (the latter from the adopted gravity given in Table \ref{tab:physpara}).  The model spectrum was rebinned to $R=60$ in the same way as the observed SED, and then the fluxes were attenuated for interstellar extinction using the IDL Code {\it fmrcurve.pro} for a ratio of total-to-selective extinction of 3.1 \citep{Fitzpatrick1999}. \textbf{Adopting an $R$ value greater than 3.1 would result in the model fit predicting a larger angular diameter. Recent extinction fits by \citet{Maiz2018} give an $R$ value close to 3.1 for all our stars with the exception of $\alpha$ Cam with an $R = 4.0$.} Finally the model spectra were rescaled according to the assumed angular diameter and interpolated to the observed wavelength points for direct comparison with the observed SED. 

In cases where there is a close, bright companion (Table \ref{tab:companions}), the extra flux from the companion was included in our fitting. The effective temperature and surface gravity of the companion were used to calculate a model companion spectrum in the same way as above using either the TLUSTY OSTAR2002 \citep{Lanz2003} or BSTAR2006 models \citep{Lanz2007}, depending on the temperature of the companion.  The companion flux was then rescaled according to the $V$-band magnitude difference $\triangle m_V$ (Table \ref{tab:companions}) and added to the model flux calculated for the primary.  This combined flux spectrum was then used for  spectrophotometric fitting of the observed SED.

We began the SED modeling process by creating a grid of angular size and effective temperature, the latter over a range and step size chosen to match the models in the TLUSTY grid. For each pair of $\theta$ and $T_{\rm eff}$, models were calculated over a grid of reddening values, and the best fit $E(B-V)$ was found by making a parabolic fit around the reddening grid point with the lowest value of reduced $\chi^2_\nu$.  The best-fit $E(B-V)$ and $\chi^2_\nu$ values were stored in matrices as functions of $\theta_{LD}$ and $T_{\rm eff}$.

Contour maps were created by plotting the $\chi^2_\nu$ matrix for each star, and these are shown in Figure \ref{fig:contours}. Overplotted as vertical lines are the angular size obtained from our interferometry with $1\sigma$ error margins, and horizontal lines show the average literature temperature from Table \ref{tab:physpara} with 1$\sigma$ error margins. The shape of the contours shows that there is generally a valley of low $\chi^2_\nu$ where the effective temperature has an inverse square dependence on angular size as expected from our discussion in Section 1.  The contours for the best-fit $E(B-V)$ values (not shown) are parallel lines that follow the curve of the minimum in $\chi^2_\nu$ space.  Therefore, any temperature and angular size combination chosen along the minimum valley will have a very similar associated $E(B-V)$ value.

We first consider what angular size is predicted from the $\chi^2_\nu$ diagrams for estimates of $T_{\rm eff}$ from published studies of the line spectra (Table \ref{tab:physpara}).  We examined the horizontal line associated with the average literature temperature for each star, and we found the angular size where the line crossed the minimum $\chi^2_\nu$.  The derived angular sizes for these best fits are listed in the final column of Table \ref{tab:diameters} under the heading $\theta _{LD} (T_{\rm eff})$. A comparison of our interferometric angular size and the model predicted size is shown in Figure \ref{fig:compare} and discussed in Section 4.2.  We assume for simplicity that there is a 4\% uncertainty associated with $\theta _{LD} (T_{\rm eff})$, which is an approximate estimate based upon flux uncertainties of 3\%, reddening uncertainties of 2\%, and line-based temperature uncertainties of 3\%.

We next consider estimates for effective temperature and reddening that rely on the observed interferometric sizes and the $\chi^2_\nu$ the contour plots.  We followed the vertical line along the value of our observed angular size and found the temperature where the lowest $\chi^2_\nu$ was attained (marked by a diamond symbol).  These interferometrically determined temperatures and the associated reddening estimates are given in Table \ref{tab:temps}, along with comparisons to previously determined values.  The range in acceptable values of $T_{\rm eff}$ can be large because the fractional errors in temperature are about twice as large as the fractional errors in angular size (which may be significant).  

The spectral energy distributions (SED) and the model best fits are shown in Figure \ref{fig:sed}. The symbols represent the observed spectral data while the solid green line shows the model SED for our interferometric size and best fit temperature and reddening. For comparison, the dashed line (often overlapping the solid line) shows the model SED based on the literature temperature and the angular size corresponding to the local minimum of the $\chi^2_\nu$ contours.

\begin{deluxetable*}{cccccc}[htb]
\tabletypesize{\footnotesize}
\tablecolumns{10}
\tablewidth{0pt}
\tablecaption{Effective Temperature and Reddening Estimates
\label{tab:temps}}
\tablehead{
\colhead{Star} & \colhead{$T_{\rm eff}$(Best Fit)} & \colhead{$T_{\rm eff}$(Literature)} & \colhead{$E(B-V)^a$} & \colhead{$E(B-V)^b$} & \colhead{$E(B-V)^c$} \\
 \colhead{Name} & \colhead{(kK)} & \colhead{(kK)}& \colhead{(mag)} & \colhead{(mag)} & \colhead{(mag)}}
\startdata
$\xi$ Per       & $<$40        & 34.3$\pm$0.8 & 0.291 & 0.25 & 0.278$\pm$0.007\\
$\alpha$ Cam    & 28.0$\pm$1.5 & 29.4$\pm$1.0 & 0.298 & 0.26 & 0.262$\pm$0.006\\
$\lambda$ Ori A & 36.0$\pm$0.9 & 34.5$\pm$0.8 & 0.107 & 0.12 & 0.177$\pm$0.011\\
$\zeta$ Ori A   & $<$28        & 29.5$\pm$1.0 & 0.067 & 0.08 & 0.044$\pm$0.007\\
$\zeta$ Oph     & 33.5$\pm$1.3 & 32.1$\pm$1.3 & 0.350 & 0.29 & 0.297$\pm$0.006\\
10 Lac          & 40.0$\pm$1.3 & 35.5$\pm$0.5 & 0.096 & 0.08 & 0.077$\pm$0.006\\
\enddata
\tablecomments{
$E(B-V)$ estimates:
a.\ Best fit, b.\ \citet{Savage1977}, c.\ \citet{Maiz2018}.
}
\end{deluxetable*}
\begin{deluxetable*}{cccccccccc}
\tabletypesize{\footnotesize}
\tablecolumns{10}
\tablewidth{0pt}
\tablecaption{Distance and Radius Estimates
\label{tab:distances}}
\tablehead{
\colhead{Star} & \colhead{$d_1$} & \colhead{$d_2$} & \colhead{$d_3$} & \colhead{$d_4$} & \colhead{$d_5$} & \colhead{$d_6$} & \colhead{$R_{\rm GAIA}$} & \colhead{$<d>$} & \colhead{$R$}\\
\colhead{Name} & \colhead{(pc)} & \colhead{(pc)} & \colhead{(pc)} & \colhead{(pc)} & \colhead{(pc)} & \colhead{(pc)} & \colhead{($R_\odot$)} & \colhead{(pc)} & \colhead{($R_\odot$)}} 
\startdata 
 $\xi$ Per &  398 &\nodata &  486$\pm$57   &  416$\pm$158   &\nodata & \nodata & \nodata & $ 433 \pm  46$ & $ 10.1 \pm 1.3$ \\
 $\alpha$ Cam & 1010 & 1175$\pm$118   & 1607$\pm$275   &\nodata &  821   &  731$\pm$175 & $20.1\pm4.9$ & $1068 \pm 346$ & $29.4 \pm 9.7$ \\
 $\lambda$ Ori A &  501 &  398$\pm$40   &\nodata &  361$\pm$90   &  438   &  417$\pm$10 & $10.0\pm0.6$ & $ 423 \pm  51$ & $10.2 \pm 1.3$ \\
 $\zeta$ Ori A &  501 &  350$\pm$35   &  297$\pm$45   &  239$\pm$48   &  391   &  381$\pm$10 & $22.8\pm1.3$ & $ 359 \pm  89$ & $21.5 \pm 5.4$ \\
$\zeta$ Oph &  154 &  188$\pm$19   &  222$\pm$22   &  112$\pm$11   &  145   &  172$\pm$31 & $8.5 - 10.0$ & $ 165 \pm  37$ & $ 8.9 - 10.5$  \\
10 Lac &  603 &  631$\pm$63   &  579$\pm$76   &  542$\pm$108   & \nodata &  478$\pm$10 & $5.7\pm1.0$ & $ 566 \pm  59$ & $ 6.7 \pm 1.4$ 
\enddata 
\tablecomments{Distance references:
1.\ \citet{Shull1985}, 
2.\ \citet{Underhill1979}, 
3.\ \citet{Megier2009}, 
4.\ \citet{vanLeeuwen2007} and \citet{Maiz2008}, 
5.\ \citet{deZeeuw1999} and \citet{Kharchenko2005},
6.\ \citet{Gaia2018}.
}
\end{deluxetable*}

%%%%%%%%%%%%%%%%%%%%%%%%%%%%%%%%%%%%%%%%%%%%%%%%%%%%%%%%%%%%%%%%%%%%%%%%%%%
\section{Discussion}
%%%%%%%%%%%%%%%%%%%%%%%%%%%%%%%%%%%%%%%%%%%%%%%%%%%%%%%%%%%%%%%%%%%%%%%%%%

\subsection{Notes on Individual Stars} 

{\sl $\xi$~Per} (HD~24912) is one of three O-type runaway stars in the list (along with $\alpha$ Cam and $\zeta$~Oph), and these are usually single stars \citep{Gies1986}.  The observed angular size is smaller than expected (Table \ref{tab:diameters}; Fig.\ \ref{fig:contours}), however, given that the measurements are preliminary and that the minimum $\chi^2_\nu$ lies at a diameter only $1.4\sigma$ larger than measured, we do not consider the difference significant.  The $\chi^2_\nu$ minimum valley has a steep slope in Figure \ref{fig:contours}, so the derived uncertainty in $T_{\rm eff}$ is large and we list only an upper limit in Table \ref{tab:temps}. 

{\sl $\alpha$ Cam} (HD~30614) presents a case where the observed angular size is somewhat larger than expected \textbf{from model predictions} (Table \ref{tab:diameters}), but $\theta_{LD}$ and $\theta_{LD}(T_{\rm eff})$ agree within the uncertainties.  

{\sl $\lambda$ Ori A} (HD~36861) represents an example with excellent agreement between $\theta_{LD}$ and $\theta_{LD}(T_{\rm eff})$ (Table \ref{tab:diameters}).  During the data analysis one of the stars used as a calibrator, HD~35149, was found to be a binary and was thus rejected from the calibration process. The final data for $\lambda$~Ori~A are calibrated with only one calibrator star, and the visibilities are slightly more noisy than otherwise expected.

{\sl $\zeta$ Ori A} (HD 37742) has a larger angular size than expected \textbf{from model predictions}  (Table \ref{tab:diameters}; Fig.\ \ref{fig:contours}). The star is the brightest component Aa of a triple system with a close companion Ab, which had a predicted separation of 33 mas at the time of our observations \citep{Hummel2013}.  The light from this companion will affect the spectrum and the visibility curve of $\zeta$~Ori~A, and we accounted for the extra flux of the companion for both the fits of the visibilities and the SED (Section 2).  We calculated a binary model for the visibilities, shown in Figure \ref{fig:zetaori}, using the predicted position angle, separation, and radius ratio $R/R_\odot=0.365$ given by \citet{Hummel2013}.  The model shows fast and low amplitude oscillations of the visibility curve that are roughly consistent with the observations, and the binary fit yields $\theta_{LD} = 0.556\pm0.029$ mas. $\zeta$~Ori~A was recently observed with the FRIEND beam combiner at the CHARA Array (M.-A. Martinod, private communication) yielding angular diameters of $0.54 \pm 0.01$ and $0.45 \pm 0.12$ mas for Aa and Ab, respectively.  Their angular diameter measurement agrees within errors with ours for $\zeta$ Ori Aa. 
The estimate of $T_{\rm eff}$ associated with the observed angular diameter attains a $\chi^2_\nu$ minimum at the lower boundary of the OSTAR2002 grid (Fig.\ \ref{fig:contours}), so only an approximate upper limit is given in Table \ref{tab:temps}.

{\sl $\zeta$ Oph} (HD 149757) is a well-known, rapidly rotating star. According to the analysis by \citet{Howarth2001}, the star has an equatorial, angular rotational velocity that is $90\%$ of the critical value, and it is viewed in an equatorial orientation with a spin inclination axis angle of $70^\circ$.  Consequently, its shape should appear oblate due to its rapid rotation.  We observed $\zeta$~Oph on different baselines with different position angles (S2W1 at $-42\fdg8$ and S2E2 at $17\fdg9$) to measure directly this deviation from a perfect sphere. Our measurements are insufficient to make a full ellipsoidal model of the size variation, but we can make a restricted fit by setting the position angle of the rotational axis from earlier polarimetric work.  $\zeta$ Oph sometimes appears as a \textbf{Oe} star with H Balmer line emission from a circumstellar disk.  \textbf{At the times of observation (July 2016 and June 2017)}, it did not show H$\alpha$ emission in its spectra, so it is reasonable to assume that any disk gas has dissipated and we are measuring the angular size of the star itself.  However, during a past epoch when a disk was present, \citet{Poeckert1979} used spectropolarimetry to determine the position angle of the disk minor axis as $132.5 \pm 6.0$ deg east from north.  We assume that the circumstellar disk axis is parallel to the stellar rotation axis, so that this is also the position angle of the projected minor axis of the star's shape.  
We made a fit of the limb-darkened angular diameter for each $V^2$ measurement, and then we fit an ellipsoid to the diameter as a function of position angle in the sky with the position angle of the minor axis set from the spectropolarimetric result.  
This gives a major axis of 0.56 mas and a minor axis of 0.48 mas. The fitted ellipse is shown in Figure \ref{fig:ellipse}.  The ratio of minor to major axis of 0.86 is similar to the polar to equatorial radius ratio of $7.5 R_\odot / 9.1 R_\odot = 0.82$ found by \citet{Howarth2001}, although the absolute radii we find are slightly larger because we adopt a larger distance (Table \ref{tab:distances}).  This star is a good target for future interferometric imaging to determine better the rotational distortion and the associated gravity and limb darkening \citep{Che2011}. $\zeta$~Oph is a runaway star that may have been launched from a supernova explosion in a binary. Its path has been traced back with that of a pulsar that was ejected during the supernova \citep{Hoogerwerf2000}.

{\sl 10 Lac} (HD 214680) has the smallest angular diameter ($\theta_{LD} = 0.11 \pm 0.02$~mas) that has been measured with CHARA to date. Because this is below the nominal resolution limit of PAVO of 0.2~mas, we need to check carefully the associated uncertainty of our measurement. Three calibrators were used for the observations of 10~Lac. The fitted angular size is very sensitive to the error in the calibrator size when the target star is smaller in angular size than the calibrator, as in this case. To account for this, we varied each calibrator in size by plus and minus one sigma, then refit the data with the new calibrator size. We then took the range between the fits as part of our error budget. This method was subsequently applied to all stars in the sample (Section 3.1). 
%The result is a fairly large error ($\approx 20\%$) for the angular diameter, but our result agrees with the SED fitting outcome (Table \ref{tab:diameters}).  
Working close to the resolution limit of PAVO resulted in some non-physical ($V^2>$1) visibility estimates. We retained these points except in one case where an entire bracket yielded visibilities greater than unity.  Both spectroscopic observations and our interferometric results arrive at relatively high effective temperatures, 40.0 kK and \textbf{35.5} kK, respectively, that are well above the 31.9 kK temperature associated with its classification of O9~V \citep{Martins2005}.  \textbf{It has been noted in other studies \citep{SimonDiaz2014,Holgado2018} that the $T_{\rm eff}$ values from the calibrations in \citet{Martins2005} are too low for O9 V stars.} However, the extended minimum valley in the $\chi^2_\nu$ diagram (Fig.\ \ref{fig:contours}) does encompass the lower temperature (as does the uncertainty region of measured angular diameter). 

\subsection{Consistency with SED Fits from Model Atmospheres} 

\begin{figure*}[htb]
\centering
  \begin{tabular}{@{}cc@{}}
    \includegraphics[trim=2cm 2cm 2cm 4cm,width=.35\textwidth, angle=90]{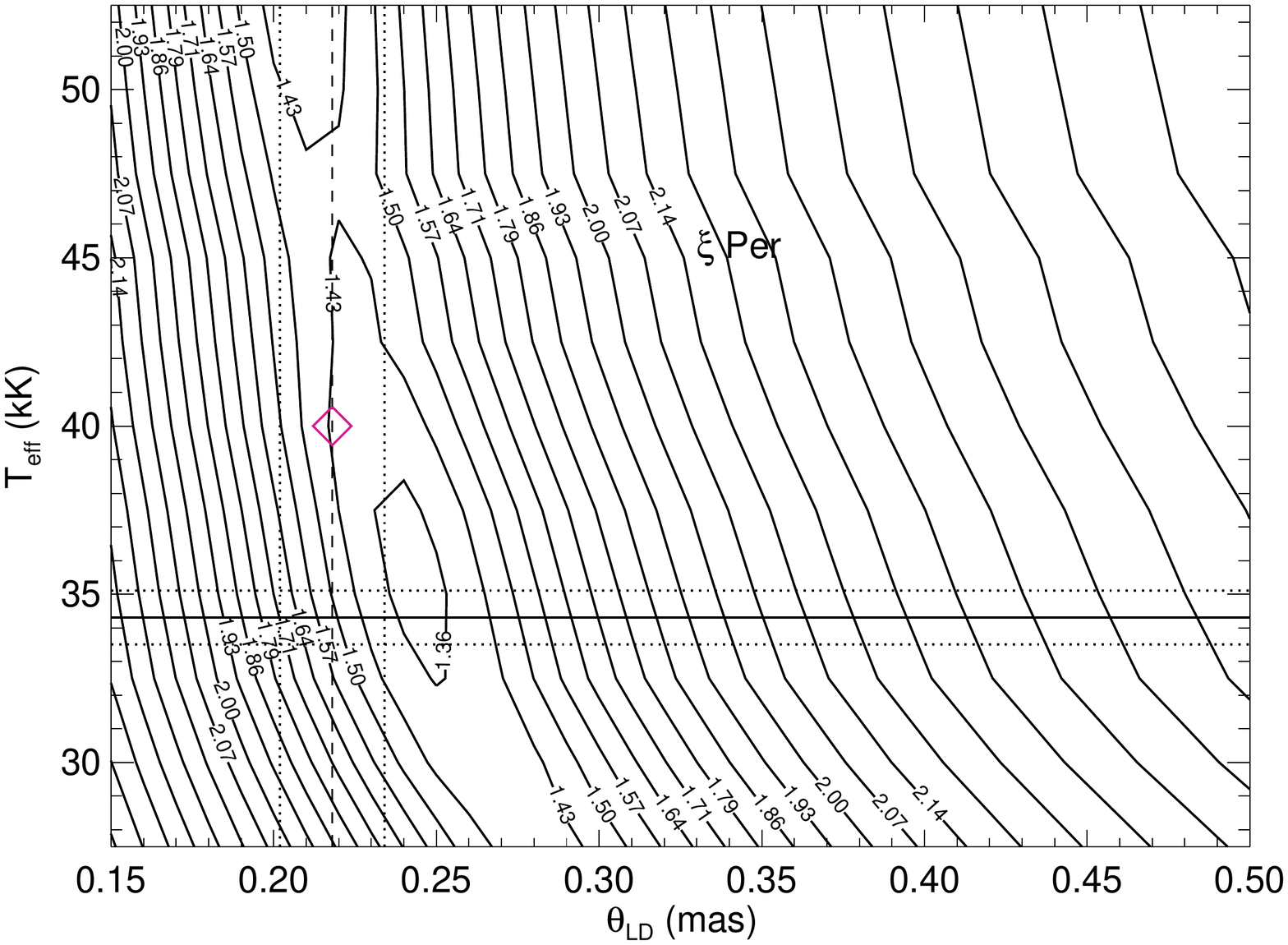} &
    \includegraphics[trim=2cm 2cm 2cm 4cm,width=.35\textwidth, angle=90]{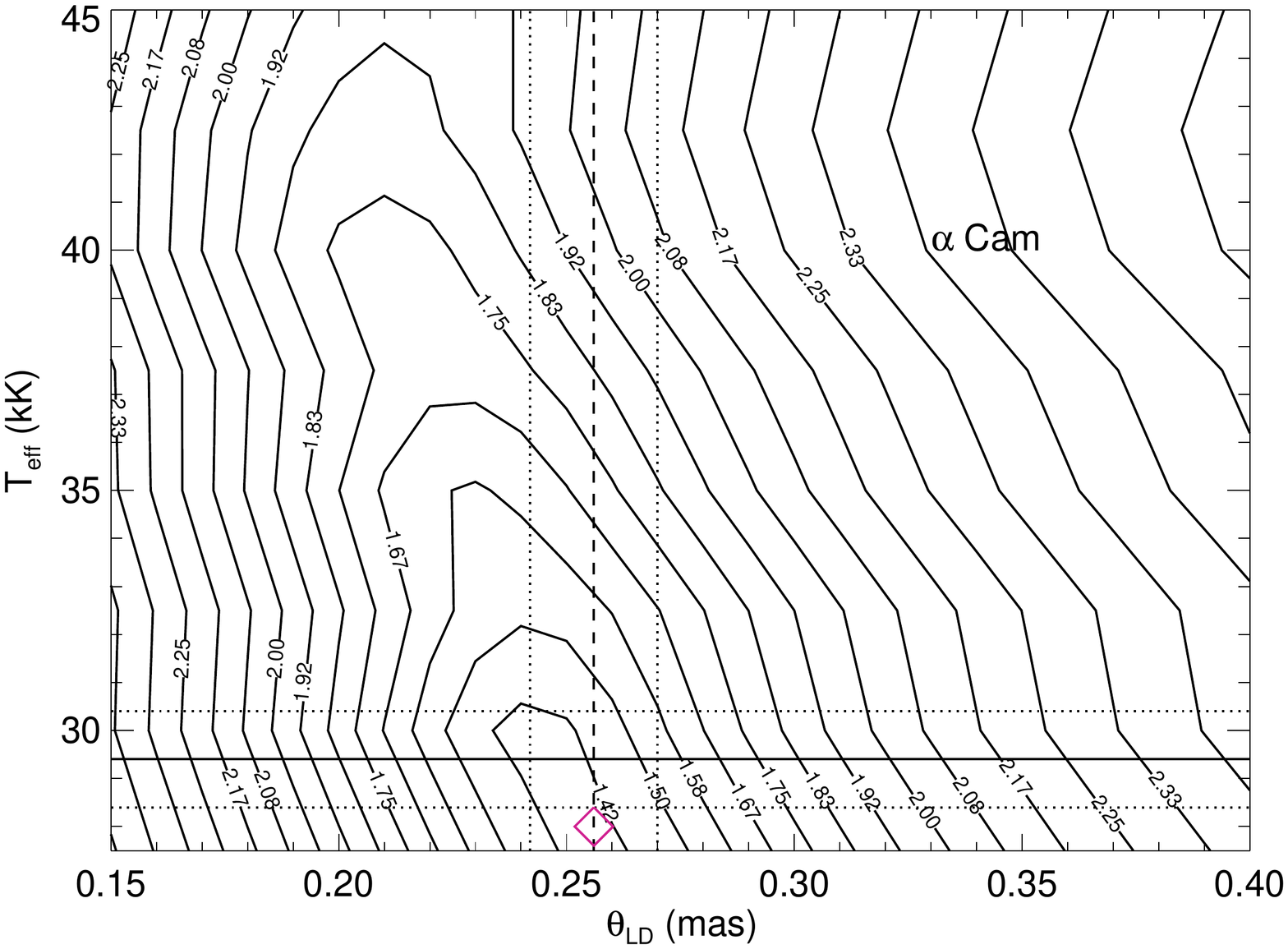} \\
    \includegraphics[trim=2cm 2cm 2cm 4cm,width=.35\textwidth, angle=90]{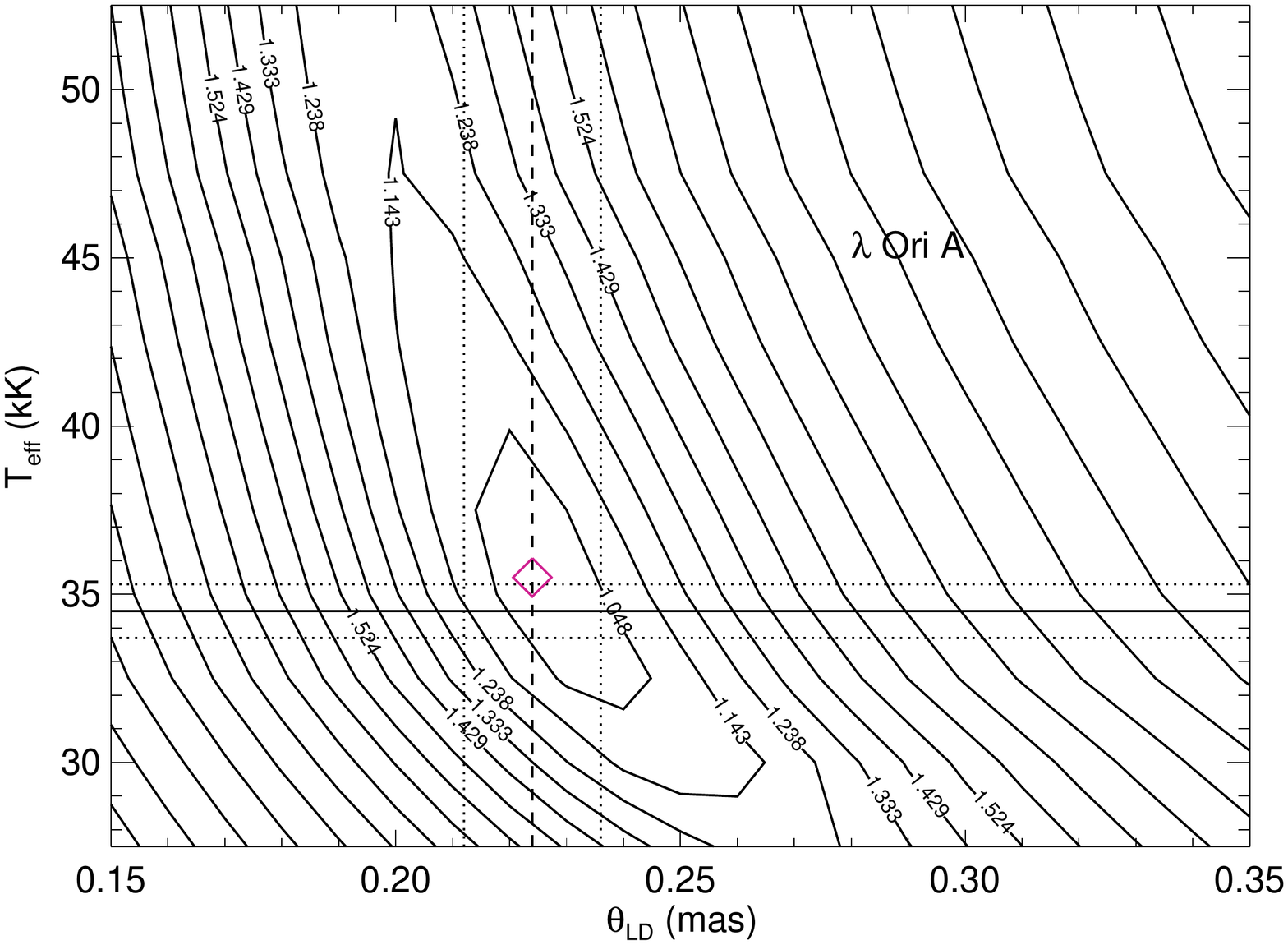} &
    \includegraphics[trim=2cm 2cm 2cm 4cm,width=.35\textwidth, angle=90]{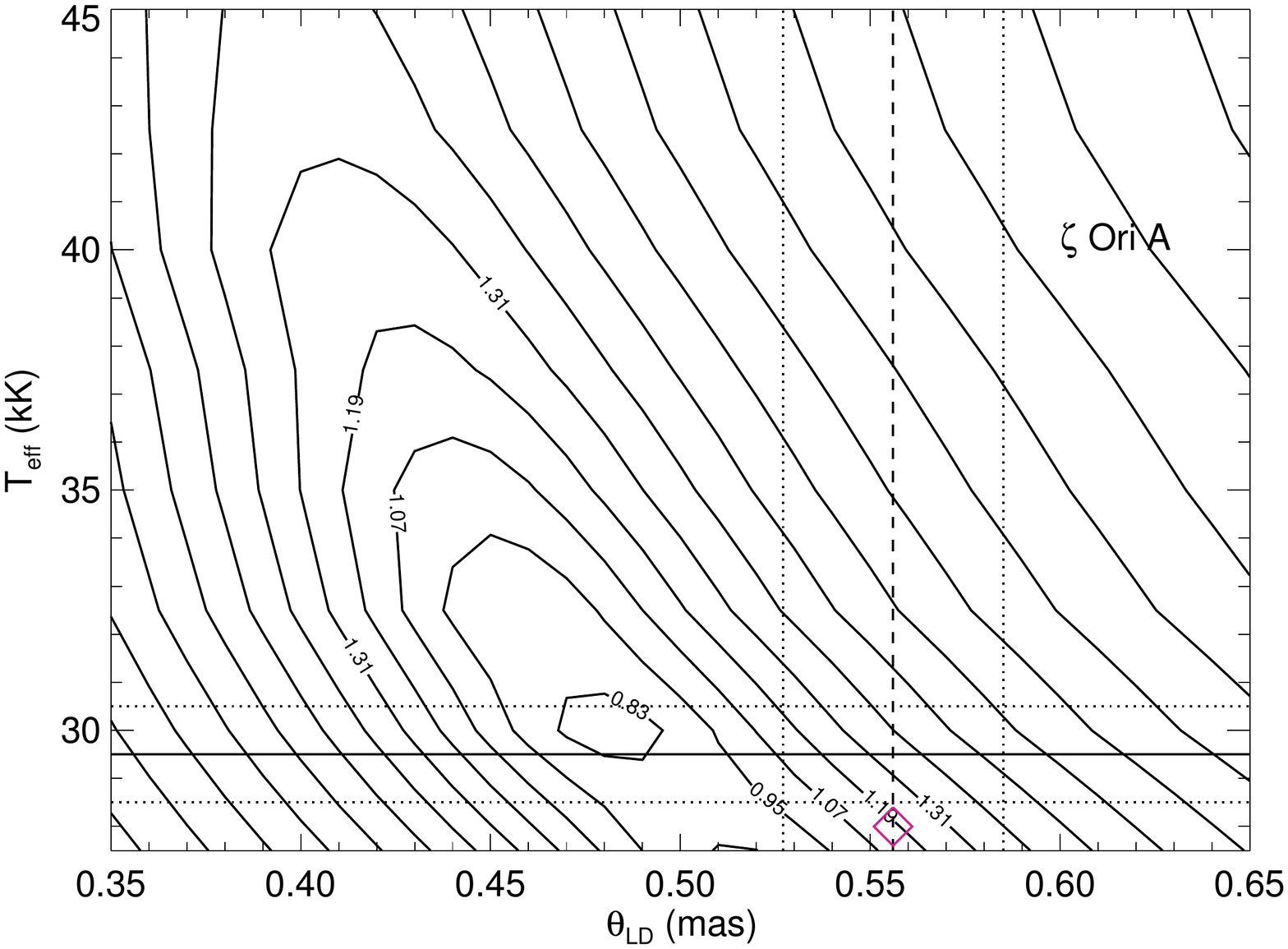}   \\
    \includegraphics[trim=2cm 2cm 2cm 4cm,width=.35\textwidth, angle=90]{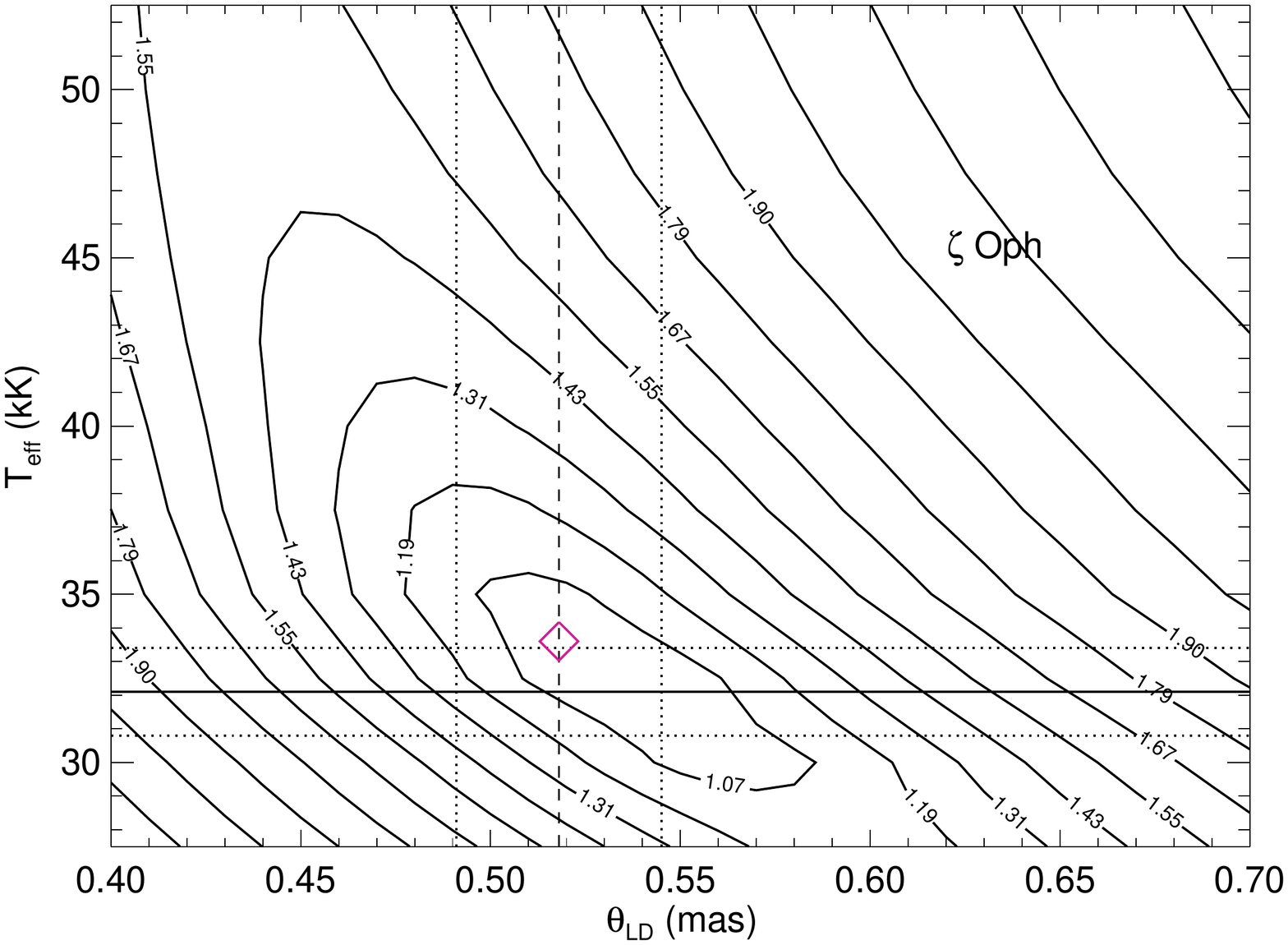} &
    \includegraphics[trim=2cm 2cm 2cm 4cm,width=.35\textwidth, angle=90]{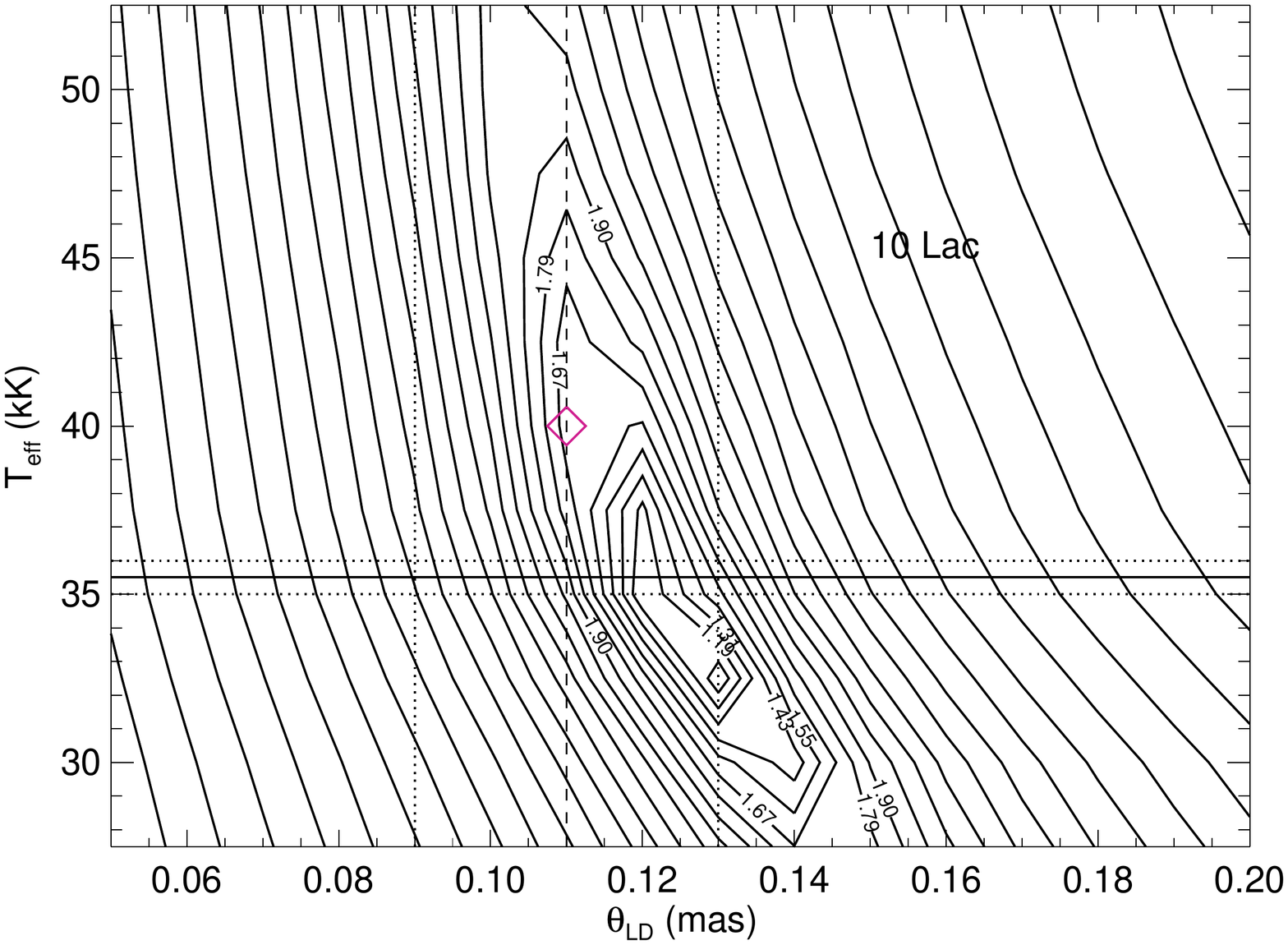}
  \end{tabular}
  \caption{Contour maps of the $\chi^2_\nu$ differences between fitted TLUSTY stellar atmosphere model and observed spectra for each star. Contours are drawn at equal intervals of $\triangle \chi^2_\nu$.  Overplotted are vertical lines showing angular size obtained from our interferometry and horizontal lines showing the average literature temperature. Dotted lines show an error margin of 1$\sigma$ for the angular size and temperature. Diamonds indicate the best fit model temperature for our directly determined angular size.}
  \label{fig:contours}
\end{figure*}

\begin{figure*}[htb]
\centering
  \begin{tabular}{@{}cc@{}}
    \includegraphics[trim=2cm 2cm 2cm 4cm,width=.35\textwidth, angle=90]{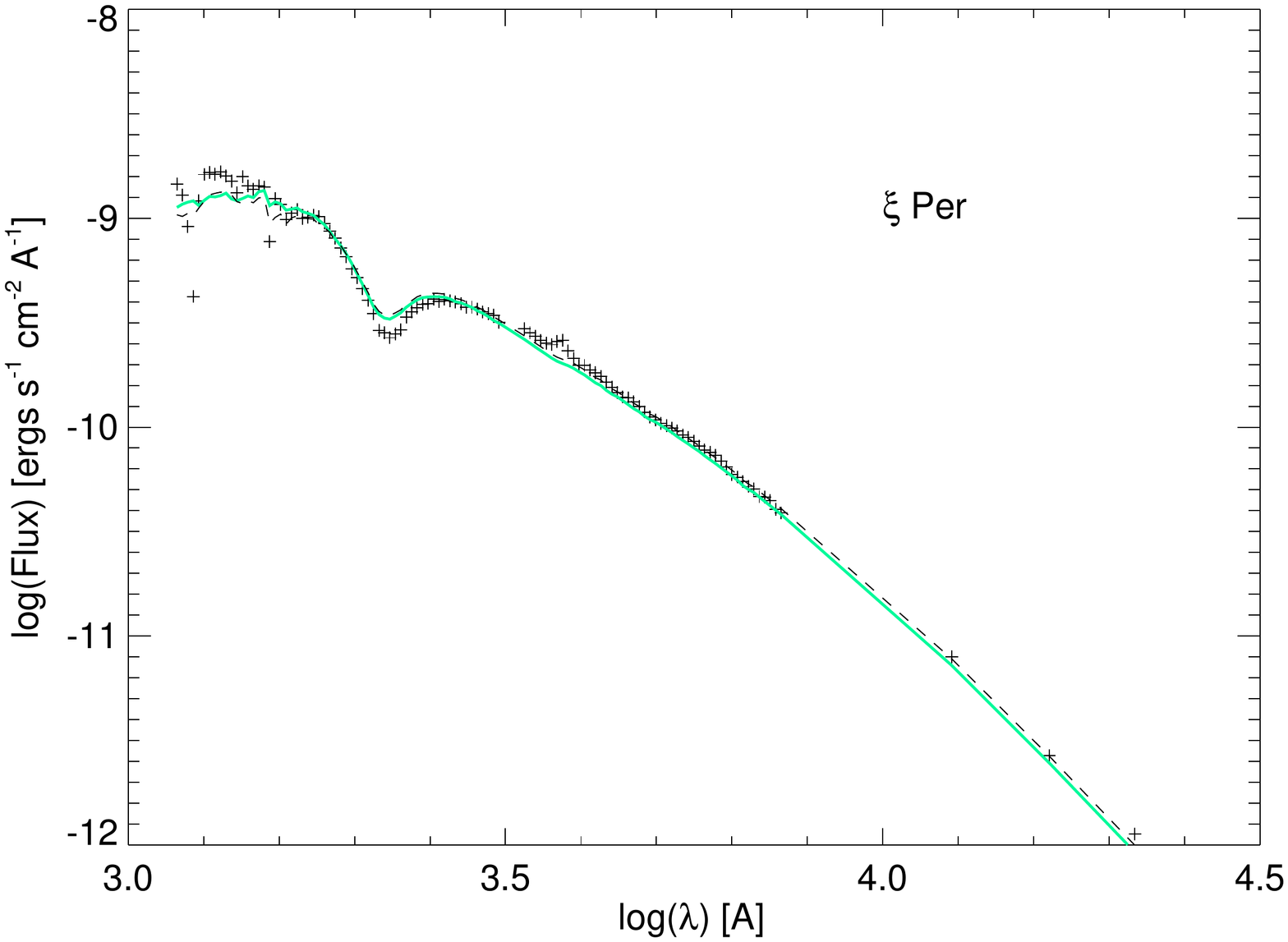} &
    \includegraphics[trim=2cm 2cm 2cm 4cm,width=.35\textwidth, angle=90]{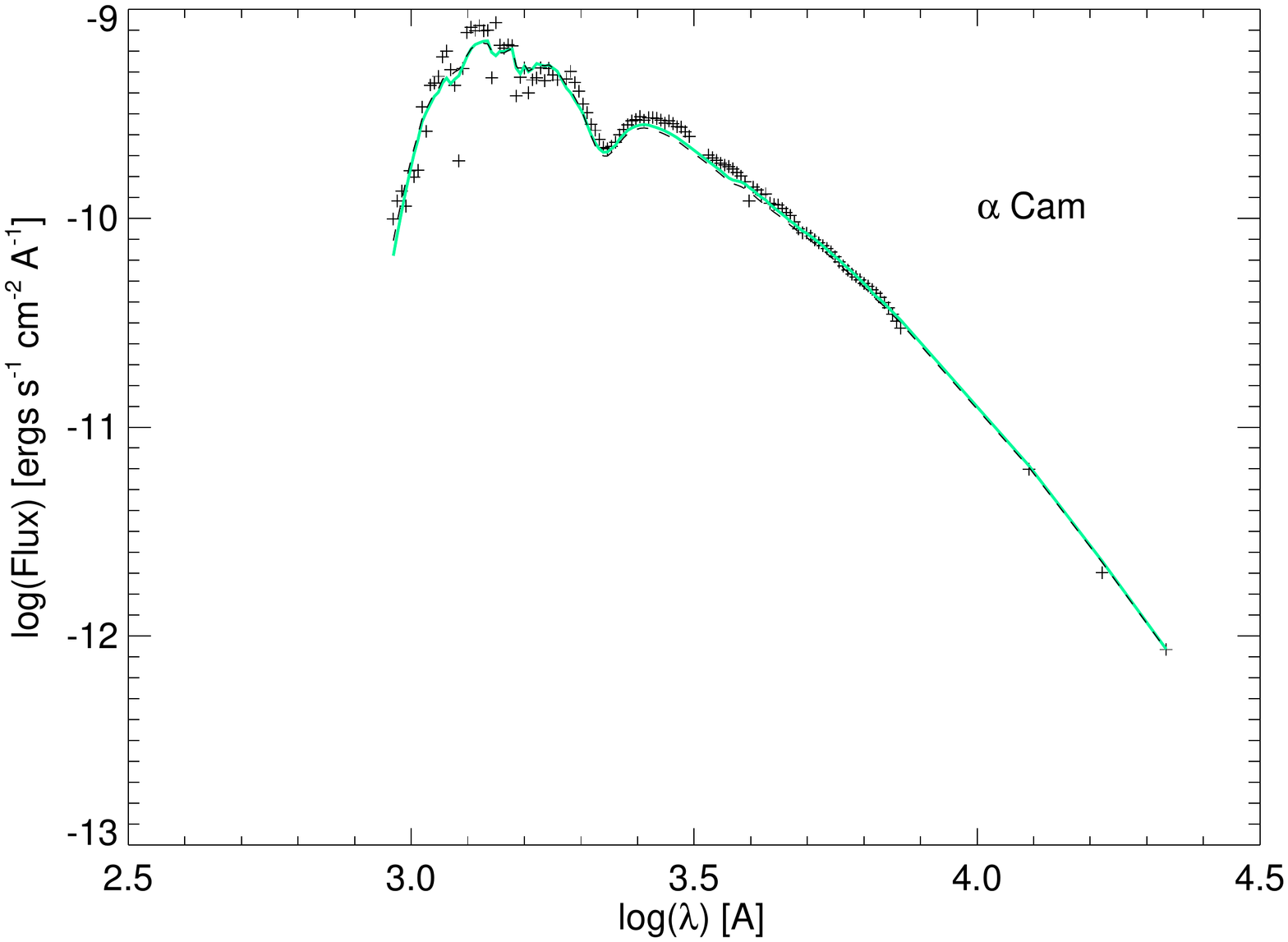} \\
    \includegraphics[trim=2cm 2cm 2cm 4cm,width=.35\textwidth, angle=90]{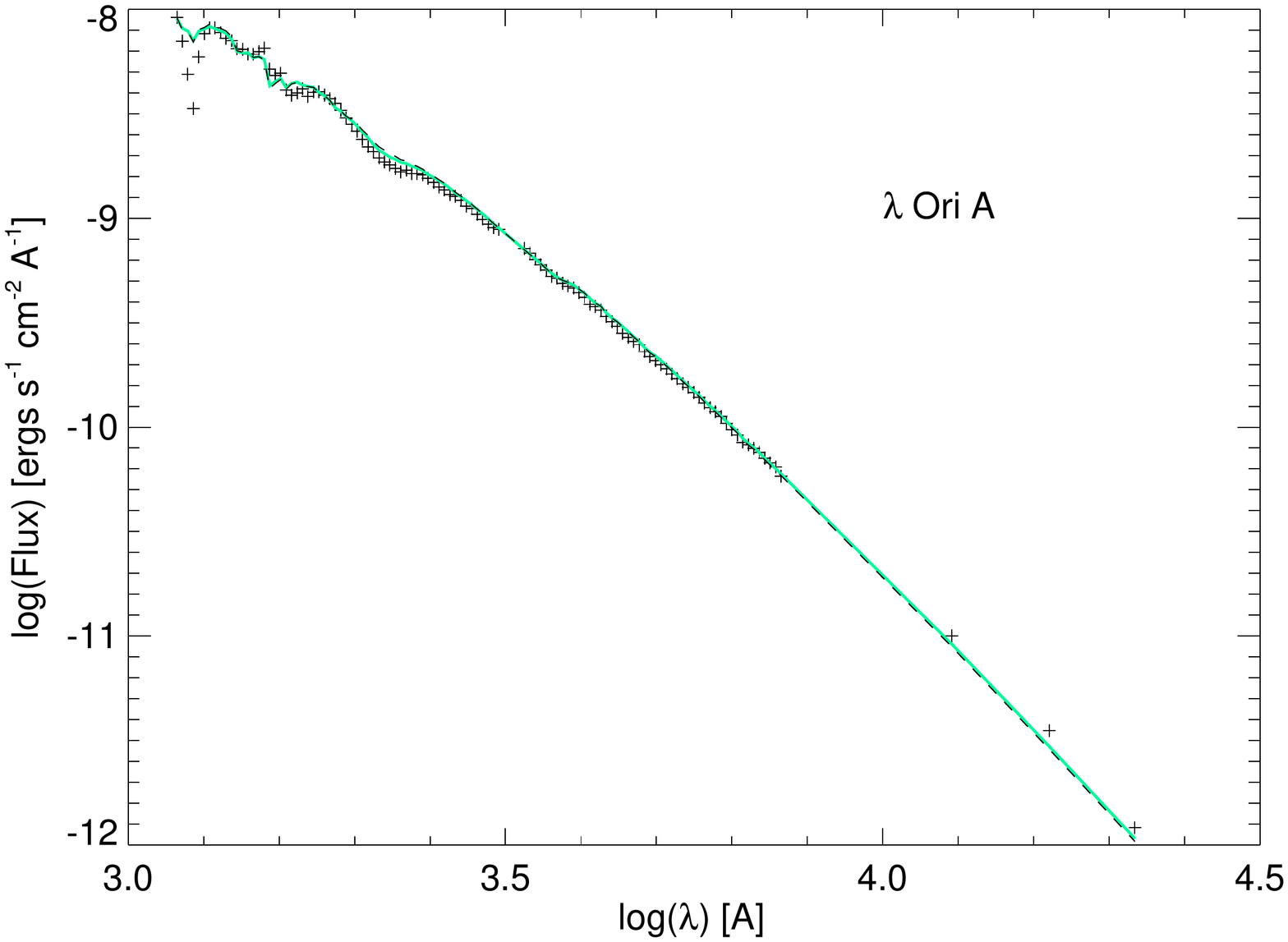} &
    \includegraphics[trim=2cm 2cm 2cm 4cm,width=.35\textwidth, angle=90]{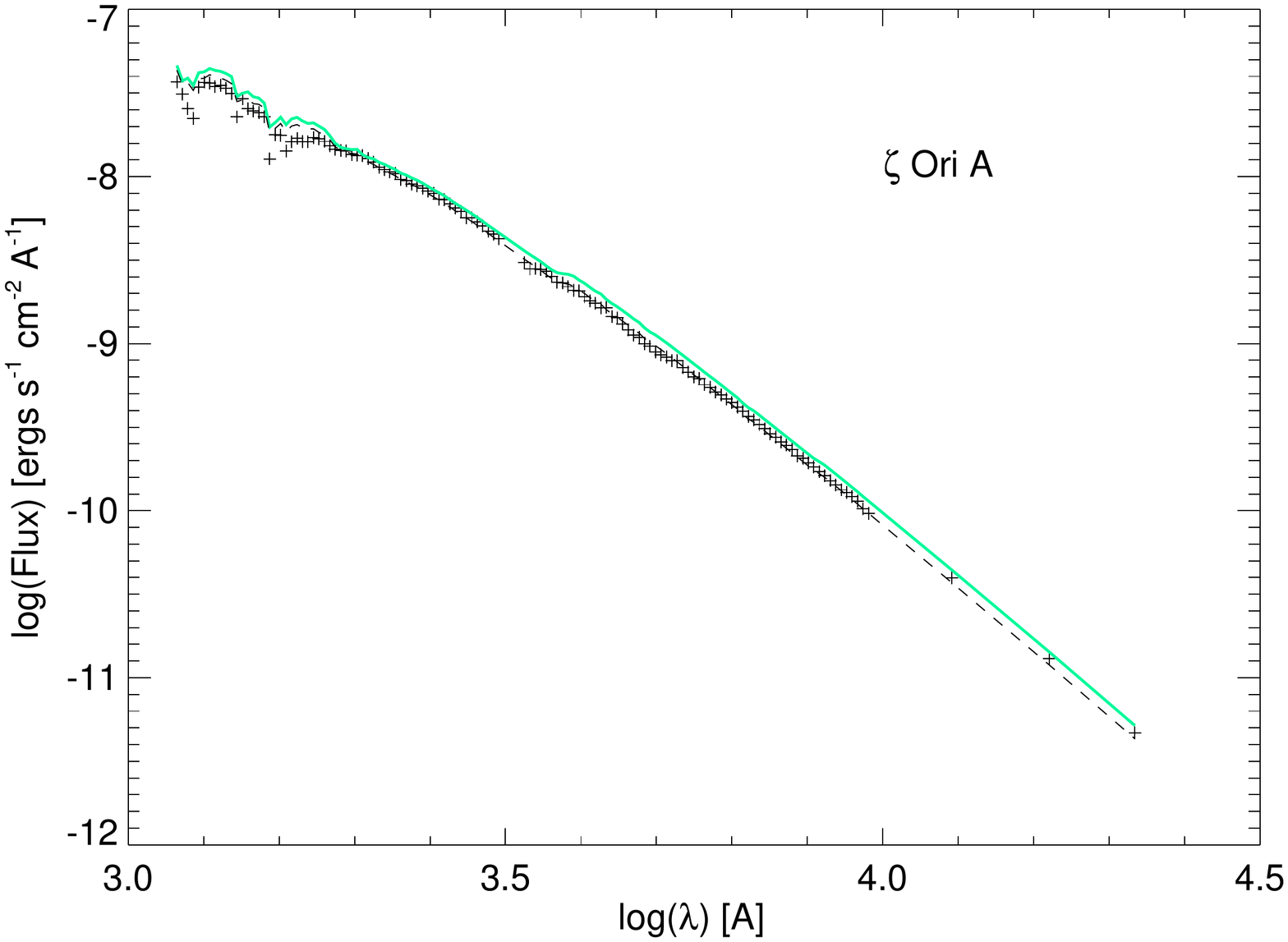}   \\
    \includegraphics[trim=2cm 2cm 2cm 4cm,width=.35\textwidth, angle=90]{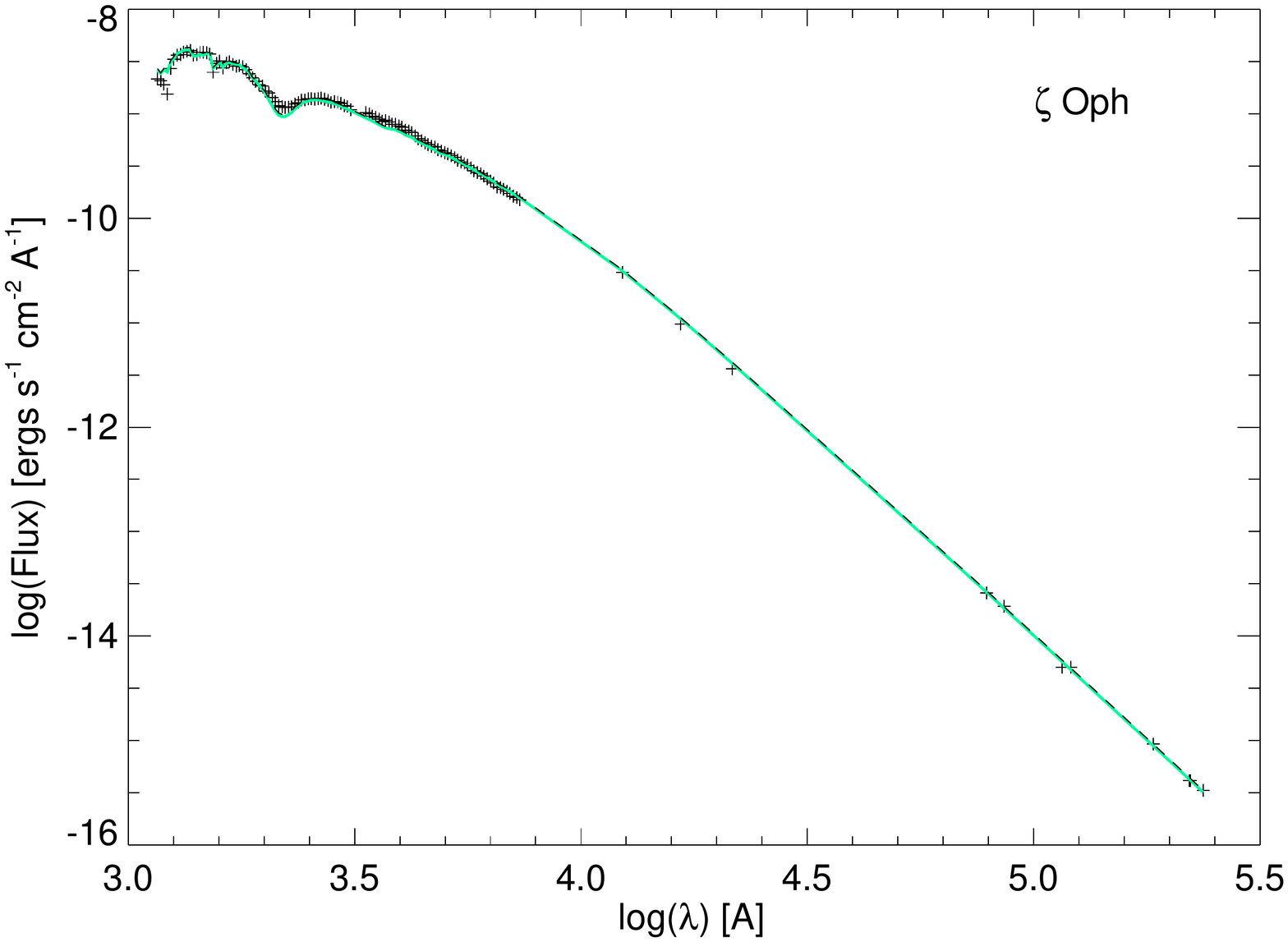} &
    \includegraphics[trim=2cm 2cm 2cm 4cm,width=.35\textwidth, angle=90]{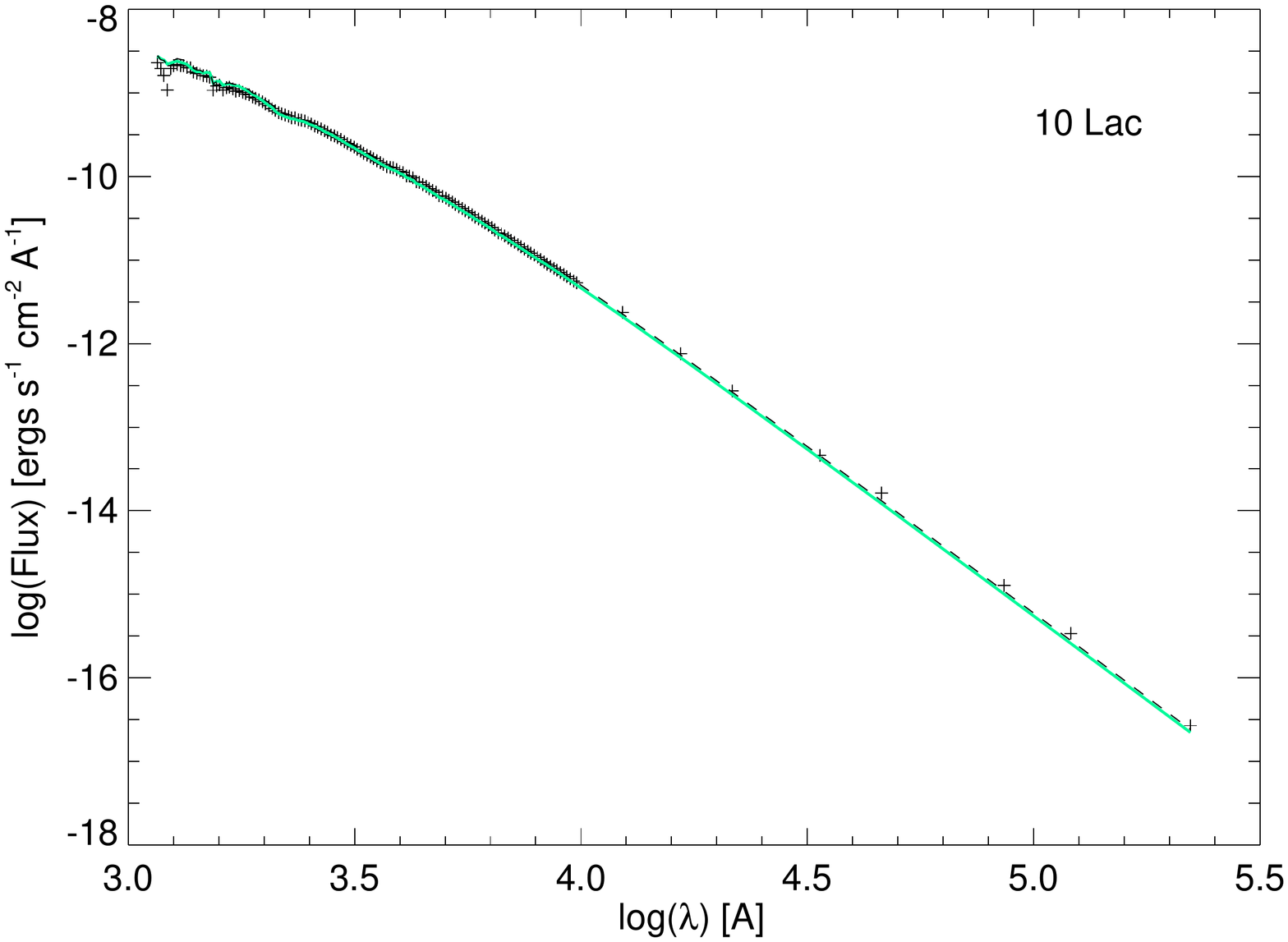}
  \end{tabular}
  \caption{Spectral energy distributions for targets stars with the best fit model shown in the solid green line. Best fit models were chosen using our observed angular diameter from interferometry combined with the best fit temperatures and reddening values found from our contour maps (Fig.\ \ref{fig:contours}). The dashed line indicates the SED derived by setting the average published $T_{\rm eff}$ and determining the best fit angular size $\theta_{LD}(T_{\rm eff})$.}
  \label{fig:sed}
\end{figure*}

\begin{figure}[h]
\centering
 \includegraphics[trim=2cm 2cm 2cm 0cm,width=.55\textwidth,angle=0]{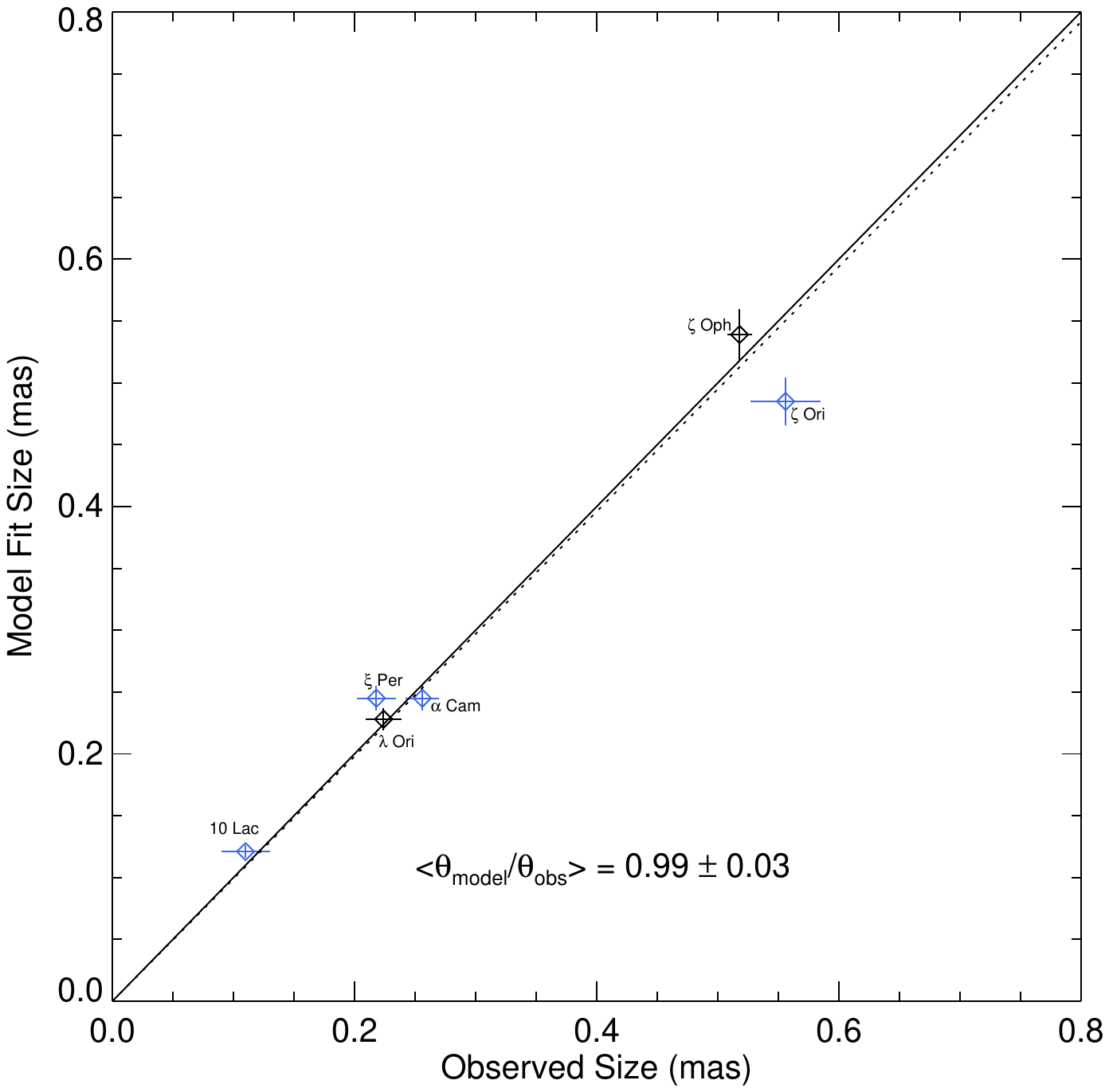}
\vspace*{-65mm}
 \caption{Observed angular size $\theta_{LD}$ compared to the angular size $\theta_{LD}$($T_{\rm eff}$) derived from the published $T_{\rm eff}$ and fit to the SED. The solid line shows a line with a slope of unity for reference, and the dashed line shows the trend for the mean ratio of these diameters. Blue points indicate diameter estimates based upon 
only a single data bracket or an extremely small angular size in the case of 10 Lac.}
 \label{fig:compare}
\end{figure}

\begin{figure}[h]
\centering
 \includegraphics[trim=2cm 2cm 2cm 4cm,width=.55\textwidth,angle=90]{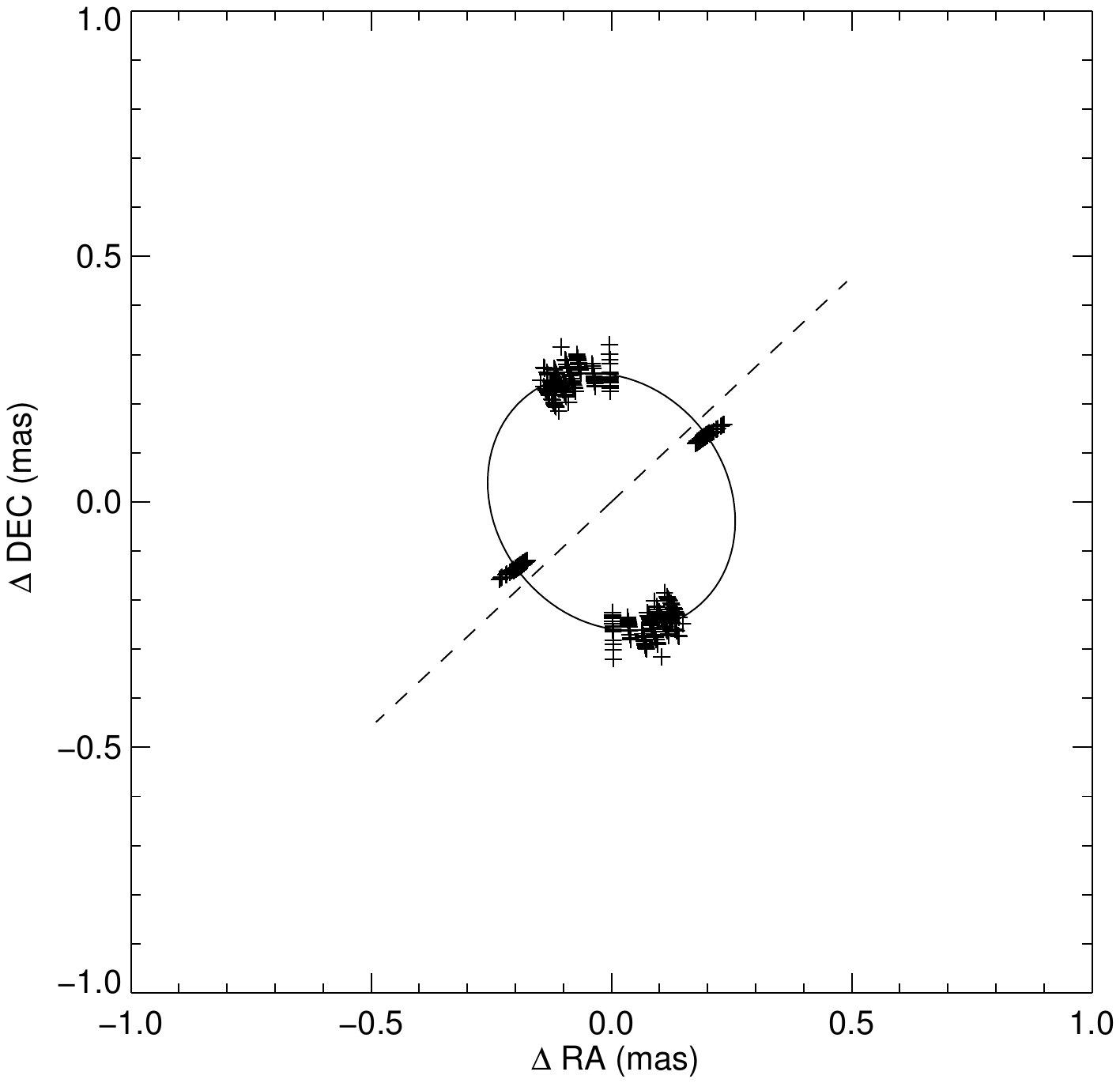}
\vspace*{-3mm}
 \caption{A simple ellipse fitted to our interferometric data for $\zeta$ Oph. Each symbol shows the derived angular size of a limb-darkened star whose visibility equals the observed value, and each is plotted at a position angle derived from the $(u,v)$ spatial frequencies of the observation.  The fit was made of the major and minor axes with the position angle of the minor axis set by published polarimetry. The dashed line shows the adopted rotational axis of the star at a position angle of $132\fdg5$.}
 \label{fig:ellipse}
\end{figure}

Our primary goal was to test whether the angular diameters, spectroscopically determined temperatures, and spectral energy distributions (SEDs) led to a consistent set of stellar parameters. If so, then the intersection of the observed angular diameter (vertical lines) and estimated effective temperature (horizontal line) would cross near the minimum $\chi^2_\nu$ contours in Figure \ref{fig:contours}. We find that the angular diameters estimated from the published effective temperatures and fits of the SEDs ($\theta_{LD}(T_{\rm eff})$ given in the final column of Table \ref{tab:diameters}) are generally in good agreement with the interferometric angular diameters.  A comparison of observed $\theta_{LD}$ and the predicted angular size $\theta_{LD}(T_{\rm eff})$ is shown in Figure \ref{fig:compare}. This figure shows that the average ratio of $\theta_{LD}(T_{\rm eff})$ to $\theta_{LD}$ is approximately 0.99 $\pm$ 0.03. Note that the three stars with the greatest discrepancy in angular size are $\xi$~Per, $\zeta$~Ori~A, and $\zeta$ Oph. 
However, the results for the first two of these stars, $\xi$ Per and $\zeta$ Ori A, are preliminary because the diameters are based upon one data bracket. $\zeta$ Oph is rotationally distorted, so gravity darkening will complicate the meaning of $\theta_{LD}$ and the comparison to a model predicted size may not be entirely valid. 
The targets with many observations, $\lambda$ Ori A and 10 Lac, show good agreement between the interferometric and spectroscopically derived angular diameters. 

%Three out of our six O stars have derived effective temperatures from our interferometric observations that exceed the average literature values. However, due to limited data coverage we consider the effective temperature for $\xi$ Per to be an upper estimate. The two stars with more data coverage that have higher than expected effective temperatures, $\lambda$ Ori A and $\zeta$ Oph, exceed literature temperature estimates by an average of 1.4 kK or 4\%.

We can use the angular size and distance to obtain the stellar radius. 
In Table \ref{tab:distances} we list distance estimates from six sources.  The columns 
labeled by $d_1$ and $d_2$ give distances based upon a calibration of 
absolute magnitude and spectral classification from \citet{Shull1985}
and \citet{Underhill1979}, respectively.  The next estimate $d_3$
is based upon the interstellar \ion{Ca}{2} line strengths \citep{Megier2009}. 
The fourth estimate $d_4$ is derived from the {\it Hipparcos} parallax
\citep{vanLeeuwen2007} with a correction term for the Lutz-Kelker bias 
\citep{Maiz2008}. The next estimate $d_5$ is the distance to 
the host cluster \citep{Kharchenko2005} or association \citep{deZeeuw1999} if the target is a known member.
The final value $d_6$ is derived from the parallaxes from Gaia DR2 \citep{Gaia2018}.  
{\bf The DR2 coverage is incomplete for bright stars (no measurements for $\xi$~Per and $\zeta$~Ori~A) and the errors are relatively large and probably underestimated \citep{Luri2018}.  Consequently, we adopted the parallax of fainter, physical companions whenever possible. }
The DR2 parallaxes for $\lambda$~Ori A, B have large errors, so we adopted the mean of the parallaxes for components C and D.  There is no parallax given for $\zeta$~Ori~A, so we used the listing for the C component.  We applied the parallax of component B for 10~Lac, because of the much larger error associated with component A. \textbf{There was no faint companion available to use for $\alpha$ Cam, so the formal Gaia DR2 parallax error is large.}
The third to last and final columns give the stellar radius corresponding to the Gaia (column 7) and mean distance (column 9), respectively, as derived from our angular size measurements (Table \ref{tab:diameters}).  
Note that the standard deviation of the mean distance may underestimate the total uncertainty if some subsets rely upon the same spectroscopic calibration of spectral type and $M_V$. 
The radius range listed for $\zeta$~Oph (HD~149757) relates to the smaller and larger angular sizes found at different position angles.   
The derived radii generally agree within the range associated with their spectral classifications as given by \citet{Martins2005}, except for the two giants, $\xi$~Per and $\lambda$~Ori~A, which have radii closer to those of main sequence stars.  

Our interferometric survey of O-star diameters yields results that are comparable to the expected diameters based upon the published temperatures and model fits of the SED (Table \ref{tab:diameters}; Fig.\ \ref{fig:compare}), except in the cases of two evolved stars, $\xi$~Per and $\zeta$~Ori~A, where our data coverage is quite limited.  Additional interferometric observations are needed to confirm the smaller and larger diameters found in these two cases, respectively.  
There are several possible explanations for such discrepancies.  First, it is possible that some of our calibrator stars have yet undetected binary companions.  If such companions added incoherent flux to the visibility measurements, then the calibrator visibilities would be depressed and we would infer smaller diameters for the targets.  Additional observations with different calibrators would test this idea.  Secondly, it is possible that the quoted diameters of the calibrators have larger than expected uncertainties.  The JSDC diameter estimates \citep{Chelli2016} rely on relations based upon colors and spectral classification that are set by measured angular sizes.  However, there are relatively few measurements among the B-type and early A-type stars that form our set of calibrators (Table \ref{tab:cals}), so the relations may need more thorough testing.  

Finally, it is possible that the model atmospheres that we used to fit the SEDs underestimate or overestimate the flux over the observed range (1200 \AA ~to 2 $\mu$m), so that the angular diameters inferred from the SED fits are too large or small, respectively.  All our fits rely on the TLUSTY grid of models \citep{Lanz2003} that assume a plane-parallel geometry and that neglect stellar winds.  We made a spot check of the models against the predictions of the non-LTE code CMFGEN \citep{Hillier1998} that can account for spherical geometry and outflows.  The initial comparison was made assuming $T_{\rm eff} = 34.4$ kK, $\log g = 3.5$, solar abundances, and a microturbulent velocity of 10 km~s$^{-1}$.  A comparison of the predicted fluxes between a plane-parallel model from CMFGEN and one from TLUSTY showed better than $1\%$ agreement in flux ratio over the observed range.  This confirms earlier results showing excellent agreement between TLUSTY and CMFGEN \citep{Hillier2001}.  Additional tests indicated that the mean flux over the observed range was about $1\%$ higher for a CMFGEN spherical model with wind loss and was about $2\%$ higher in models with the microturbulent velocity increased to 20 km~s$^{-1}$. Thus, while model differences may account for angular size discrepancies of a few percent, the larger difference seen in the supergiant $\zeta$~Ori~A may require other explanations.  Interferometric observations at longer wavelengths would help test models that include winds, because of the increasing proportion of wind flux into the IR-regime.  We will extend this investigation in a subsequent paper to the cooler B-type stars to further explore the differences between interferometrically and spectroscopically determined parameters for massive stars. 

\acknowledgments
This work is based upon observations obtained with the Georgia State University Center for High Angular Resolution Astronomy Array at Mount Wilson Observatory.  The CHARA Array is supported by the National Science Foundation under Grant No.\ AST-1211929, AST-1411654, AST-1636624, and AST-1715788.  Institutional support has been provided from the GSU College of Arts and Sciences and the GSU Office of the Vice President for Research and Economic Development. KDG expresses her thanks for the financial support of a NASA Georgia Space Grant Fellowship.  
This research has made use of the Jean-Marie Mariotti Center JSDC catalog, the VizieR catalogue access tool and SIMBAD database (CDS, Strasbourg, France), and data from the European Space Agency (ESA) mission
{\it Gaia} (\url{https://www.cosmos.esa.int/gaia}), processed by the 
{\it Gaia} Data Processing and Analysis Consortium (DPAC,
\url{https://www.cosmos.esa.int/web/gaia/dpac/consortium}). 
Funding for the DPAC
has been provided by national institutions, in particular the institutions
participating in the {\it Gaia} Multilateral Agreement.

\software{TLUSTY \citep{Hubeny1995,Lanz2003,Lanz2007}, CMFGEN \citep{Hillier1998}}

% Fig. 1 Visibilities

%Fig. 2 Zeta Ori binary fit inset

% Fig. 3 chi-squared fits of the SED as functions of size and temperature

% Fig. 4 SED fits
\newpage

% Fig. 5 comparison of angular sizes from interferometry and SED

% Fig. 6 UD ellipsoidal fit for zeta Oph

\clearpage

\bibliography{apj-jour,Ostarref}
%%%%%%%%%%%%%%%%%%%%%%%%%%%%%%%%%%%%%%%%%%%%%%%%%%%%%%%%%%%%%%%%%%%%%%%%%%%%%%
\end{document}